\documentclass[journal]{IEEEtran}
\usepackage{tipa}
\usepackage{bbm}
\usepackage{mathrsfs}
\usepackage{url,subfigure,epsfig,graphicx}
\usepackage{amssymb,amsmath}
\usepackage{indentfirst}
\usepackage{algorithmic}
\usepackage{algorithm}
\usepackage{pdfpages}
\usepackage{times,verbatim,amsfonts,amsmath,color}
\usepackage{pifont}
\usepackage{multirow}
\usepackage{booktabs}
\usepackage{adjustbox}
\usepackage{cite}
\usepackage{tikz}
\usepackage[edges]{forest}
\usetikzlibrary{arrows.meta, shapes.geometric, calc, shadows}
\usetikzlibrary{arrows.meta}
\definecolor{hidden-draw}{RGB}{20,68,106}
\definecolor{hidden-pink}{RGB}{255,245,247}
\usepackage{threeparttable}

\usepackage{threeparttable}
\usepackage{pifont} % For checkmark and cross symbols
\usepackage{amssymb} % For \circ and \bullet symbols
\usepackage{wasysym} % For \LEFTcircle symbol
\usepackage{array}
\usepackage{tabularx}
\usepackage{tikz}
\newcommand*\emptycirc[1][0.7ex]{\tikz\draw (0,0) circle (#1);} 
\newcommand*\halfcirc[1][0.7ex]{%
	\begin{tikzpicture}
		\draw[fill] (0,0)-- (90:#1) arc (90:270:#1) -- cycle ;
		\draw (0,0) circle (#1);
\end{tikzpicture}}
\newcommand*\fullcirc[1][0.7ex]{\tikz\fill (0,0) circle (#1);} 

\usepackage{wasysym} % For \LEFTcircle symbol
\usepackage{xcolor} % 加载xcolor包
\definecolor{mycustomcolor}{RGB}{193,16,16} 
\usepackage{hyperref}
\hypersetup{
	colorlinks=true,
	linkcolor=blue,  % 引用数字颜色
	citecolor=mycustomcolor,  % 引用颜色
	urlcolor=blue    % URL颜色
}

\usepackage[switch,pagewise]{lineno}

\usepackage[commandnameprefix=always, defaultcolor=blue]{changes}

\IEEEoverridecommandlockouts
\makeatletter
\def\ps@headings{\def\@oddhead{\mbox{}\scriptsize\rightmark \hfil \thepage}\def\@evenhead{\scriptsize\thepage \hfil \leftmark\mbox{}}\def\@oddfoot{}\def\@evenfoot{}}
\makeatother 

\usepackage{geometry}
\usepackage[pagewise]{lineno}
\usepackage[switch,pagewise]{lineno}

\begin{document}
%\linenumbers
	\title{A Survey on Semantic Communication Networks: Architecture, Security, and Privacy}

 \author{
    \IEEEauthorblockN{
    Shaolong~Guo\IEEEauthorrefmark{2},
    Yuntao~Wang\IEEEauthorrefmark{2}, 
    Ning~Zhang\IEEEauthorrefmark{3},
    Zhou~Su\IEEEauthorrefmark{2}\IEEEauthorrefmark{1}, 
    Tom H. Luan\IEEEauthorrefmark{2},
    Zhiyi~Tian\IEEEauthorrefmark{4},
    and Xuemin (Sherman) Shen\IEEEauthorrefmark{5}
    }\\
    
    \IEEEauthorblockA{
    \IEEEauthorrefmark{2}
    School of Cyber Science and Engineering, Xi'an Jiaotong University, Xi'an, China\\
    \IEEEauthorrefmark{3}
    Department of Electrical and Computer Engineering, University of Windsor, Windsor, Canada\\    
    \IEEEauthorrefmark{4}
    School of Computer Science, University of Technology Sydney, Sydney, Australia\\
    \IEEEauthorrefmark{5}
    Department of Electrical and Computer Engineering, University of Waterloo, Waterloo, Canada\\
    \IEEEauthorrefmark{1}Corresponding Author: zhousu@ieee.org
    }}
    
        \maketitle

	\begin{abstract}
        With the rapid advancement and deployment of intelligent agents and artificial general intelligence (AGI), a fundamental challenge for future networks is enabling efficient communications among agents.
        Unlike traditional human-centric, data-driven communication networks, the primary goal of agent-based communication is to facilitate coordination among agents. Therefore, task comprehension and collaboration become the key objectives of communications, rather than data synchronization. Semantic communication (SemCom) aims to align information and knowledge among agents to expedite task comprehension. {While significant research has been conducted on SemCom for two-agent systems, the development of semantic communication networks (SemComNet) for multi-agent systems remains largely unexplored.} In this paper, we provide a comprehensive and up-to-date survey of SemComNet, focusing on their fundamentals, security, and privacy aspects. We introduce a novel three-layer architecture for multi-agent interaction, comprising the control layer, semantic transmission layer, and cognitive sensing layer. We explore working modes and enabling technologies, and present a taxonomy of security and privacy threats, along with state-of-the-art defense mechanisms. Finally, we outline future research directions, paving the way toward intelligent, robust, and energy-efficient SemComNet. This survey represents the first comprehensive analysis of SemComNet, offering detailed insights into its core principles as well as associated security and privacy challenges.
	\end{abstract}


	\section{Introduction}
        \begin{figure}[!t]\setlength{\abovecaptionskip}{0.0cm}%\vspace{-3mm}
		\centering
		% \hspace{0.2cm}
		\includegraphics[width=9.5cm]{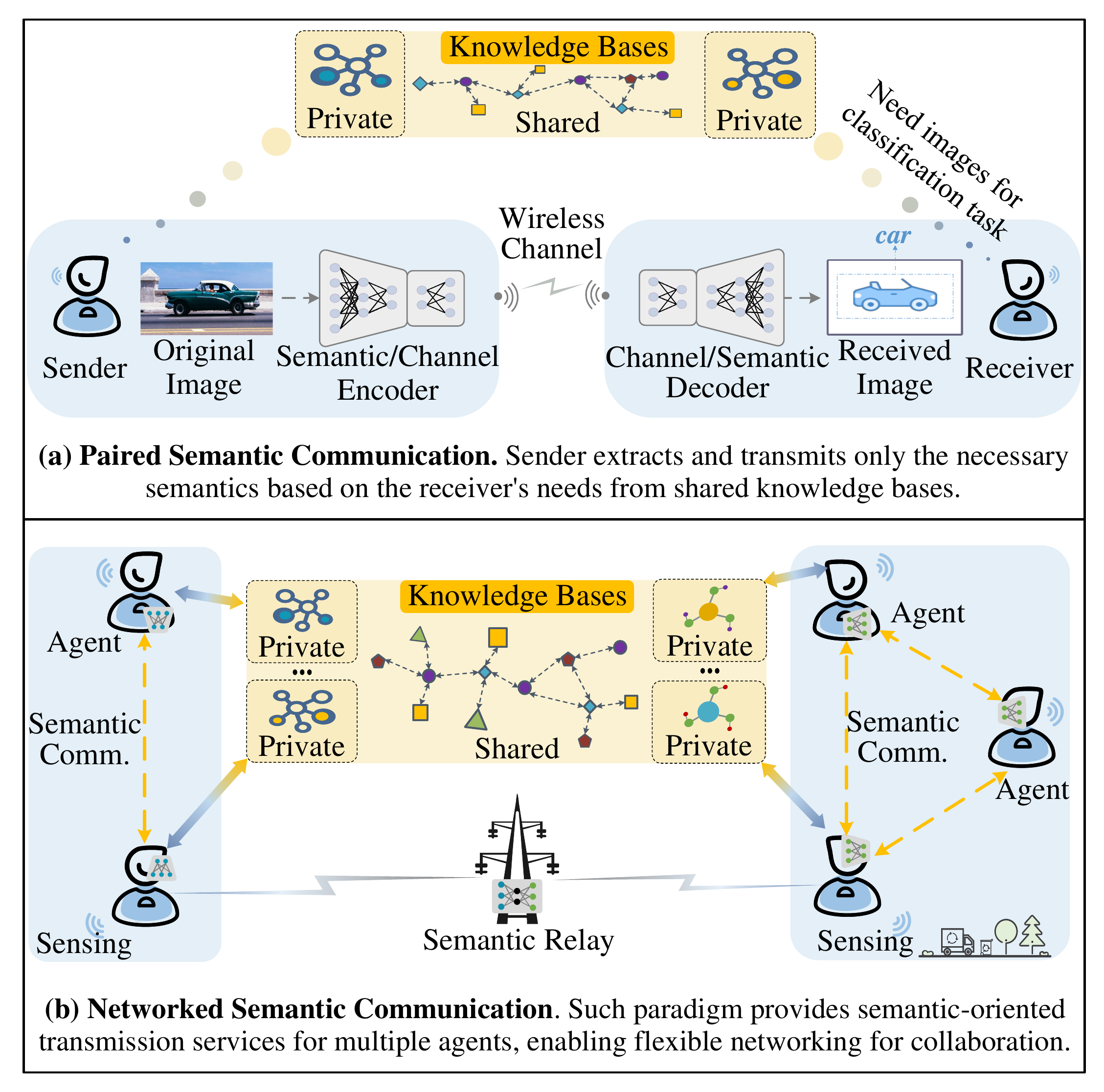}
		\caption{{Illustration of paired semantic communication and networked semantic communication.}}\label{fig:Comparsion} \vspace{-3mm}
	\end{figure}
With the surge of advances in artificial intelligence (AI), our era is evolving from digitization (i.e., connected things) to intellectualization (i.e., connected intelligence).
Under this evolution, communication devices are gradually empowered with human-like intelligence and reasoning capabilities \cite{zhang2022toward, luo2022semantic}. As predicted by Research\&Market \cite{zhang2022toward}, by 2030, over 125 billion AI devices (known as agents\footnote{Beyond the basic communication functions, these agents are equipped with advanced AI algorithms \cite{luo2022semantic}.}) will connect to the Internet, indicating their pivotal role in future communication applications. 

\subsection{Motivation of Semantic Communication}\label{subsec:Challenges}  
{Unlike traditional Internet users requiring data communications, the goal of agent-based communications is to facilitate the task executions of agents.} As a result, the agent-based communication presents three salient features:
\begin{itemize}
\item \textit{Computing-oriented communications}, that the communications between agents should be sufficient to meet the computing requirements for task executions,
\item \textit{Persistant communications}, that communications among agents are typically persistent until all tasks are finished,  
\item \textit{Memory-based communications}, that new information transmitted may depend on the previous information as the new task of agents may depend on previous ones. 
\end{itemize}

{Due to the features of agent-based communications, the traditional data communication framework may no longer fit, which motivates the new paradigm of semantic communication (SemCom) \cite{xie2021deep}.} {SemCom targets computing-oriented communications by first identifying the communication purposes (i.e., computing requirements) of agents; only the desired semantic meanings of data sufficient for computing are transmitted.} To extract the semantic meaning of data, SemCom enables agents to exchange valuable and prior information (called the knowledge bases (KBs) as illustrated in Fig.~\ref{fig:Comparsion}), which is continuously updated to fit the persistent and memory-based features of agent-based communications. In this manner, SemCom can significantly reduce redundant data, facilitating collaborative agents to perform complex tasks.

\begin{table*}[!t]%\scriptsize
	\begin{center}\setlength{\abovecaptionskip}{0cm}
		\caption{Summary of Important Abbreviations in Alphabetical Order}\label{table-abbr}
		\begin{tabular}{ll|ll|ll}
			\toprule
			\textbf{Abbr.}  &\textbf{Definition}            & \textbf{Abbr.} &\textbf{Definition}           &\textbf{Abbr.} &\textbf{Definition} \\
			\midrule
			AC   &Access Control                         &AEs &Adversarial Examples       &AI   &Artificial Intelligence   \\
			AuthN &Authentication                        &Comm. &Communication             &CSI  &Channel State Information \\
			DL &Deep Learning                           &DoS  &Denial of Service          &DP  &Differential Privacy \\
			GAI  &Generative Artificial Intelligence     &GANs &Generative Adversarial Networks &IoT &Internet of Things \\
			IP   &Intellectual Property                  &JSCC &Joint Source-Channel Coding &KBs &Knowledge Bases \\
			KG   &Knowledge Graph                        &ML   &Machine Learning           &PLA &Physical Layer Authentication \\
			PLS &Physical Layer Security                &QKD &Quantum Key Distribution   &QoE  &Quality-of-Experience \\
			RIS &Reconfigurable Intelligent Surface     &RL &Reinforcement Learning     & {SemCom} & {Semantic Communication} \\
			SemComNet &Semantic Communication Networks  &SI  &Semantic Information       &SNR  &Signal-to-Noise \\
			\bottomrule
		\end{tabular}
	\end{center}
\end{table*}
 
Based on the topology, SemCom can be further divided into two categories:

     \textit{Paired Semantic Communication:} As illustrated in Fig.~\ref{fig:Comparsion}(a), the paired SemCom only considers the semantic-aware communication between two agent nodes. {Compared to traditional communication, paired SemCom embraces various benefits, including ultra-high efficiency, enhanced reliability, and strong compatibility.}
     
     {First, SemCom exhibits ultra-high transmission efficiency \cite{luo2022semantic, xie2021deep}, in which pervasive AI and shared KBs allow agents to achieve \emph{understand-before-transmit}. This capability enables SemCom to excel at extracting and transmitting compact semantic information (SI) while filtering out unnecessary data, thereby alleviating the transmission burden \cite{han2023semanticpreserved, xie2021deep}.}
    
    {Second, SemCom demonstrates enhanced transmission reliability, even under harsh communication channel conditions (e.g., high bit error rate) \cite{han2023semanticpreserved, wang2022taskoffloading, zeng2024USV}. Impaired data from noisy wireless channels can be effectively corrected through semantic-level reasoning, guided by accumulated knowledge \cite{huang2023semantica, peng2022robustb}. Besides, the reconstruction of SI in SemCom relies on the matched semantic decoder, which enhances data security by making it more challenging for unauthorized entities to interpret or misuse the transmitted data.} 
    
    {Third, SemCom exhibits strong compatibility with existing communication infrastructures. Note that SemCom does not entirely replace traditional communication techniques. Instead, they can complement and reinforce each other. On the one hand, traditional communication technologies, such as reconfigurable intelligent surface (RIS) \cite{zhao2022semkey} and orthogonal frequency division multiplexing (OFDM) \cite{nan2023physicallayera}, can be integrated into SemCom to facilitate stable and reliable SI transmission. On the other hand, SemCom can be potentially integrated with existing communication infrastructures to enhance the performance of current systems. Researchers have established a field test network for 6G communication, demonstrating that SemCom could achieve 6G transmission capabilities on existing 4G links \cite{ChineseEngineers6G}.}

	\textit{Networked Semantic Communication:}
     The semantic communication networks (SemComNet) {represent} a network of agent nodes \cite{zhang2022wisdomevolutionary, shi2021semanticd} using SemCom. As illustrated in Fig.~\ref{fig:Comparsion}(b), SemComNet could provide semantic-oriented transmission services for multiple agents, enabling efficient and flexible networking \cite{zhang2022toward}. As such, they allow seamless collaboration among agents with shared intents and objectives to execute complex tasks \cite{zhang2022toward}. With SemComNet, a vast array of emerging intelligent applications are supported across key domains, including smart homes, autonomous driving, smart factories, digital twins, virtual reality (VR), Metaverse, and Tactile Internet.
    
    %\vspace{-2mm}
    However, extending paired SemCom to a networked environment serving multiple agents presents inherent challenges, including but not limited:

    \begin{itemize}
        \item {\textit{Training burden}. SemCom benefits from powerful AI models for semantic processing \cite{qin2024ai}, but training these AI-driven semantic models requires high-performance computing and extensive labeled datasets. In addition, these models are typically task-specific and tailored to pairs of transceivers. As the number of agents in SemComNet increases, the resource demands for training and deploying semantic models grow exponentially, particularly when communicating with different agents and handling various tasks. This scaling challenge can quickly overwhelm the available resources of individual agents.}

        \item \textit{KBs consensus}. In paired SemCom, a critical prerequisite and challenge is establishing consensus on KBs between two agents \cite{tian2023asynchronous}. This becomes more complex in SemComNet involving multiple agents, as SemComNet should ensure semantic compatibility and continuously maintain multiple KBs among agents. As new knowledge emerges or existing knowledge becomes outdated, the relevant KBs need to be upgraded. It is unrealistic for any single agent to update and coordinate these KBs independently, necessitating coordinated and collaborative efforts.
        \item {\textit{Topology dynamics}. In SemComNet, nodes can dynamically join and depart from the networks. As a result, both the semantic training models of individuals and the KBs consensus may need to be continuously updated across the entire network to accommodate the network dynamics.} 
    \end{itemize}

    % \\textcolor{red}{In summary, it is necessary to build a unified networking architecture for efficient, semantic-oriented interaction among multiple agents, which covers the efficient knowledge/model sharing, collaborative environment perception, and overall resource scheduling.} 
	
	% \begin{figure}[!t]
	% 	\centering\setlength{\abovecaptionskip}{-0.0cm}
	% 	\includegraphics[width=8.4cm]{fig/Application.pdf}%width=6.8
	% 	\caption{The potential applications of the SemComNet span across various fields, including a) autonomous driving, b) smart homes, and c) the Metaverse. In these applications, multiple collaborative agents can efficiently extract and transmit semantic information from massive information via semantic communication.} \label{fig:Application} \vspace{-2mm}
	% \end{figure}

	 % massive knowledge sharing, the delivery of privacy-sensitive SI, to the AI-driven SI interpretation	
	\subsection{Motivation of Securing SemComNet}\label{subsec:Challenges}     
Due to the new structure of communications, the SemComNet present distinct challenges in security and privacy issues, which should be addressed foremost before its deployment \cite{shen2023secure, peng2024semantic}.
SemComNet face a variety of security risks and privacy breaches across its procedures:
\begin{itemize} 
\item Hardware fragility: Compared to traditional networks, intelligent SemComNet heavily rely on computing resources (e.g., GPUs or other AI accelerators) for various AI-driven semantic codecs training. However, hackers may launch sponge examples attack \cite{shumailov2021sponge} to vanish the acceleration hardware strategies. Such a threat may cause excessive energy consumption and severe performance degradation, and even result in denial of service (DoS) to other agents.
\item Communication risk: SemComNet rely on conveying and interpreting concise SI for efficiency improvement. However, semantics ambiguity (i.e., semantic noise) \cite{peng2022robustb, du2023rethinking, li2022crossmodal} and the fragility nature of semantic models \cite{wang2024privacy, du2023rethinking} expose huge security risks. For instance, hackers could exploit these vulnerabilities via semantic adversarial attacks to mislead transceivers, as validated in \cite{nan2023physicallayera, kang2023adversariala}. Specifically, by adding imperceptible noise to transmitted data and injecting modified data into semantic codecs at either transmitter or wireless channel sides \cite{wang2023surveyc}, hackers could manipulate SI extraction or interpretation drastically. Besides, the subtle modification to transmitted data makes such attacks covert and hard-to-detect \cite{shen2023secure}.
\item KBs breach: Through proactive sensing and knowledge discovery, agents can establish and enrich their private KBs \cite{cheng2024knowledge}. These private KBs typically contain sensitive information (e.g., individual preferences and behaviors), which opens new avenues for privacy-focused hackers \cite{shen2023secure}. For instance, malicious agents may attempt unauthorized access and retrieval of private information through brute-force attack \cite{li2023secure}, which poses a serious threat to confidentiality of SemComNet.
\end{itemize}

Apart from the discussed threats, existing security countermeasures may be ineffective and lacking in adaptability due to intrinsic features of SemComNet, such as \textit{heterogeneous components}, \textit{autonomous intelligence}, and \textit{large-scale structure}.
Specifically, 
1) the integration of heterogeneous components within SemComNet, including various communication modes, data modalities, and agents \& KBs types, presents enormous interoperability difficulties \cite{qiu2018how}.
% \cite{wang2023surveyd}
2) Since SemComNet comprise numerous intelligent agents with high-level autonomy, monitoring and managing their behaviors within the decentralized SemComNet become a challenge. Moreover, agents beyond effective control may emerge as potential security vulnerabilities, leading to exploratory attacks and data breaches.
3) The key prerequisite for reliable SemCom is the frequent querying of KBs to synchronize context and establish consensus between sender and receiver \cite{tian2023asynchronous}, which is challenging in the large-scale and time-varying SemComNet among diverse participants. Besides, attackers can exacerbate this challenge through attacks such as desynchronization of KBs and KBs poisoning attacks \cite{shen2023secure}.
    %4) The flexible SemComNet structure allows agents adapt their roles and modes to fulfill different tasks, which presents intricacies in ensuring trust management and security \cite{wang2022surveyb}.
    %4) As the interactions in SemComNet are task-driven (i.e., agents adapt their interaction roles and modes to fulfill different tasks), which needs a flexible and adaptive network structure. Such dynamic SemComNet presents intricacies in ensuring trust management and security \cite{wang2022surveyb}. 
    % 4) As the interactions in SemComNet are complex (including explicit interaction with other entities and implicit interaction with KBs) and task-driven (i.e., agents adapt their interaction roles and modes to fulfill different tasks), there is a need for flexible and adaptive network structure, which in turn lead to a dynamic topology. In the task-driven SemComNet, agents adapt their interaction modes and roles to fulfill different collaborative tasks.
	Consequently, it is imminent to develop a secure, trustworthy, and green SemComNet to overcome these challenges.
	
\begin{table*}[!t]
	\centering
    \renewcommand{\arraystretch}{1.2} % Adjust row spacing
	\resizebox{\textwidth}{!}{%
		\begin{threeparttable}
			\caption{{A Comparison of Contribution Between Our Survey and Relevant Surveys}}
			\label{contribution}
        \begin{tabular}{|c|c|l|c|c|c|c|c|c|}
            \hline
            \textbf{Year} & \textbf{Refs.} & \textbf{Contribution} & {\textbf{A}} & {\textbf{B}} & {\textbf{C}} & {\textbf{D}} & {\textbf{E}} & {\textbf{F}} \\ \hline
            
            2021 & \cite{lan2021whatd} & \parbox{13.5cm}{Survey on wireless SemCom including enabling technologies, components, and design approaches.} & \fullcirc & \halfcirc & \fullcirc & \emptycirc & \emptycirc & \emptycirc \\ \hline
            
            2021 & \cite{shi2021semanticd} & \parbox{13.5cm}{Discussions on the semantic-aware network framework including its architecture and open problems.} & \fullcirc & \fullcirc & \emptycirc & \halfcirc & \emptycirc & \emptycirc \\ \hline
            
            2022 & \cite{luo2022semantic} & \parbox{13.5cm}{Discuss recent advancements in workflow, use cases, and key issues of end-to-end SemCom.} & \fullcirc & \fullcirc & \fullcirc & \emptycirc & \emptycirc & \emptycirc \\ \hline
            
            2022 & \cite{qin2022semantic} & \parbox{13.5cm}{Explore the principles and challenges of DL-driven SemCom for transmitting multi-modal data.} & \fullcirc & \fullcirc & \halfcirc & \emptycirc & \emptycirc & \emptycirc \\ \hline
            
            2022 & \cite{uysal2022semantica} & \parbox{13.5cm}{Review on semantic-aware communication from a data significance aspect.} & \fullcirc & \emptycirc & \fullcirc & \emptycirc & \emptycirc & \emptycirc \\ \hline
            
            2022 & \cite{zhang2022wisdomevolutionary} & \parbox{13.5cm}{Discuss the key components and architecture of wisdom SemCom and review application scenarios and open issues.} & \fullcirc & \halfcirc & \fullcirc & \fullcirc & \emptycirc & \emptycirc \\ \hline
            
            2023 & \cite{gunduz2023transmitting} & \parbox{13.5cm}{A systematic taxonomy in semantic and task-oriented communication from an information-theoretic perspective.} & \fullcirc & \fullcirc & \emptycirc & \emptycirc & \emptycirc & \emptycirc \\ \hline
            
            2023 & \cite{yang2023semanticb} & \parbox{13.5cm}{Comprehensive survey on fundamentals, potential applications, and open issues of SemCom-driven 6G systems.} & \fullcirc & \fullcirc & \fullcirc & \halfcirc & \emptycirc & \emptycirc \\ \hline
            
            2024 & \cite{lu2023semanticsempowered} & \parbox{13.5cm}{A tutorial-cum-tutorial on AI-driven SemCom from the ecosystem, frameworks, techniques, to application.} & \fullcirc & \fullcirc & \fullcirc & \emptycirc & \emptycirc & \emptycirc \\ \hline

            2024 & \cite{trevlakis2024natively} & {\parbox{13.5cm}{Discuss paired SemCom design, theoretical framework, as well as SemCom networking architectures.}} & \fullcirc & \fullcirc & \emptycirc & \fullcirc & \emptycirc & \emptycirc  \\ \hline

            % 2024 & \cite{chaccour2024less} & {\parbox{13.5cm}{Tutorial on pillars technologies, metrics, and roadmap towards reasoning-driven end-to-end SemCom systems.}} & \fullcirc & \halfcirc & \emptycirc & \halfcirc & \emptycirc & \emptycirc  \\ \hline

            Now & \textbf{Ours} & \parbox{13.5cm}{Comprehensive survey of the fundamentals, security, and privacy of SemComNet, discussions on the general architecture, working modes, application scenarios, and security/privacy threats of the SemComNet, review on critical challenges, potentially advanced solutions, and research directions in building future SemComNet.} & \fullcirc & \fullcirc & \fullcirc & \fullcirc & \fullcirc & \fullcirc \\ \hline
        \end{tabular}
			\begin{tablenotes}
			{	\item \textbf{\textit{A}}: Paired SemCom Design  \  \textbf{\textit{B}}: Multi-modal Data  \ \textbf{\textit{C}}: Use Cases \  \textbf{\textit{D}}: Networked SemCom Architecture   \textbf{\textit{E}}: Security \& Privacy Threats  \ \textbf{\textit{F}}: Security \& Privacy Solutions
                    \item  \fullcirc \ : fully included \ \halfcirc \ : partially included \ \emptycirc  \ : not included}
			\end{tablenotes}
		\end{threeparttable}
	}
\end{table*}
	
	\subsection{Related Works}\label{subsec:Contributions}
	The field of SemCom has garnered considerable research attention, giving rise to several surveys that explore its different aspects to date.
	For instance,
	% Kalfa \emph{et al.} \cite{kalfa2021goalorientedc} investigate the applications and future challenges from the goal-oriented semantic signal processing perspective.
	Lan \emph{et al.} \cite{lan2021whatd} review the key components, enabling technologies, and design approaches of wireless SemCom.
	Shi \emph{et al.} \cite{shi2021semanticd} introduce the classic SemCom and a semantic-aware network model including its architecture and open problems.
	% Chaccour \emph{et al.} \cite{chaccour2022lessa} discuss the pillars technologies, key performance indicators, and application of novel reasoning-driven SemCom.
	Luo \emph{et al.} \cite{luo2022semantic} present recent advancements in DL-based end-to-end SemCom, covering various use cases and future trends.
	Qin \emph{et al.} \cite{qin2022semantic}
	comprehensively survey the principles and challenges of DL-driven SemCom systems for multi-modal data transmission. 
	Uysal \emph{et al.} \cite{uysal2022semantica} provide a view of semantic-aware networked architecture from the data importance aspect.
	Zhang \emph{et al.} \cite{zhang2022wisdomevolutionary} investigate an AI-native SemCom-empowered network and discuss the prototyping, potential application scenarios, and key challenges.
	Gunduz \emph{et al.}  \cite{gunduz2023transmitting} provide a holistic review of semantic and task-oriented communication from an information-theoretic perspective.
	Yang \emph{et al.} \cite{yang2023semanticb} systematically review the fundamentals of SemCom, potential applications, and open issues in 6G communication systems.
	Lu \emph{et al.} \cite{lu2023semanticsempowered} present a survey-cum-tutorial on AI-empowered SemCom technology from the ecosystem, frameworks, techniques, to application. 
    Trevlakis \emph{et al.} \cite{trevlakis2024natively} provide a comprehensive analysis of SemCom in 6G networks which covers semantic knowledge, timeliness, and information theory aspects, along with proposing a networked SemCom architecture.
    % Chaccour \emph{et al.} \cite{chaccour2024less} discuss the pillars technologies, key performance indicators, and comprehensive roadmap towards end-to-end reasoning-driven SemCom system.}
    {However, as shown in Table~\ref{contribution}, existing surveys \cite{lan2021whatd, luo2022semantic, qin2022semantic, uysal2022semantica, gunduz2023transmitting, lu2023semanticsempowered} primarily focus on constructing SemCom between paired agents, neglecting the networked paradigm for collaborative agents. Although the works 
    \cite{shi2021semanticd, zhang2022wisdomevolutionary, yang2023semanticb, trevlakis2024natively} explore the networked SemCom, they fail to cover critical aspects such as the working modes comprehensively, use cases of SemComNet, and its security and privacy considerations.} To bridge these research gaps, our paper provides a systematic review of the fundamentals of SemComNet, including an in-depth discussion of its three-tier architecture, working modes, enabling technologies, and practical use cases. {Besides, with the growing significance of SemComNet, studying its security and privacy issues becomes essential for real-world deployment, which has been largely overlooked in previous surveys. Our survey offers a taxonomy of security and privacy threats across SemComNet's three layers and explores both existing and potential defense mechanisms}. Table~\ref{contribution} provides a summary of our work's contributions about previous surveys in the field of SemCom. 

    % Different from them, our work represents the first comprehensive survey of networked SemCom and its security aspect. As shown in Table~\ref{contribution}, we cover  We expand the scope from mere individual node interactions to a more complex, interconnected network structure.Particularly, this paper focuses on a comprehensive investigation of SemComNet from the perspectives of architecture, working modes, characteristics, and supporting technologies. Furthermore, with the emergence of SemComNet, it is necessary and urgent to study its security in future communication systems. This survey overviews the security threats, as well as existing and potential defense solutions for constructing a secure SemComNet, in which little survey effort has been made. 
	
	\begin{figure}[!t]
		\centering \setlength{\abovecaptionskip}{-0.1cm}
        \includegraphics[width=9.5cm]{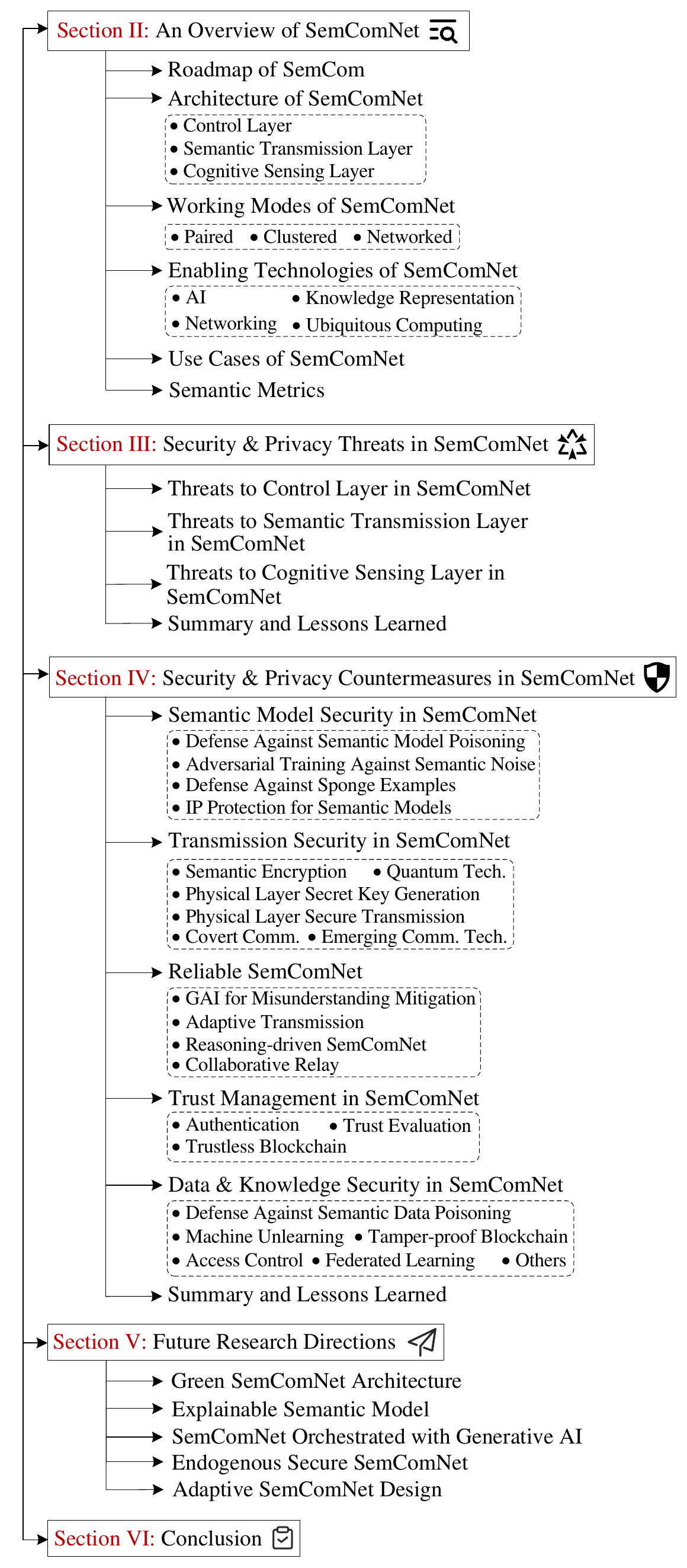}
		\caption{{Organization structure of this paper.}}\label{fig:organization}\vspace{-5mm}
	\end{figure}
	
	In this paper, we provide a systematic review of the general architecture, as well as the potential security/privacy threats, state-of-the-art countermeasures, and future trends of SemComNet.
    To the best of our knowledge, this is the first survey paper in the literature presenting the fundamentals of SemComNet for multi-agent interaction, as well as an in-depth analysis of the security and privacy aspects.
	The main target of our work is
	1) to provide readers with a general understanding of SemComNet, as well as the scope and implications of security threats and challenges in SemComNet, 
	and 2) to highlight effective/potential paths and methods to prevent these threats, thereby fostering the secure SemComNet for various intelligent applications. The contributions of this work are four-fold.
	\begin{itemize}
		\item \textit{Overviews of SemComNet (Section II):} 
		A three-tier SemComNet architecture including the \emph{control layer}, the \emph{semantic transmission layer}, and the \emph{cognitive sensing layer} is first introduced. Then we elaborate on working modes (i.e., \emph{paired}, \emph{clustered}, and \emph{networked}), along with its use cases, enabling technologies, and evaluation metrics.
		\item \textit{Security Threats and Challenges (Section III):} We present a taxonomy of security and privacy threats in the SemComNet across three layers and investigate the critical challenges to address them.
		\item \textit{State-of-the-Art Countermeasures (Section IV):} We review the existing defense solutions from both academia and industry, as well as assess their potential for establishing the secure, trustworthy, reliable, and privacy-preserving SemComNet.
		\item \textit{Future Research Directions (Section V):} We discuss open research opportunities and outline future directions of SemComNet aiming to facilitate new starters conducting research in this field.	
	\end{itemize}
	
	Finally, we draw conclusions in Section VI. The organizational structure of this work is depicted in Fig.~\ref{fig:organization}, and Table~\ref{table-abbr} summarizes a list of key acronyms.

	\section{An Overview of Semantic Communication Networks}\label{sec:OVERVIEW}
	In this section, we first clarify the roadmap of semantic communication (SemCom) and the motivation of the proposed semantic communication networks (SemComNet). Then, we introduce an envisaged architecture for SemComNet, in which working modes and enabling technologies are elaborated.

        \subsubsection{Roadmap of SemCom}
	{
        Research in SemCom follows three major trends: the shift from single-modal to multi-modal data transmission, the evolution from single-task to multi-task execution, and the progression from paired SemCom to networked SemCom.}
        \begin{itemize}

        \item 	{
\emph{From Single-modal to Multi-modal Data Transmission.}	
        SemCom research initially concentrated on single-modal data, such as text, images, audio, or video, as shown in Table~\ref{tab:datasets}. While efficient, single-modal systems have limitations in fully capturing the diversity of communication needs. The shift to multi-modal transmission integrates data from different sources (e.g., combining text with images or audio), which enhances the semantic richness. However, this comes with increased system complexity, particularly in SI extraction, data alignment, and fusion across modalities.}
        \item	{
         \emph{From Single-task to Multi-tasks Execution.}
         SemCom systems have demonstrated exceptional performance in single-task scenarios such as image classification \cite{xie2021deep} and speech recognition \cite{han2023semanticpreserved} in both efficiency and accuracy. However, as communication scenarios become increasingly complex, the focus of SemCom is shifting towards the simultaneous handling of multiple tasks \cite{zhang2024unified, tian2023asynchronous}.
         For instance, in SemCom-empowered autonomous driving, multiple tasks such as environmental perception and object detection need to happen concurrently with minimal latency. 
         Nevertheless, transitioning to multi-task SemCom introduces several challenges. For instance, task interference may arise when different tasks require varying features, potentially leading to performance degradation \cite{zhang2024unified}. }
         
        \item 	{
        \emph{From Paired SemCom to Networked SemCom}.
        In the future, SemCom is evolving from a paired paradigm \cite{xie2021deep, luo2022semantic, zhang2024unified} to a networked paradigm \cite{shi2021semanticd}.  This transition to SemComNet introduces inherent challenges (e.g., training burden, KBs consensus, and topology dynamics, as discussed in the Introduction), which significantly increase system complexity and require effective interactions and high-level coordination among multiple agents.
        For instance, a critical prerequisite for the success of SemComNet is the establishment and continuous updating of shared KBs among these agents \cite{shi2021semanticd}, which demands collaborative efforts to gather environmental semantics \cite{qin2023generalized}. Despite the complex and long-term challenges of achieving networked SemCom, the potential benefits for multi-agent communication scenarios are substantial.}

        \end{itemize}

\begin{table}[!t]
\centering
\renewcommand{\arraystretch}{1.2} % Adjust row spacing
\caption{{Common and typical datasets used by SemCom divided by modality type}}
\label{tab:datasets}
\begin{tabular}{|c|p{7cm}|} % Define column widths and alignment
\hline
\textbf{\begin{tabular}[c]{@{}c@{}}{Data} \\ {Modality}\end{tabular}} & \textbf{{Typical Datasets}} \\
\hline
{Text} & {European Parliament \cite{xie2021deep}; WMT 2018 News \cite{xie2022taskoriented}} \\
\hline
{Image} & \begin{tabular}[c]{@{}l@{}} {MNIST, CIFAR-10, SVHN, and USPS \cite{zhang2023deep}; Stanford Online} \\ {Products, CUB-200-2011, Cars1, and In-Shop Clothes \cite{xie2022taskoriented}} \end{tabular} \\
\hline
{Audio} & {Librispeech \cite{han2023semanticpreserved}; LJSpeech \cite{weng2023deep}} \\
\hline
{Video} &  {CamVid \cite{gong2023adaptive}; Vimeo90K, HEVC test dataset, and UVG \cite{wang2023wireless}} \\
\hline
\end{tabular}
\end{table}

	% \subsection{Motivation of SemComNet}
	%Paired SemCom proves effective for direct and intelligent interactions between two agents \cite{qin2022semantic,luo2022semantic,xie2021deep}. However, in practical scenarios such as Metaverse and collaborative autonomous driving, where multiple agents need effective interactions and high-level coordination (including resources and tasks).
	%Furthermore, a critical prerequisite for successful SemCom is the establishment and continuous updating of common KBs between the sender and receiver \cite{shi2021semanticd}. This necessitates collaborative efforts in collecting environmental semantics \cite{qin2023generalized} and updating shared knowledge when unknown semantic entities emerge, which task exceeds the capabilities of a single agent.
	%Additionally, the accuracy of semantic understanding is influenced by various contextual elements, such as user emotions and intents \cite{shi2021semanticd}. Capturing these elements from a single agent is insufficient, while employing multi-user perception within a networked system may be effective.
	%In summary, it is necessary to build a unified networking architecture for efficient, semantic-oriented interaction among multiple agents.
	
	In this paper, we extend the concept of paired SemCom to networked SemCom \cite{shi2021semanticd,uysal2022semantica} and propose the state-of-the-art SemComNet. As illustrated in Fig.~\ref{fig:Comparsion}, it represents a semantic-oriented networking paradigm that could better support connected multi-agent intelligent interaction \cite{lu2023semanticsempowered} via SemCom.
	In the proposed architecture (refer to Sect.~\ref{subsec:Architecture}), by leveraging KBs maintenance, task scheduling, resource allocation, SI delivery, as well as environmental perception and agent cognition, SemComNet effectively enhance the collaborative semantic transmission abilities of agents \cite{lu2023semanticsempowered}. As such, SemComNet enable multiple agents to efficiently realize semantic interactions and collectively accomplish complex tasks for 6G and beyond applications \cite{zhang2022wisdomevolutionary}.
	
	According to the executed tasks, we further classify the SemComNet into \emph{clustered SemCom} and \emph{networked SemCom}. The former allows multiple agents collaboratively to accomplish a common task, while the latter is expected to handle more complex application scenarios that comprise various transmission tasks \cite{zhang2024unified} and to support coordination across these tasks (refer to Sect.~\ref{subsec:workmode} for more details).

	\subsection{Architecture of SemComNet}\label{subsec:Architecture}
\begin{figure*}[!t]\setlength{\abovecaptionskip}{-0.0cm}
		\centering
		\includegraphics[width=0.75\textwidth]{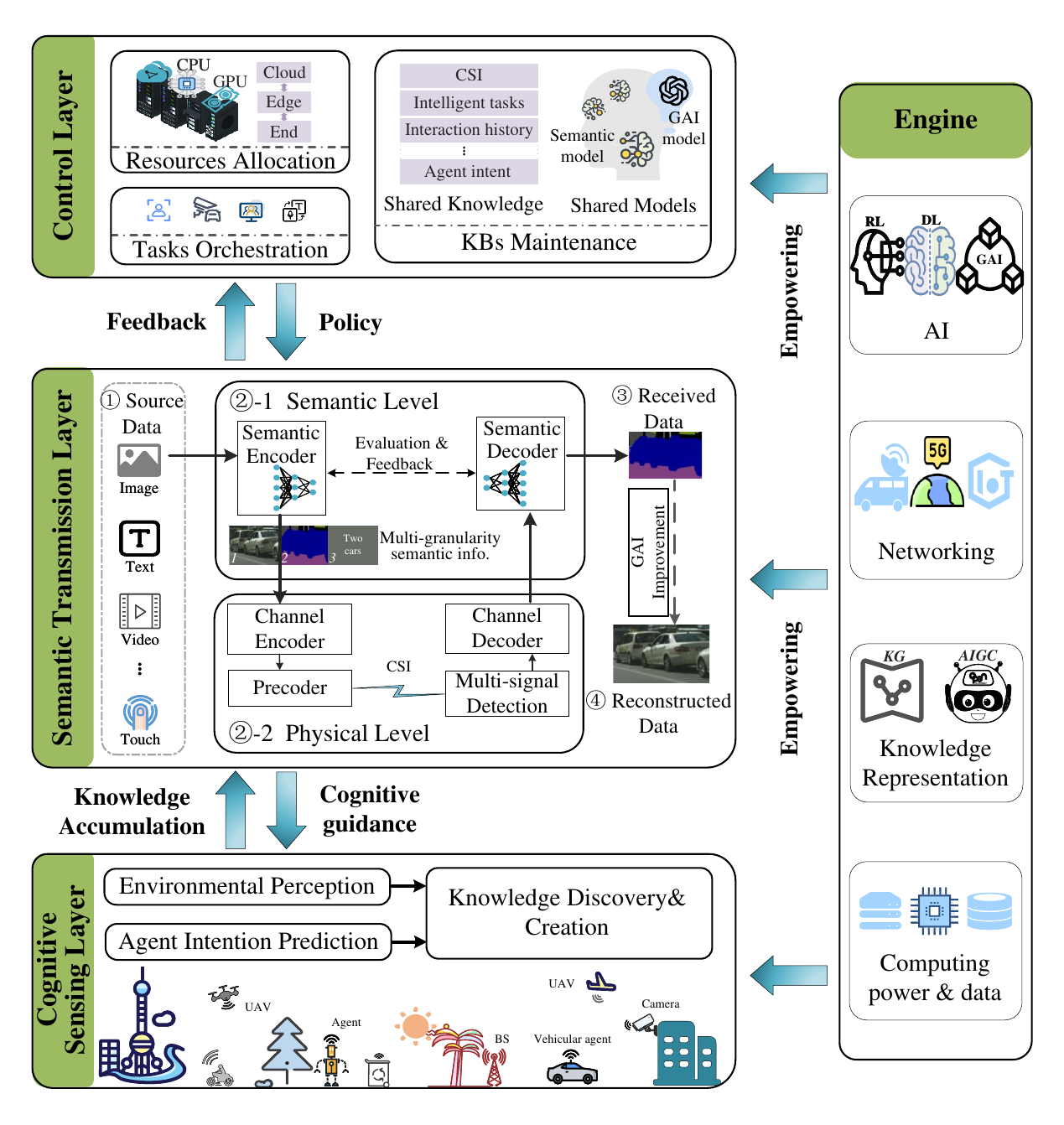}
		{\caption{{Architecture of the SemComNet.}}\label{fig:Arch}}\vspace{-5mm}
	\end{figure*}

	In this paper, we propose a general SemComNet architecture for information sharing comprising multiple agents. 
	Based on previous studies \cite{al-fuqaha2015internet, qiu2018how, shi2021semanticd}, we have identified three layers of functionality, which are shown in Fig.~\ref{fig:Arch} and described as follows:
	
	\begin{itemize}
		\item \emph{Control layer}:
        It serves as the management and orchestration component, responsible for managing shared KBs, scheduling tasks, and dynamically allocating resources across the cloud-edge-end architecture within SemComNet.
		
		\item \emph{Semantic transmission layer}: 
        This layer ensures efficient interaction among collaborative agents by providing semantic-oriented information delivery services.
		
		\item \emph{Cognitive sensing layer}: 
        It upgrades the sensing layer by integrating cognition ability, which is tasked with environment sensing, agents' intent inference, and structuring accumulated data into contextual knowledge to enrich private KBs.

	\end{itemize}
	
		% [width=0.9\textwidth,height=15.40cm]

	\subsubsection{Control Layer} \label{control_layer}
	As the top-level management and orchestration component, this layer is mainly responsible for three functions, i.e., \emph{KBs maintenance}, \emph{task scheduling}, and \emph{overall resources coordination}, covering key aspects of network management including data \& knowledge, operation, and resources.
	
	\textit{(i) Maintenance of knowledge bases:} 
	This layer manages multiple shared KBs to enhance the semantic understanding and reasoning abilities of agents in SemComNet. 
	To support diverse SemComNet applications, these KBs contain diverse shared semantic models (i.e., semantic codecs) and comprehensive contextual knowledge (e.g., source data, CSI, and communication task requirements). 
	On the one hand, agents need to deploy diverse semantic models for efficient SI extraction and reconstruction in various SemComNet applications. However, training these models is time-consuming and costly for agents. In response, this layer provides various shared semantic models, enabling agents to transition quickly from traditional to semantic-oriented service provisioning. Consequently, the shared models alleviate the necessity to train desired models from scratch.
	On the other hand, this layer offers extensive prior knowledge to assist agents during semantic reasoning and decoding phases, thereby improving the efficiency and reliability (e.g., eliminating semantic ambiguity) of interaction. 
	The knowledge can be represented in various forms including knowledge graph (KG) \cite{li2022crossmodal}, structured databases, and parameters in GAI models \cite{xu2024unleashing, min2023recent}, as detailed in Sect.~\ref{subsec:knowledge}.

	\textit{(ii) Intelligent communication tasks scheduling:} 
	Task scheduling within the control layer is essential for orchestrating network operations, especially in a SemComNet with heterogeneous agents and ever-increasing tasks (e.g., 3D video conferencing and VR transmission).
	It ensures tasks are executed at optimal times, in the correct sequence, and by the appropriate agents to maximize network performance \cite{shi2021semanticd}.
    On the one hand, it involves aligning task requirements with the capabilities of different agents, optimizing resource utilization, and ensuring timely task assignments.
    On the other hand, it evaluates and monitors system performance metrics such as response times, quality-of-experience (QoE), and agent feedback, with a focus on ensuring the accuracy, relevance, and contextual coherence of transmitted SI.

	\textit{(iii) Adaptive overall resource allocation:} 
	To enhance the performance of SemComNet, the resources including communication, and computational power (including storage) need to be carefully allocated to facilitate efficient interaction within large-scale SemComNet. 
	Given that SemComNet require significant computing power to train various semantic codecs and update KBs frequently, the cloud-edge-end hierarchical framework is utilized for on-demand resource utilization \cite{xiao2023imitation}.
	Specifically, the cloud offers massive-scale high-performance computing resources (e.g., powerful CPUs, GPUs, and memory) that are suitable for global KBs maintenance and shared semantic model training. By leveraging edge resources located at the network edge (e.g., base stations and access points), the data transfer latency and privacy leakage can be reduced \cite{pan2024cloudedge}, thereby facilitating real-time processing and access of shared KBs for agents. Pervasive end-agents are equipped with limited computing capabilities, allowing them to handle simpler tasks such as environment sensing \& processing, updating privacy-sensitive KBs, and conducting semantic-oriented information delivery.

	\subsubsection{Semantic Transmission Layer}   \label{transmission_layer}
This layer ensures efficient and reliable delivery of SI through two key stages: preparation and semantic-aware transmission. The preparation stage lays the foundation for effective semantic transmission by focusing on both knowledge synchronization and model training. In terms of knowledge, agents align with shared KBs to establish a common understanding. Regarding model training, the objective is to equip agents with semantic models to handle diverse tasks. Current methods for training semantic codecs can be categorized into three types: private learning, federated learning, and centralized learning \cite{zhao2023data}. 
% Lightweight semantic codecs may also be derived through knowledge distillation \cite{chen2023trustworthya} from pre-trained models stored in shared KBs.}

%{This layer ensures efficient and reliable delivery of SI through two key stages: preparation and semantic-aware transmission. This stage establishes the foundation for effective semantic transmission, focusing on both knowledge and model aspects. For the knowledge aspect, agents synchronize their backgrounds and align with shared KBs to establish a common ground. For the model aspect, the objective is to train semantic models for agents to handle various tasks. Current semantic codecs training methods could be categorized into private learning, federated learning, and centralized learning \cite{zhao2023data}. These agents may obtain lightweight semantic codecs through knowledge distillation \cite{chen2023trustworthya} from pre-trained models stored in shared KBs.

During the semantic-aware transmission stage, the process operates through four phases: (1) multi-source data fusion, (2) SI extraction, (3) physical-layer reliable transmission, and (4) semantic recovery and enhancement.

	\textit{(i) Multi-source data fusion.} This initial phase involves the collection and fusion of multi-modal data sources (e.g., image, point cloud, and video) \cite{meng2020survey, li2022crossmodal, xie2022taskoriented}.

	\textit{(ii) SI extraction.}
	This phase is a critical operation within this layer which employs a semantic encoder to extract meaningful and desired information from informative multi-modal source data \cite{li2022crossmodal, luo2022multimodal} while filtering out redundant and known knowledge for the receiver.
	Guided by prior knowledge from shared KBs, the semantic encoders can adaptively extract multi-level SI on demand. 
	For instance, for image transmission with various task requirements, high-level SI containing a general summary of an image (e.g., two cars) suffices for classification tasks, especially for resource-limited transmitters. Meanwhile, middle-level SI provides a more specific understanding of objects, regions, or specific structures, making it suitable for scene comprehension tasks. Furthermore, low-level SI involves pixel-level details, significantly enhancing precise scene reconstruction tasks, even with significant knowledge gaps among transceivers. 
	Additionally, the transmitter dynamically adjusts its strategy according to the recipient's feedback and channel conditions, enhancing adaptability to transceiver needs and environmental variations.

	\textit{(iii) Physical-layer reliable transmission.}
	To maximize SI transmission efficiency and mitigate interference among multiple agents, this layer employs precoding techniques (e.g., beamforming and spatial multiplexing) to optimize transmitted signals at the source. The precoder leverages spatial diversity to enhance signal quality and enable simultaneous data transmission. 
	Then, by leveraging channel coding, the extracted semantic data streams are reliably delivered through physical channels. 
	Besides, to offer better performance gains, the DL-based joint source-channel coding (JSCC) paradigm \cite{bourtsoulatze2019deep, xu2023deepa} can map the source data to channel symbols for enhanced efficiency and flexibility while well-settle cliff-effect by training in an end-to-end manner.

	\textit{(iv) Semantic recovery and enhancement. }
	After receiving the transmitted SI, the receiver agent first performs multi-user signal detection to identify and separate signals related to itself. This process involves employing intelligent algorithms such as intelligent radio \cite{luo2022semantic}, to estimate the CSI, and detect and differentiate between the various signals.
	Then, guided by shared KBs matched by the control layer, the receiver performs semantic decoding and provides feedback to the transmitter.
	Besides, to conduct semantic reasoning and parsing with auxiliary knowledge and context, the receiver can rectify transmission errors, such as content blurring or partially missing. For instance, injecting blurred SI into GAI models to accomplish content correction and enhancement \cite{xia2023generative}, thereby improving the accuracy and reliability of SemComNet.
	
	 	\begin{figure*}[!t]\setlength{\abovecaptionskip}{-0.1cm}
		\centering
		%\hspace{-0.75cm}
		\includegraphics[width=1\textwidth]{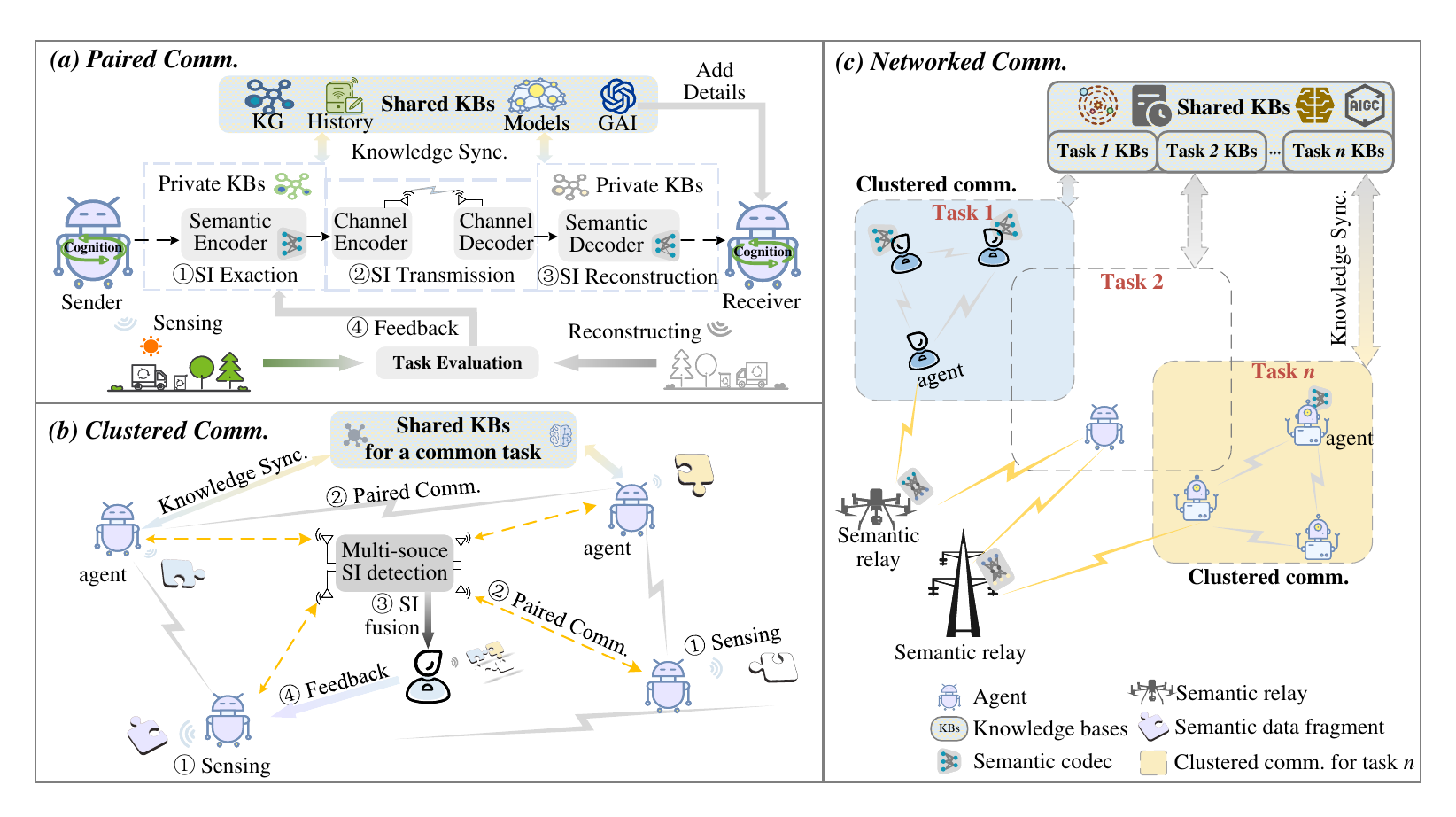}
		{\caption{{SemComNet working modes. (a) Paired SemCom: interaction between two agents via SemCom. (b) Clustered SemCom: each agent collaborates with others via paired SemCom within a cluster for a common task.
        (c) Networked SemCom: multiple clustered agents interact for different tasks, where intra-cluster agents interact via clustered SemCom and inter-cluster communication may assisted by semantic relay.}}\label{fig: WorkMode}}\vspace{-2mm}
	\end{figure*}

	\subsubsection{Cognitive Sensing Layer} \label{cognitive_layer}
    This layer represents an upgraded version of the traditional sensing layer \cite{qiu2018how} by incorporating human-like cognitive processing capabilities into system design \cite{wu2014cognitive}. Specifically, unlike the traditional sensing layer, this layer not only has direct interfaces with the physical environment to actively perceive the environment through sensors but also integrates the cognitive abilities of agents. For human beings, cognition involves acquiring understanding and knowledge through thought, experience, and senses \cite{seo2023semanticsnative}. Similarly, for intelligent agents, the cognitive abilities include understanding sensor data, reasoning about agents' intents, and discovering knowledge to enrich their private KBs \cite{wu2014cognitive, seo2023semanticsnative}.
    The cognitive sensing layer involves three features: \emph{i) perceiving the surrounding environment}, \emph{ii) inferring the intents of agents}, and then \emph{iii) organizing the accumulated information into knowledge} for the SemComNet \cite{qin2023generalized}.
	A detailed discussion of these features follows:                
        
	\textit{(i) What do the agents have?}
	The pervasive agents equipped with on-body sensors can actively sense and collect environmental semantics such as light, sound, and CSI from the surroundings \cite{qin2023generalized}. For instance, vehicular agents leverage various sensors (e.g., GPS and cameras) to collect data about other vehicles and traffic conditions. 
    Besides, these intelligent agents within SemComNet possess the capability of semantic understanding \cite{zhang2022toward} to interpret environmental data including identifying specific events or entities. For instance, they could recognize specific patterns in sound or objects in images, thereby understanding events occurring in the environment. Subsequently, this layer processes the raw data into meaningful descriptions regarding environmental and contextual aspects, such as ``vehicle traffic is growing".

	\textit{(ii) What do the agents want?}
    Identifying the agents' intents is crucial for meeting their different transmission requirements, thereby improving agents' QoE and unleashing the potential of SemCom \cite{zhang2022wisdomevolutionary, seo2023semanticsnative}.
    For instance, in the SemComNet-empowered smart transportation scenarios, the intent of vehicular agents for communication may be either ``warn" of imminent hazards (e.g., sudden stops ahead), ``coordinate" actions among vehicles (for safe intersection crossing), or ``optimize" traffic flow through intelligent traffic signal management. 
    Understanding these intents is crucial for effective environment perception and semantic noise reduction \cite{zhang2022toward}.    
	To accurately infer agents' intent, this layer analyzes these agents' behaviors (e.g., habit preferences, bodily movements, emotional states) and interaction histories, employing statistical tools and AI technologies for predictive insights. The AI-based implementation approaches can be categorized into four types \cite{zhao2023tutorial}: (1) traditional machine learning (ML)-based methods; (2) DL-based methods; (3) RL-based methods; and (4) cognitive model-based methods.
	Specifically, traditional ML algorithms such as Bayesian networks and hidden Markov models excel in performing pattern recognition from collected data in an explainable manner.
	When dealing with extensive historical data, DL approaches such as recurrent neural networks and long short-term memory (LSTM) models excel in analyzing action sequence dependencies and inferring underlying intents from extensive data.
	Furthermore, to adapt to unseen environments, RL techniques empower agents to learn optimal inference strategies through trial and error. %In the realm of RL, inverse RL \cite{xiao2023imitation} enables the deduction of underlying reward functions from the observed expert behaviors, thereby uncovering their intents and goals.
	Moreover, in scenarios with scarce training data, the cognitive model \cite{bourgin2019cognitive, wu2021unified, seo2023semanticsnative} leverages the theories from cognitive science (e.g., theory of mind \cite{rabinowitz2018machinea}), which could predict the agents' decision-making processes and intents. 
	For instance, inspired by human cognitive processes, Rabinowitz \emph{et al.} \cite{rabinowitz2018machinea} propose the machine theory of mind network, which can learn other agents' behaviors to model their mental states including desires, beliefs, and intents from limited data.

\begin{figure}[!t]
  \setlength{\abovecaptionskip}{-0.0cm}
  \centering
  %\hspace{-0.8cm} % Adjust this value to move the image more to the left
  \includegraphics[width=10cm]{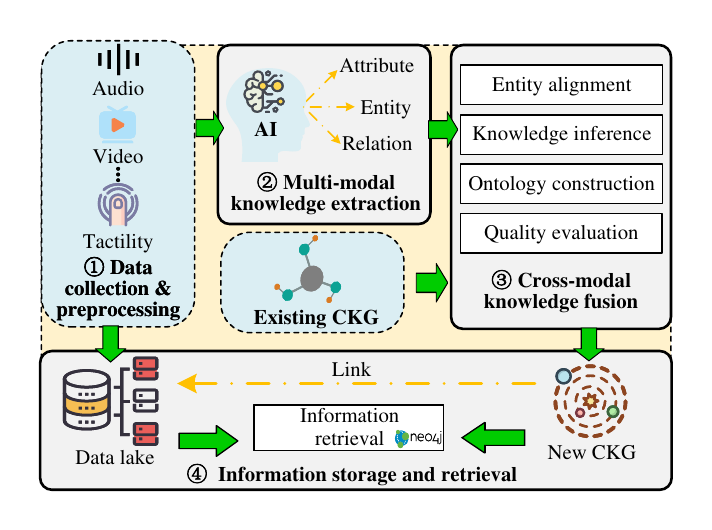}
  \caption{{Illustration of the construction of cross-modal knowledge graph (CKG) in \cite{li2022crossmodal}, which involves four key steps:
  1) data collection and preprocessing, 2) multi-modal knowledge extraction, 3) cross-modal knowledge fusion that combines extracted knowledge and existing CKG, and 4) information storage and retrieval, supported by graph databases (e.g., neo4j).}}\label{fig:KB_construction}
  \vspace{-2mm}
\end{figure}

	\textit{(iii) How do the agents acquire knowledge?}
	After environmental perception and intent inference processes, this layer could effectively transform the derived sensing data and agents' intents into structured forms of knowledge. For instance, KGs and large GAI model-based KBs store knowledge in the form of graph structures and model parameters \cite{xu2024unleashing,xia2023generative}, respectively. {In \cite{li2022crossmodal}, Li \emph{et al.} illustrate how to construct a cross-modal knowledge graph (CKG) as the KBs, as depicted in Fig.~\ref{fig:KB_construction}. The transformation of raw multi-modal data into structured knowledge includes four phases: 1) data collection and preprocessing to standardize input, 2) multi-modal knowledge extraction to derive entities, relations, and attributes, 3) cross-modal knowledge fusion to integrate data and expand the KBs, and 4) information storage and retrieval using graph databases for efficient knowledge access. As such, it enriches private KBs, enabling agents to enhance their understanding and contextual reasoning within SemComNet.} Furthermore, an essential interplay occurs between private KBs of agents and shared KBs of the control layer for continuous knowledge updating and alignment \cite{xiao2023imitation}. This interaction ensures the integration of individual knowledge with globally shared knowledge, such as aggregating new environmental semantics \cite{qin2023generalized} to shared KBs and removing obsolete knowledge. Consequently, this process ensures the consistency of multiple KBs in SemComNet and boosts the accuracy of SemCom \cite{shi2021semanticd}. {However, agents may be reluctant to share their private KBs due to privacy concerns \cite{shi2021semanticd}. Federated learning (FL) provides a promising solution \cite{wang2022social}, allowing agents to collaboratively construct shared KBs without exposing sensitive information, as detailed in Sect.~\ref{subsec:FL_privacy_preservation}.}

	\subsection{Working Modes of SemComNet} \label{subsec:workmode}
	As shown in Fig.~\ref{fig: WorkMode}, the SemComNet have three working modes according to the communication modes: (1) \textit{paired SemCom} where SI is directly transmitted between two agents rather than raw data streams, (2) \textit{clustered SemCom} enabling collaborative interactions and the exchange of SI among agents within clusters for the same task, and (3) \textit{networked SemCom} facilitating connections for different clustered agents engaged in diverse tasks, either through direct intra-cluster connections or indirectly assisted by \emph{semantic translators}\footnote{{Semantic translators (a.k.a semantic relays \cite{guo2024distributed, tang2024cooperative, luo2022autoencoderbaseda,lin2024semanticforward}) act as intelligent relays that facilitate the semantic meaning forwarding between the transceivers. Their role is similar to that of human translators who bridge interactions between individuals from different cultural backgrounds. For instance, when the source and destination have mismatched KBs or incompatible semantic codecs, the semantic translator decodes the SI using its shared knowledge with the source and re-encodes it based on its shared knowledge with the destination.}}. In Table~\ref{workingmode}, we summarize the main differences between the above three working modes in SemComNet.
 
		\begin{table*}[!t]
		\centering
		\caption{A Summary of Three Working Modes in the SemComNet}\label{workingmode} 
		\renewcommand{\arraystretch}{1.2} % 调整行高
		\begin{tabular}{l|ccc}\toprule
			\textbf{} & \textbf{Paired Semantic Comm.}
			& \textbf{Clustered Semantic Comm.} & \textbf{Networked Semantic Comm.} \\ \midrule
			\textbf{Number of Tasks}  & \begin{tabular}[c]{@{}c@{}}1\end{tabular}
			& \begin{tabular}[c]{@{}c@{}}1 \end{tabular}
			& \begin{tabular}[c]{@{}c@{}} $\geq 2$ tasks across clusters \end{tabular} \\ \midrule

			\textbf{Number of Participants}  & \begin{tabular}[c]{@{}c@{}}2 agents\end{tabular}
			& \begin{tabular}[c]{@{}c@{}}$> 2$ agents form a cluster \end{tabular}
			& \begin{tabular}[c]{@{}c@{}}Multiple clusters  \end{tabular} \\ \midrule

			\textbf{Communication Modes} & \begin{tabular}[c]{@{}c@{}} End-to-end \\ semantic comm. \end{tabular}
			& \begin{tabular}[c]{@{}c@{}} Paired semantic \\ comm. within a cluster \end{tabular}
			& \begin{tabular}[c]{@{}c@{}} Clustered semantic comm. and\\ inter-cluster via semantic translators \end{tabular} \\ \midrule
			
			\textbf{Application Scenarios}  & \begin{tabular}[c]{@{}c@{}}Collaborative task \\ execution   between two agents \end{tabular}
			& \begin{tabular}[c]{@{}c@{}}Collaborative task \\  execution within a cluster of agents \end{tabular}
			& \begin{tabular}[c]{@{}c@{}}Scenarios requiring  \\ coordination across tasks \end{tabular} \\ \midrule
			
			{\textbf{Types of KBs}}  & {\begin{tabular}[c]{@{}c@{}}Private and shared \\ KBs between two agents \end{tabular}}
			& {\begin{tabular}[c]{@{}c@{}}Private and task-special \\ KBs among clustered agents\end{tabular}}
			& {\begin{tabular}[c]{@{}c@{}}Multi-layered comprising \\ public, task-specific, and private KBs\end{tabular}} \\ \bottomrule
		\end{tabular}
	\end{table*}

\subsubsection{Paired Semantic Communication}	
As illustrated in Fig.~\ref{fig: WorkMode} (a), this mode allows two agents to interact and aims to deliver the key meaning behind the source data, instead of transmitting the raw bit streams. 
	Before the interaction, both communicating entities undergo the preparation stage. Here, agents engage in synchronizing background knowledge and aligning their shared KBs to establish a common ground and context for communication. To alleviate the computation burden of agents, lightweight semantic codecs can transfer from large, powerful models within shared KBs via knowledge distillation\footnote{Knowledge distillation (KD) is an ML model compression technique that transfers knowledge from a large, complex teacher model to a smaller student model. In SemComNet, KD allows agents to create lightweight yet effective semantic codecs from larger models (e.g., pre-trained semantic models), benefiting resource-constrained agents with limited computational power.}.

	The transmission involves three principal phases, i.e., SI exaction, SI transmission, and SI reconstruction.
	\textit{i) SI exaction phase.} In this phase, the transmitter uses the semantic encoder to extract compact SI from original multi-modal data streams, while simultaneously filtering out irrelevant information. 
	The work \cite{yang2023semanticb} categorizes semantic encoder designs into four main approaches: DL-based, RL-based, KB-assisted, and semantic-native approaches. Among these, various DL-based solutions \cite{xie2021deep, luo2022semantic, zhang2024unified} are identified as the mainstream direction currently.
	\textit{ii) SI transmission phase.} After extraction, the transmitting agents transmit processed SI to the receiver. The channel codecs are responsible for ensuring error-free transmission of SI, capable of combating noise, interference, and other challenges related to the physical layer. During channel coding, mechanisms such as error correction, compression, and encryption can be employed to ensure the integrity, efficiency, and security of the SI.
	\textit{iii) SI reconstruction phase.}
	Upon receiving the transmitted SI, the receiver utilizes a semantic decoder to interpret and reconstruct the SI into a format relevant to its tasks.
	The semantic decoder design should be tailored to specific downstream tasks, which can be classified as pragmatic task execution (e.g., image classification) and observable information reconstruction (e.g., video transmission) \cite{zhang2023deep}. During reconstruction, the receiver interacts with KBs to ensure the SI aligns with the contextual background.
	
	Finally, a quality assessment is conducted to verify the reconstructed information maintains reliability and suitability at the semantic level. Based on the assessment and CSI, the transmission strategy is adaptively adjusted to mitigate semantic noise and channel-related disturbances.
	
	 \subsubsection{Clustered Semantic Communication}\label{subsec:clustered}
	This mode involves multiple agents to collaboratively accomplish a common task, as depicted in Fig.~\ref{fig: WorkMode} (b). 
	Agents are organized into clusters based on common interests in tasks such as cooperative navigation and real-time object detection \cite{abbasi2007survey}. Within each cluster, sharing domain-specific knowledge and task-specific semantic models enables efficient collaboration and SI exchange. Achieving clustered SemCom requires the following four steps.
	\textit{i) Perception and information collection.} 
	At first, agents utilize their sensors to actively collect data from the surrounding environment \cite{qin2023generalized}. Then, each agent independently cognitive sensing processes this data locally and extracts relevant semantic fragments\footnote{{Semantic fragment (a.k.a. semantic seb \cite{zhang2022wisdomevolutionary}) refers to the smallest unit of meaningful information (i.e., SI) in a SemCom system. Generally, a semantic seb consists of a feature or set of features extracted from the source data, which vary depending on the method used to extract them and the specific communication needs or tasks. Moreover, the same content's semantic sebs may be interpreted inconsistently across agents with different backgrounds. Note that, the definition and standardization of semantic sebs remain evolving, necessitating further research.}} that contribute to the shared task, such as collaborative object perception.
	\textit{ii) Semantic fragment sharing via paired SemCom.}
	Within the cluster, agents share the locally extracted semantic fragments with other agents via paired SemCom. 
	\textit{iii) SI fusion.}
	Upon receiving SI from other agents, the recipient applies multi-signal detection and multi-modal data fusion \cite{luo2022multimodal}. These processes aim to reconstruct a more comprehensive and holistic semantic representation.
	\textit{iv) Re-transmission for missing information. }
	During the semantic reconstruction phase, the receiver identifies missing or incomplete information via verification. Subsequently, feedback or requests are sent to the respective agents, prompting them to re-transmit the necessary information fragments \cite{xiao2023imitation}. This process aims to achieve a complete semantic understanding and fulfill the task requirements.

	\subsubsection{Networked Semantic Communication}
	This paradigm is expected to accommodate complex application scenarios comprising various tasks \cite{xie2022taskoriented} and to support coordination across tasks \cite{shi2021semanticd}.
	In this paradigm, connected agents are organized into different clusters based on their involved tasks. 
	Within each task, agents efficiently exchange SI via clustered SemCom. Meanwhile, for inter-cluster communication, agents belonging to different clusters can be assisted or coordinated by \emph{semantic translators}, a.k.a semantic relay nodes \cite{guo2024distributed, tang2024cooperative, luo2022autoencoderbaseda,lin2024semanticforward}.
	These translators, as depicted in Fig.~\ref{fig: WorkMode} (c), could be acted by powerful unmanned aerial vehicles and base stations, which possess cross-domain knowledge and diverse semantic models. Their role extends beyond basic SI forwarding, emphasizing semantic translation to guarantee accurate interpretation and exchange of SI at the semantic level \cite{guo2024distributed}.
	Consequently, semantic translators not only mitigate knowledge background disparities among transceivers across clusters, but also alleviate agents from the necessity of training multiple semantic models across tasks. As such, it boosts connectivity and efficiency for networked agents.

	The networked SemCom mode relies on multi-layered KBs \cite{xiao2023imitation}, including public, task-specific, and private KBs. Specifically, public KBs provide common-sense knowledge and general models accessible to networked agents. Within each cluster, agents share task-specific KBs, which concentrate on domain expertise and task-related knowledge (e.g., diseases, symptoms, and treatments in medical diagnosis tasks). Additionally, private KBs owned by individual agents mainly contain personalized and sensitive knowledge accumulated by each agent (e.g., preferences, interests, and communication context). 
	These multi-layered KBs provide essential context and knowledge for SI extraction, delivery, and reconstruction for agents.

 \begin{table*}[htbp]
\centering
\renewcommand{\arraystretch}{1.4} % Adjust row spacing
\caption{{Enabling Technologies and Their Roles in SemComNet}}
\label{tab:enabling_tech}
\begin{tabular}{|c|p{8.8cm}|p{4.8cm}|} % Center first column, other columns unchanged
\hline
\textbf{{Enabling Tech.}} & \textbf{{Role}} & \textbf{{Covered Tech.}} \\
\hline
{AI} & {Semantic-aware data processing and decision making} & {ML, DL, RL, and GAI} \\
\hline
{Networking Technology} & {Ensure efficient SI transmission and reliable connectivity for all agents} & {5G and beyond, LoRa, BLE, and IoT} \\
\hline
\parbox{3.5cm}{\centering {Knowledge Representation}} & {Manage and represent knowledge for efficient retrieval and KBs construction} & {KGs, ontologies, and scene graphs} \\
\hline
{Ubiquitous Computing} & {Provide on-demand and sufficient computing resources for SemComNet} & {Mobile, edge, and cloud computing} \\
\hline
\end{tabular}
\end{table*}

	\subsection{Enabling Technologies of SemComNet}
	\subsubsection{AI} \label{subsec:AI}
	Serving as the foundation of SemComNet, AI techniques including DL, RL, and GAI significantly enhance the ability of SI extraction and overall efficiency in dynamic communication conditions and diverse task demands.
	Specifically,
	DL models such as LSTM and Transformer \cite{xie2021deep, luo2022semantic} allow semantic models to robustly abstract key features and capture long-range dependencies within data. 
    Given the frequent need for model updates in SemComNet, which can lead to service interruptions and significant time consumption, transfer learning \cite{xie2021deep} facilitates the reuse of existing models and knowledge for new tasks, accelerating training and improving data efficiency.
    Moreover, to mitigate ``catastrophic forgetting" \cite{delange2022continual} in transfer learning, continual learning becomes essential in SemComNet which helps semantic models adapt to new tasks without forgetting learned knowledge.
	To address the issue of scarce training samples in the SemComNet environment, GAI models such as generative adversarial networks (GANs) and diffusion models prove beneficial \cite{xia2023generative, pan2024cloudedge}. These models could create extensive, high-quality, and personalized data samples resembling real-world scenarios for training diverse semantic models.
	Besides, RL empowers agents with autonomous implicit semantic reasoning \cite{xiao2023imitation} and real-time decision-making in SemComNet tasks such as autonomous driving and drone navigation.
	
	\subsubsection{Networking Technology}
	In SemComNet, networking technology {such as IoT, Bluetooth Low Energy (BLE), Long Range (LoRa), as well as 5G and beyond (B5G)} plays a crucial role in enhancing its performance and efficiency, achieving reliable network connections, and real-time SI transmission between agents. 
	The increasing affordability and advancing intelligence of IoT devices \cite{yin2023multidomain} have made it feasible to deploy numerous intelligent agents. Simultaneously, the abundant deployment of sensors in the IoT provides SemComNet with rich perceptual data, facilitating knowledge accumulation and SI comprehension.
    Moreover, BLE offers short-range and low-latency wireless networking, ideal for intra-cluster communication, enabling seamless data exchange with minimal energy consumption \cite{leonardi2022lora}. LoRa enhances SemComNet by enabling long-range, reliable communication for inter-cluster agents spread across wide areas, even in challenging environments.
    Besides, 5G and beyond \cite{zhang2022wisdomevolutionary} support high-speed data transmission, increased network capacity, and global communication coverage, including remote areas and oceans, fulfilling the real-time response and efficient data transfer requirements of SemComNet.

	\subsubsection{Knowledge Representation Technology}
	\label{subsec:knowledge}
	Acting as ``memory" of SemComNet, the knowledge representation such as knowledge graph (KG) \cite{wang2023knowledgeb,  yang2023semanticb, li2022crossmodal}, {ontologies \cite{trevlakis2024natively, qiu2020survey}}, and scene graphs \cite{zhang2023optimization} can transform contextual information and experience into a comprehensible format for agents. They also empower SemComNet to manage, search, and reason over vast amounts of knowledge.
	For instance, KGs can be integrated into SemComNet to organize and aggregate vast unstructured data into a structured knowledge format from diverse domains \cite{wang2023knowledgeb}, presented in graphical form. % This method fosters a deeper understanding of correlations among various elements of knowledge and also provides contextual background for supporting semantic extraction, and recovery.
    {
    By formally defining the concepts and their relationships, ontologies \cite{trevlakis2024natively} offer a structured representation that aids in identifying entities (e.g., people, places, and events) within data.}	Besides, Zhang \emph{et al.} \cite{zhang2023optimization} use scene graphs to capture and store knowledge about the objects and their relationships in the original image.

	\subsubsection{Ubiquitous Computing}
	SemComNet require significant computing power to sustain AI model training and storage, as well as multiple KBs management and synchronization. In response, ubiquitous computing \cite{wang2023surveyc} establishes an environment for SemComNet where computing resources are seamlessly and invisibly embedded in various agents (e.g., wearable devices and sensors), ensuring widespread availability. To enhance agents' QoE, the cloud-edge-end network architecture offers on-demand access to computational and storage resources. The cloud tier offers powerful computing resources for training shared semantic models, while the edge tier facilitates rapid computation processing for nearby agents. Additionally, through collaborative maintenance of multiple pre-cached KBs across the cloud-edge-end infrastructure, agents can efficiently access required knowledge, improving semantic delivery performance.

        {
        \subsubsection{Summary}
        As shown in Table~\ref{tab:enabling_tech}, AI, networking, knowledge representation, and ubiquitous computing technologies form the fundamental pillars of SemComNet. AI enhances semantic understanding and intelligent optimization, networking technology provides reliable and scalable transmission, knowledge representation structures and manages knowledge, and ubiquitous computing supplies the necessary computational resources. These four technologies collectively enable intelligent and semantic-aware interaction among agents within SemComNet.}

    \subsection{{Use Cases of SemComNet}}
    With the assistance of SemComNet, a vast array of emerging intelligent applications across numerous key domains are enabled \cite{zhang2024intellicise, yang2023semanticb}, such as autonomous driving, holographic-type communications, and the Metaverse. %while Fig. 2 depicting three representative applications of SemComNet.}
	\subsubsection{{Autonomous Driving}}
	{Autonomous driving depends on real-time data exchange among vehicles, roadside infrastructure, and cloud servers to make safe driving decisions. Equipped with sensors such as cameras and LiDAR, autonomous vehicles continuously generate vast amounts of data to monitor their surroundings and provide critical inputs for decision-making. However, conventional communication systems often struggle to meet the ultra-reliable, low-latency, and high-throughput transmission demands required for fast decision-making and stable coordination.
	SemComNet {represent} the novel paradigm for intelligent collaboration in autonomous driving by integrating user intent and semantics into the communication process. This enhances data exchange and decision-making capabilities among vehicles, roadside infrastructure, and other entities, enabling efficient vehicle-road collaboration while meeting the strict demands of perception accuracy and latency in smart transportation systems. }

In a SemComNet-enabled autonomous driving scenario, each vehicle maintains private KBs with sensitive data (e.g., historical routes and location coordinates) while sharing semantic KBs with an edge server. These shared KBs include background knowledge (e.g., city maps) and task-specific semantic models. For tasks requiring rapid inference, vehicles reconstruct SI from multi-modal data and shared KBs in real-time to make local decisions (e.g., acceleration, braking). By accessing the edge server’s knowledge, vehicles expand their perception range and make accurate decisions with minimal computation. In more complex scenarios, vehicular agents transmit SI to the edge server and other nearby vehicles, which collaborate to make final decisions. This collaborative process enhances task performance by leveraging the complementary nature of SI and task correlations. For instance, Feng et al. \cite{feng2024semantic} propose a multi-user SemCom system for edge intelligence-enabled autonomous driving. They employ the JSCC method and fusion module for various users and data modalities to reduce data redundancy and optimize task coordination, thus improving transmission efficiency.
 
    \subsubsection{{Holographic-type Communications}}
        {Holographic-type communication \cite{akyildiz2022holographic} leverages holographic display systems to capture individuals and their surroundings remotely and transmit them over networks. At the receiver's end, laser projections create real-time 3D holograms, enabling interaction with virtual projections. This technology provides an immersive, multi-modal experience via naked-eye holography. However, current holography requires substantial data volumes for synthesizing images and 3D data. For example, multi-angle holographic videos demand terabit-per-second (Tbps) transmission rates and sub-1 ms latency \cite{akyildiz2022holographic} to achieve full immersion, which remains challenging.}

         {
        Instead of transmitting all raw data, SemComNet focus on delivering only critical SI, reducing the data volume for transmission. For instance, in holographic video conferencing, this might involve transmitting key features such as facial expressions instead of full volumetric data. In  \cite{cheng2023enriching}, Cheng \emph{et al.} propose a semantic-driven framework, named SemHolo, for enhancing holographic telepresence in 6G networks. By prioritizing task-relevant aspects (e.g., key gestures and facial expressions) over full volumetric data, SemHolo enables the efficient delivery of 3D content. A proof-of-concept implementation demonstrates the feasibility of real-time interactive telepresence while ensuring visual quality and reconstruction fidelity.        
        Besides, the multi-modal data transmission capabilities of SemComNet could link different modalities through a common semantic framework. This allows for expressing the same semantic meaning across various modalities, enabling simultaneous recovery of multi-modal information from a single semantic context. By focusing on semantic representations, SemComNet address the bandwidth challenges of holographic communication while improving efficiency.
}
        
    \subsubsection{{Metaverse}}
	The Metaverse, envisioned as a next-generation Internet, offers a fully immersive and high-fidelity virtual universe where users interact in computer-generated environments \cite{wang2022survey}. For instance, Metaverse concerts provide large online audiences with a 360-degree panoramic view and an interactive experience, allowing digital avatars to engage with both the venue and other participants.
    However, large-scale multi-user activities in the Metaverse demand significant resources from both backend servers and user devices (e.g., VR headsets). Specifically, for the Metaverse servers, rendering detailed virtual environments with numerous users requires immense computing and transmission resources. As for Metaverse users, they face the challenge of constantly receiving and rendering potentially thousands of other participants' positions and actions.

    In practice, not all content needs to be transmitted or rendered to maintain a seamless and immersive experience \cite{hsu2024socialaware}. SemComNet address this by transmitting only semantically meaningful data, reducing storage and transmission loads. For instance, rather than transmitting full 3D models of a user’s gestures, SemComNet transmits only essential information, such as key movements that trigger interactions. This reduces data load, accelerates processing, and improves system performance \cite{yang2023semanticb}. To ensure trustworthy SemCom in the Metaverse, Chen \emph{et al.} \cite{chen2023trustworthya} propose a secure multi-user SemCom system based on intelligent radio and GANs, while FL balances privacy and communication accuracy.

	\begin{figure}[!t]\setlength{\abovecaptionskip}{0.0cm}%\vspace{-3mm}
	% \centering
	\hspace{-0.4cm}
	\includegraphics[width=10cm]{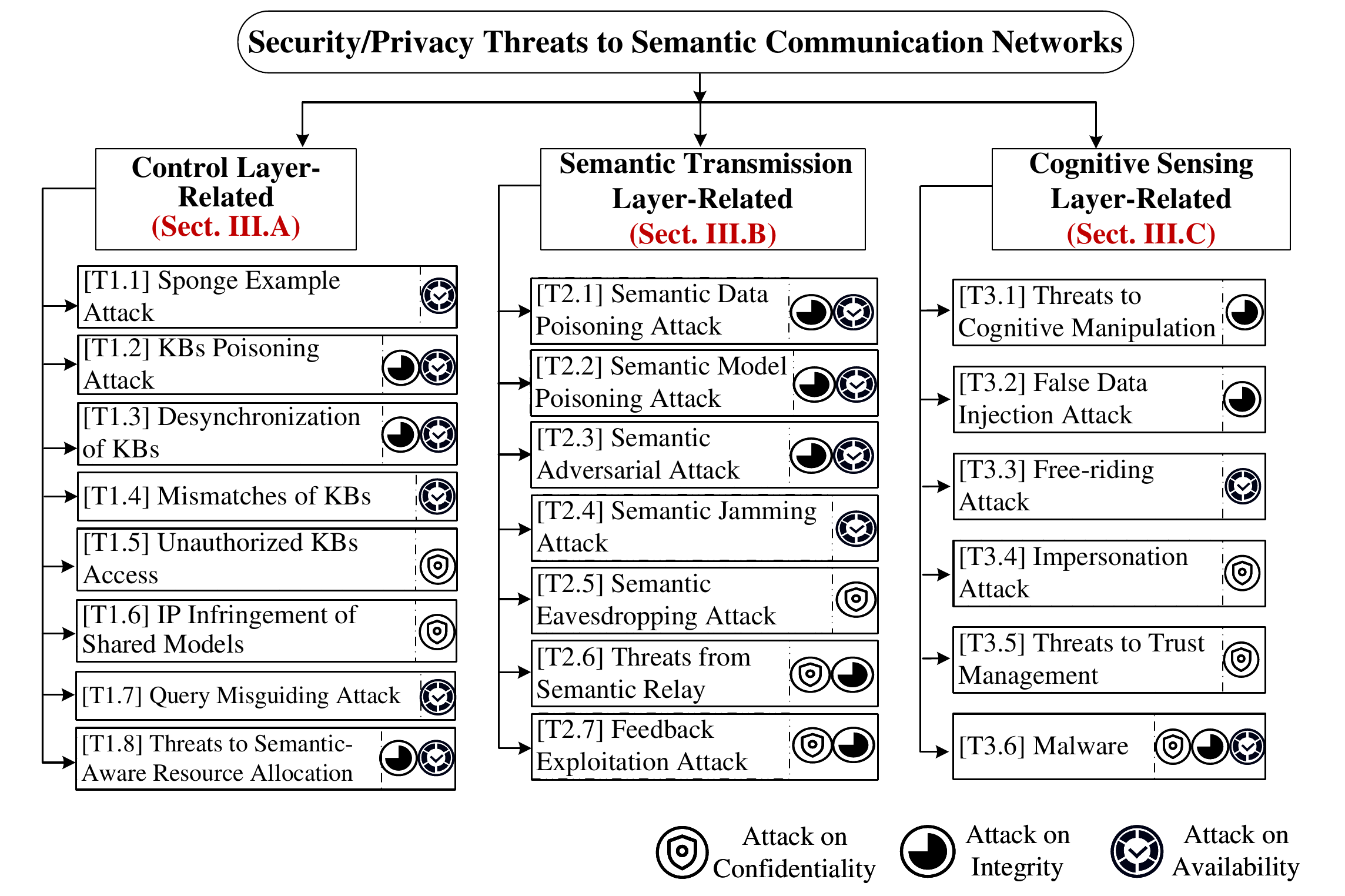}
	{\caption{{The taxonomy of security/privacy threats to SemComNet from the three functional layers (i.e., control layer, semantic transmission layer, and cognitive sensing layer).}}\label{fig:threatstaxonomy}}\vspace{-5mm}
	\end{figure}	

    \subsection{{Semantic Metrics}}
    {Unlike traditional communication systems that use metrics such as \textit{symbol-error rate} (SER) or \textit{bit-error rate} (BER), semantic metrics in SemComNet focus on the significance and meaning of information \cite{trevlakis2024natively}. These metrics can be classified into three categories \cite{zhang2024intellicise}, i.e., \textit{significance-oriented}, \textit{meaning-oriented}, and \textit{combined} metrics. Specifically, significance-oriented metrics evaluate the value and relevance of information based on its timing, timeliness, and impact on subsequent tasks \cite{trevlakis2024natively, yang2023semanticb}. These metrics help prioritize information transmission in resource-constrained scenarios. For example, the \textit{age of information} (AoI) \cite{popovski2022perspective} and \textit{value of information} (VoI) \cite{uysal2022semantica} are two representative metrics used for significance evaluation in SemComNet \cite{trevlakis2024natively}. AoI quantifies the timeliness of information at the destination by measuring the time elapsed since the latest data was generated at the source, while VoI considers both the timeliness and quality of the transmitted information \cite{trevlakis2024natively}. Besides, the meaning-oriented metrics assess the similarity between the transmitted and received SI, prioritizing semantic fidelity over pixel or bit-level accuracy. Lastly, combined metrics (e.g., semantic QoE \cite{kadam2024semantic, zhang2023intelligent}) integrate aspects of semantic fidelity, communication costs, and user experience, providing a more holistic evaluation.}

    {
    Besides, the above semantic metrics can significantly influence the security and privacy aspects of SemComNet. Take significance-oriented metrics as an example, they can be used to prioritize time-sensitive information and detect timing anomalies that may indicate malicious activities. For instance, an attacker may exploit vulnerabilities (e.g., delaying or replaying outdated or irrelevant data) to confuse or mislead the receiving agent. By monitoring metrics such as AoI or VoI, SemComNet can identify and discard data that is excessively old or delayed beyond acceptable thresholds, thereby mitigating the risk of timing-based attacks.}

	\section{Security and privacy threats in SemComNet}
	Despite the promising prospects of SemComNet, it faces significant security and privacy issues, which have not been widely discussed. As critical systems, SemComNet's security and privacy concerns must encompass the confidentiality, integrity, and availability (CIA) of resources (including data, shared knowledge, and semantic models). Specifically, confidentiality issues primarily involve unauthorized access to sensitive resources, as well as potential privacy and intellectual property (IP) infringement. Integrity is compromised by attacks that alter or corrupt data/knowledge (e.g., launch false data injection, poisoning, and adversarial attacks) which compromise its accuracy and semantic meaning. Availability focuses on threats disrupting services, such as DoS and malware attacks, affecting SemComNet's operational functionality for legitimate users. 	

\begin{figure}[!t]\setlength{\abovecaptionskip}{0.0cm}%\vspace{-3mm}
		\centering
		% \hspace{0.2cm}
		\includegraphics[width=9.5cm]{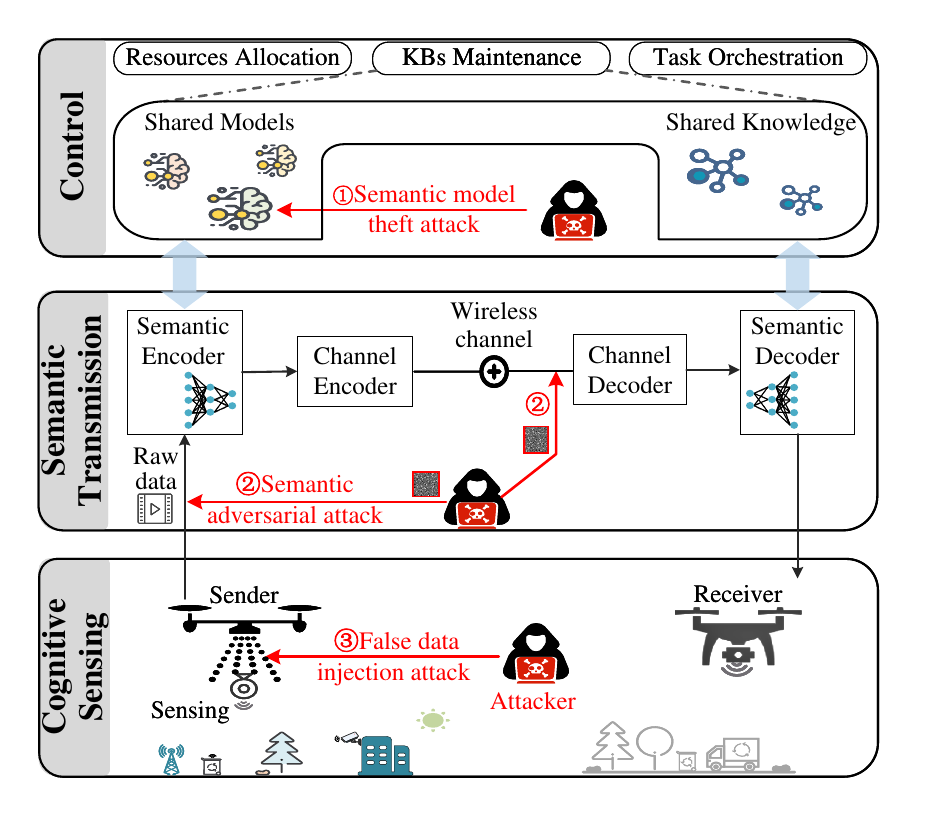}
		{\caption{An illustrative example of semantic model theft, semantic adversarial, false data injection attacks in the SemComNet.}\label{fig:attack_layered}}\vspace{-5mm}
	\end{figure}

Besides, to effectively and comprehensively identify these threats, we conduct a layer-by-layer analysis.
As shown in Fig.~\ref{fig:threatstaxonomy}, we categorize a broad scope of security/privacy threats in SemComNet (from Sect.~\ref{subsec:threat1} to Sect.~\ref{subsec:threat3}) from its three layers of functionality, i.e., control, semantic transmission, and cognitive sensing layers. Fig.~\ref{fig:attack_layered} provides three illustrative examples of attacks within these layers.

	\subsection{Threats to Control Layer in SemComNet}\label{subsec:threat1}
	As outlined in Sect.~\ref{control_layer}, the control layer in SemComNet is tasked with top-level management and orchestration for tasks, resources, and shared KBs. 
	However, adversaries could launch specialized attacks such as denial of services (e.g. by injecting sponge examples) which aims at depleting resources and rendering SemComNet unavailable.
	Besides, compared with traditional communication networks, a distinctive feature of SemComNet is its maintenance of multiple KBs to provide plenty of shared knowledge and semantic models for collaborative agents. 
	However, this feature incurs vulnerabilities such as KBs poisoning, unauthorized KBs access, and theft of shared semantic models. Below, we enumerate the typical threats related to the control layer in SemComNet.

	\begin{itemize}

		\item \emph{Sponge Examples Attack [T1.1]}.
            {In traditional networks, denial-of-service (DoS) attack \cite{mitev2023whata, shumailov2021sponge} disrupt normal operations by overwhelming systems with excessive traffic, preventing legitimate users from accessing network services. 
            In SemComNet, a variant known as sponge examples attack emerges \cite{yao2024survey, shumailov2021sponge} targets energy consumption and latency in DL components. The threat is severe due to SemComNet's reliance on computational resources (e.g., powerful CPUs and GPUs) for DL-driven semantic models training. }
            Hackers can inject crafted sponge examples into semantic codecs, nullifying hardware acceleration and causing excessive energy consumption and response delays.
            In \cite{shumailov2021sponge}, Shumailov \emph{et al.} introduce two methods for generating sponge examples: gradient-based and genetic-based. The former is a white-box approach requiring access to model parameters, while the latter is a black-box technique that optimizes samples based on energy or latency metrics by simply querying the model. 
            {To mitigate such threats, one solution is worst-case performance analysis \cite{shumailov2021sponge}, which establishes processing time and energy thresholds using natural examples. Inputs exceeding these thresholds are rejected, ensuring system robustness. A more comprehensive solution is discussed in Sect.~\ref{subsec:sponge_examples_defense}.}

		\item \emph{Knowledge Bases Poisoning Attack [T1.2]}.		
            Unlike conventional communication networks, SemComNet relies heavily on shared KBs to aid semantic understanding and reasoning \cite{lu2023semanticsempowered}. However, this reliance on KBs introduces new attacks that are relatively easy to exploit but challenging to detect.	Specifically, malicious entities may manipulate the storage nodes of KBs (e.g., cloud and edge servers) via unauthorized access \cite{wang2023surveyc} to influence their cached knowledge \cite{shen2023secure}. As shown in Fig.~\ref{fig:attack}, one specific attack involves KBs poisoning \cite{wang2023surveyc}, where attackers inject false, harmful, or misleading knowledge into KBs, thereby deceiving receivers and deteriorating the overall performance of SemComNet. To mitigate this, access control (AC) and blockchain technologies, as discussed in Sect.~\ref{Data_Knowledge_AC}, offer secure KBs management, ensuring tamper-proof records and preventing unauthorized modifications.

		\item \emph{Desynchronization of Knowledge Bases [T1.3]}.
            The primary target of this attack is to undermine both the integrity and availability of KBs. Attackers may induce inconsistency or desynchronization of KBs by disrupting network connections, manipulating update frequencies, or tampering with updated versions. This desynchronization within SemComNet allows adversaries to secretly delay, modify, or even destroy KBs without detection \cite{wang2023surveyc}. 
		As a result, inconsistent semantic understandings arise among agents, compromising the overall robustness and effectiveness of SemComNet.
		For instance, malicious entities might intentionally introduce conflicting or outdated knowledge, thereby desynchronizing multiple KBs and interfering with the accurate extraction of SI. {Currently, this type of attack and its corresponding defenses remain unexplored in the field of the SemCom domain, warranting further research.}

		\item \emph{Mismatches of Knowledge Bases [T1.4]}.
		This threat may affect the availability of KBs matching mechanism, leading to semantic noise between transceivers \cite{luo2022semantic}.
		Before interaction, participating agents in SemComNet should synchronize and align their prior knowledge \cite{xie2021deep, tian2023asynchronous} to maintain consistent and up-to-date contextual understanding.
		However, in realistic scenarios, agents may be reluctant to share their sensitive knowledge with others due to privacy concerns and substantial communication burdens, resulting in KBs mismatches between transmitter and receiver \cite{cheng2024knowledge}. These mismatches may incur semantic-level misunderstandings (e.g., semantic noise) at the receiver \cite{luo2022semantic}.
		Besides, the dynamic nature of the environment necessitates continual updates in KBs, further exacerbating the mismatches or disparities between them. {To effectively mitigate this risk, techniques such as semantic relay (detailed in Sect.~\ref{subsec:relay_defense}) and resilient semantic understanding schemes (discussed in Sects.~\ref{subsec:GAI_defense},\ref{subsec:reasoning_defense}) may be beneficial.}
  
		% {To mitigate privacy concerns, Cheng \emph{et al.} \cite{cheng2024knowledge} present a novel knowledge discrepancy-oriented privacy-preserving method (KDPP), which leverages knowledge mapping (align unknown knowledge with existing knowledge) and knowledge path cutting-off to mitigate privacy inference.}

		\item \emph{Unauthorized Knowledge Bases Access [T1.5]}.
            {Unauthorized access \cite{mitev2023whata} in communication systems occurs when entities gain access to systems or data without proper authorization, potentially leading to threats such as data theft, system manipulation, or service disruption.}
           {In SemComNet, such a threat expands the attack surface due to the existence of multiple privacy-sensitive KBs \cite{cheng2024knowledge}.} For instance, regionally shared KBs contain relatively personal and sensitive knowledge that is only accessible to agents within the coverage of edge servers \cite{xiao2023imitation}. Curious agents may attempt unauthorized access or retrieval of data from KBs via various approaches such as brute-force attacks \cite{li2023secure} and impersonation attacks \cite{mitev2023whata}, which pose a severe risk of privacy breaches. Moreover, attackers can tamper with the KBs with poisoned data or backdoors \cite{tian2022comprehensive} after gaining unauthorized access, resulting in deteriorated communication performance and loss of confidentiality. {More defense details can refer to Sect.~\ref{Data_Knowledge_AC} and Sect.~\ref{Data_Knowledge_Blockchain}.}

		\item \emph{IP Infringement of Shared Semantic Models [T1.6]}.
		SemComNet accelerate the shift of transmission services from traditional to semantic-oriented paradigms by providing diverse shared semantic models for authorized agents. As such, the computational burden of these agents may be significantly reduced, mitigating the need to train semantic codecs from scratch. However, significant IP threats may arise in SemComNet due to the replicability characteristic of these shared semantic models, which allow authorized entities to resell these valuable models for illegal profit without being detected \cite{zhang2018protectinga}.
		Furthermore, during the model duplication and distribution phases, these semantic models are susceptible to risks such as theft (as depicted in the upper part of Fig.~\ref{fig:attack_layered}), counterfeiting, and unauthorized imitation \cite{nie2024deep}. Considering the expensive and time-consuming nature of their training process, the above risks may pose significant economic and confidentiality losses for SemComNet. {To effectively mitigate IP infringement, techniques such as watermarking and blockchain can be beneficial, as discussed in Sect.~\ref{subsec:IP_protection}.}

		\item  \emph{Query Misguiding Attack [T1.7]}.
		{In SemComNet, agents frequently query the KBs \cite{li2023secure} to synchronize relevant knowledge. However, adversaries may manipulate KBs and alter the query behavior \cite{li2023secure} through various methods such as injecting malicious SQL code, installing malware, or phishing.}
        For instance, adversaries can impede agents from generating informative queries by introducing misleading evidence, referred to as \emph{bait evidence} \cite{xi2023security}, illustrated as Fig.~\ref{fig:attack}. For instance, when an agent queries KBs regarding the first president of the USA, an expected response should provide information about George Washington. Yet, adversaries might craft misleading queries, such as ``Was George Washington the first president of the United States?" to mislead the KBs and obstruct the retrieval of accurate knowledge. {Currently, research on defending against such threats in SemComNet is still lacking and requires further study.}

		\begin{figure}[!t]\setlength{\abovecaptionskip}{-0.0cm}%\vspace{-3mm}
		\centering
		\includegraphics[width=9cm]{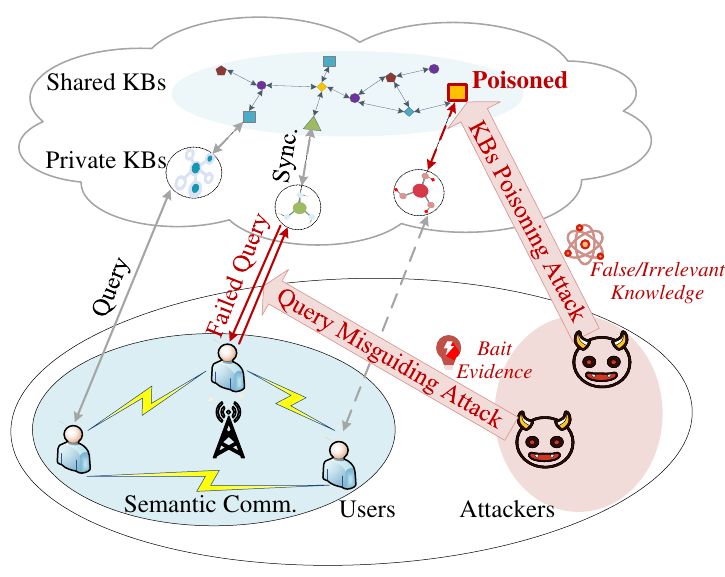}
		\caption{{An illustrative example of KBs poisoning attack and query misguiding attack in the SemComNet.}}\label{fig:attack} 
		\vspace{-2mm}
		\end{figure}
	
		\item \emph{Threats to Semantic-Aware Resource Allocation [T1.8]}.
		{SemComNet offer a semantic-aware and spectrum-efficient paradigm that enables users to transmit semantic-rich information with minimum spectrum resources \cite{zhang2023intelligent, zhang2023optimization}.}
		Especially in scenarios with limited resources (e.g., spectrum and power), allocation strategy is decided based on semantic importance. It means prioritizing the transmission of semantic-rich information using limited spectrum resources \cite{zhang2023drldrivena}. 
		However, hackers may execute the feature importance-aware attack \cite{wang2021feature} by injecting transferable adversarial examples (AEs) into semantic codecs. These examples not only divert the attention of codecs to areas with poor semantic content but also lead to the failure of scarce spectrum allocation (e.g., resources are occupied by semantic-poor information). As a result, the integrity of semantic codecs and the availability of services may be compromised. {At present, defense strategies against such threats in SemComNet remain unexplored and require further investigation.}
		
		\end{itemize}

	\vspace{-3mm}
	\subsection{Threats to Semantic Transmission Layer in SemComNet}\label{subsec:threat2}
	{This layer provides semantic transmission services tailored to the needs of agents and the semantics of sent data, enabling efficient and intelligent communication akin to human interactions \cite{xiao2023imitation}.} 
	However, SemCom, often referred to as ``AI + Communication" \cite{luo2022semantic, lu2023semanticsempowered, qin2022semantic}, inherit dual vulnerabilities. On the one hand, sensitive SI and computing tasks are exposed to broadcast channels, offering attackers the chance to launch attacks such as semantic jamming and semantic eavesdropping.
	{On the other hand, advanced AI-empowered semantic codecs are susceptible to manipulation through semantic adversarial or poisoning attacks. Such threats can alter and forge transmitted meanings, introducing semantic noise \cite{luo2022semantic} and disrupting communication services in SemComNet.} Below, we identify the typical attacks related to the semantic transmission process.
	
	\begin{itemize}
        \item \emph{{Semantic Data Poisoning Attack [T2.1].}}
        {
        {In SemComNet, adversaries can launch semantic poisoning attacks to undermine the integrity and availability of semantic models, reducing transmission accuracy \cite{du2023rethinking}. These attacks fall into two categories: data poisoning and model poisoning, which depend on whether the attacker contaminates the transceivers' training datasets or directly alters semantic models \cite{li2023secure, wang2022threats}.} In data poisoning attacks, attackers inject or modify a subset of the training data to degrade the performance of semantic codecs or induce unexpected behaviors. Such attacks are further divided into dirty-label poisoning and clean-label poisoning. {Dirty-label poisoning alters both the data and corresponding labels, making it easier to detect due to the inconsistency between the data and labels. In contrast, clean-label poisoning is subtler as only the data is manipulated.} Common defenses against data poisoning attacks include data preprocessing \cite{borgnia2021strong}, robust training \cite{wang2022threats}, and poison filtering mechanisms \cite{li2017learning} (as detailed in Sect.~\ref{subsec:semantic_data_defense}).}

        \item \emph{{Semantic Model Poisoning Attack [T2.2].}}
        {
        Different from data poisoning attacks that require access to training data, semantic model poisoning attacks directly target the model's parameters or training process. These attacks typically occur when attackers gain access to semantic codecs' learning and update procedures, which are out of the transceivers' control. {Such attacks are particularly severe in collaborative or distributed training settings, where multiple agents jointly train semantic models using local data and necessitate frequent model exchanges.} In such scenarios, attackers may control some clients and manipulate their updates during training, or implate backdoors that trigger specific behaviors under certain conditions \cite{tian2022comprehensive}.
        {Typical defensive strategies against model poisoning include model sanitization in centralized learning settings \cite{wang2019neural}, as well as robust aggregation and anomalous clients detection \cite{blanchard2017machine, zhao2021shielding} in collaborative training settings (as detailed in Sect. IV-A1)}.}
% . Model sanitization \cite{wang2019neural} involves detecting anomalies in model parameters and restoring them to their normal function, while robust aggregation and malicious clients detection \cite{blanchard2017machine, zhao2021shielding} could scrutinize model updates from clients, minimizing the impact of malicious contributions and improving the overall resilience of SemComNet

		\item \emph{Semantic Adversarial Attack [T2.3]}.
		Existing works \cite{zhang2019adversariala} show that deep neural networks (DNNs) are susceptible to AEs, i.e., subtle perturbations added to the inputs of a DL model, causing it to make incorrect decisions. Since SemComNet relies on advanced DL models, these vulnerabilities can be inherited, and the over-the-air transmission of SI further exacerbates the risk. {Specifically, as shown in the middle part of Fig.~\ref{fig:attack_layered}, the semantic adversarial attack can target both the sender and receiver sides. At the sender, attackers introduce subtle perturbations to distort semantic meaning and impede accurate SI extraction.} For instance, Hu \emph{et al.} \cite{hu2023robust} leverage iterative fast gradient sign method (FGSM) \cite{zhang2019adversariala} to generate sample-dependent semantic noise that misleads the transmitter, which may dramatically alter the semantics of data. At the receiver (potential via wireless channel), subtle perturbations to semantic decoder inputs may fail semantic decoding and misunderstanding between transceivers. 
        {To mitigate such attack, solutions within ML domain provide valuable insights, such as adversarial training \cite{hu2023robust, peng2022robustb, nan2023physicallayera}, defensive distillation \cite{zhang2019adversariala}, detection of AEs \cite{zhang2019adversariala}, and weight perturbation \cite{hu2023robust} (as detailed in Sect. \ref{subsec:adversarial_training})}.

% {Moreover, detecting such perturbations is more challenging due to inherent random noise in wireless channels.} Bahramali \emph{et al.} \cite{bahramali2021robust} discuss an optimization-based approach for adding minimal perturbations to legitimate signals in wireless DNN-based systems. These perturbations mimic natural wireless noise, misleading DNNs while ensuring their undetectability and maintaining attack effectiveness.
 
		\item \emph{Semantic Jamming Attack [T2.4]}.
		{The conventional jamming attack \cite{liu2017physical} disrupts communication by emitting interference or high-power signals \cite{zhao2024generative}, causing bit-level interruptions or degraded communication quality.} 
        {In contrast, semantic jamming attack targets the integrity of the transmitted data's semantics \cite{yang2023secure, tang2023ganinspired}, degrading the consistency and quality of semantic content and preventing accurate interpretation at the receiver \cite{du2023rethinking}.}
        For instance, Tang \emph{et al.} \cite{tang2023ganinspired} propose a semantic jammer that generates jamming data streams by randomly sampling from a Gaussian distribution. 
        However, their scheme \cite{tang2023ganinspired} assumes the jammer shares the same network structure as the sender, which may be impractical for real-world scenarios. 
        Besides, to counter such attack, Tang \emph{et al.} \cite{tang2023ganinspired} introduce a GAN-inspired framework, {where a semantic jammer (generator) and a robust receiver (discriminator) optimize strategies through adversarial gaming, thereby improving the receiver's resilience to semantic jamming. {Apart from directly countering jamming signals, techniques such as spread spectrum (detailed in Sect.~\ref{subsec:emerging_defense}) \cite{li2023ubiquitousa} and covert communication (detailed in Sect.~\ref{subsec:covertcomm_defense}) \cite{chen2023coverta, du2023rethinking, hu2024covert} can conceal the existence of SI transmission, making it difficult for jammers to detect and target SI.}}

		\item \emph{Semantic Eavesdropping Attack [T2.5]}.	
     {Eavesdropping, a common threat in traditional communication, involves intercepting transmitted data to expose sensitive information \cite{liu2017physical}.}
      In SemComNet, a semantic eavesdropping attack involves intercepting transmitted signals to infer their semantic meanings \cite{chen2023model, wang2024privacy}. 
      {To evaluate the threat, Du \emph{et al.} \cite{du2023rethinking} present the semantic secrecy outage probability (SSOP), which quantifies the success rate of intercepting and decoding SI. Defending against semantic eavesdropping is easier than traditional eavesdropping, as accurately decoding intercepted SI is highly challenging without access to paired semantic codecs and shared knowledge.}
    {However, two significant threats remain. 
    Firstly, the rapid advancement of GAI models enables attackers to use these models as general semantic decoders \cite{xie2024towards}, capable of semantic interpretation and intent understanding \cite{xu2024unleashing}.
    Secondly, during semantic codecs training, gradient sharing of models can expose sensitive data. As demonstrated in \cite{zhu2019deep}, observing gradients allows attackers to reconstruct fine-grained training data.}
    {To mitigate these risks, techniques such as semantic data encryption (detailed in Sect.~\ref{subsec:data_encryption_defense}), covert communication (discussed in Sect.~\ref{subsec:covertcomm_defense}) \cite{chen2023coverta, du2023rethinking, hu2024covert}, physical layer security technologies (discussed in Sects.~\ref{subsec:PLK_defense},~\ref{subsec:PLST_defense},~\ref{subsec:emerging_defense}), and quantum technologies (detailed in Sect.~\ref{subsec:quantum_defense}) could be employed for secure semantic transmission.}

		\item \emph{Threats from Semantic Relay [T2.6]}.
            {
            In wireless communication systems, relays extend communication range and improve data transmission reliability through cooperative strategies \cite{liu2017physical} such as amplify and forward (AF). However, untrusted relays introduce security risks, including eavesdropping, data tampering, or selectively dropping transmissions.}  In SemComNet, these threats persist and may lead to more severe consequences. Semantic relay nodes not only forward data but also translate semantic signals at the semantic level \cite{tang2024cooperative,luo2022autoencoderbaseda}, especially in scenarios where transceivers lack matched knowledge \cite{guo2024distributed,lin2024semanticforward}. 
            Typically, these relay nodes store rich background knowledge \cite{guo2024distributed}, including privacy-sensitive information about transceivers. On the one hand, curious relays may steal this private information, resulting in more severe privacy breaches. 
            On the other hand, malicious relays may exploit their advantageous position to inject viruses \cite{gaber2024malware} into SemComNet or manipulate the relayed SI via adding semantic noise \cite{kang2023adversariala}, potentially causing erroneous interpretations of semantics. 
            {However, current defense strategies against such threats in SemComNet remain unexplored and warrant further investigation.}

		\item \emph{Feedback Exploitation Attack [T2.7]}.
		In SemComNet, the transmitter adjusts its transmission strategy according to feedback from the receiver and physical channel (e.g., CSI). 
        If the receiver struggles to interpret the content, the sender transmits more detailed, low-level SI to aid understanding \cite{zhou2022adaptive}.  However, two security issues arise during this feedback process, i.e., feedback leakage and feedback tampering.
		The former occurs when malicious entities analyze feedback to infer sensitive information about agents, such as their preferences and behavior patterns, potentially constructing detailed agents' profiles and threatening confidentiality \cite{shen2023secure}. Meanwhile, the latter refers to malicious feedback modification, leading to inappropriate adjustments in transmission rates. That potentially causes communication failures and affects the availability of SemComNet. 
        {Currently, the defense against such attack in the SemComNet domain remains an open issue, warranting further research.}
		
		\end{itemize}

		% \item \emph{Threats to privacy exposure}

	\vspace{-3mm}
	\subsection{Threats to Cognitive Sensing Layer in SemComNet}\label{subsec:threat3}
	As described in Sect.~\ref{cognitive_layer}, the cognitive sensing layer in SemComNet is distinguished for its advanced cognitive capabilities, including environment perception, agent intention inference, knowledge discovery, and private KBs establishment. However, these capabilities also introduce unique vulnerabilities during the data \& knowledge perception, processing, and sharing phases.
	Various threats exist within the cognitive sensing layer, including but not limited to false perception data injection, free-riding, and cognitive manipulation via ``information bombs". Additionally, risks such as unauthorized private KBs access and private KBs poisoning attacks, mirror the vulnerabilities found in the control layer. We list typical threats to the cognitive sensing layer in SemComNet as below.
	
 \tikzstyle{my-box}=[
rectangle,
draw=hidden-draw,
rounded corners,
text opacity=1,
minimum height=1.5em,
minimum width=5em,
inner sep=2pt,
align=center,
fill opacity=.5,
line width=0.8pt,
]

% \tikzstyle{layer1}=[my-box, fill=blue!30,align=left,font=\normalsize,
% inner xsep=2pt,
% inner ysep=4pt,
% line width=0.8pt,]
% \tikzstyle{layer2}=[my-box, fill=green!30,align=left,font=\normalsize,
% inner xsep=2pt,
% inner ysep=4pt,
% line width=0.8pt,]
% \tikzstyle{layer3}=[my-box, fill=red!30,align=left,font=\normalsize,
% inner xsep=2pt,
% inner ysep=4pt,
% line width=0.8pt,]

% \tikzstyle{layer1}=[my-box, fill=cyan!20,align=left,font=\normalsize,
% inner xsep=2pt,
% inner ysep=4pt,
% line width=0.8pt]
% \tikzstyle{layer2}=[my-box, fill=violet!20,align=left,font=\normalsize,
% inner xsep=2pt,
% inner ysep=4pt,
% line width=0.8pt]
% \tikzstyle{layer3}=[my-box, fill=yellow!20,align=left,font=\normalsize,
% inner xsep=2pt,
% inner ysep=4pt,
% line width=0.8pt]

\definecolor{layercolor1}{RGB}{57,127,199}
\definecolor{layercolor2}{RGB}{241,182,86}
\definecolor{layercolor3}{RGB}{4,6,118}
\tikzstyle{layer1}=[my-box, fill=layercolor1!40,align=left,font=\normalsize,
inner xsep=2pt,
inner ysep=4pt,
line width=0.8pt]
\tikzstyle{layer2}=[my-box, fill=layercolor2!40,align=left,font=\normalsize,
inner xsep=2pt,
inner ysep=4pt,
line width=0.8pt]
\tikzstyle{layer3}=[my-box, fill=layercolor3!40,align=left,font=\normalsize,
inner xsep=2pt,
inner ysep=4pt,
line width=0.8pt]

\tikzstyle{leaf}=[my-box, minimum height=1.5em,
fill=hidden-pink!180, text=black, align=left,font=\normalsize,
inner xsep=2pt,
inner ysep=4pt,
line width=0.8pt,
]
\begin{figure*}[!t]
	\centering
	\resizebox{0.94\textwidth}{!}{
		\begin{forest}
			forked edges,
			for tree={
				grow=east,
				reversed=true,
				anchor=base west,
				parent anchor=east,
				child anchor=west,
				base=left,
				font=\large,
				rectangle,
				draw=hidden-draw,
				rounded corners,
				align=left,
				minimum width=4em,
				edge+={darkgray, line width=1pt},
				s sep=3pt,
				inner xsep=2pt,
				inner ysep=3pt,
				line width=0.8pt,
				ver/.style={rotate=90, child anchor=north, parent anchor=south, anchor=center},
			},
			where level=1{text width=5.4em,font=\normalsize,}{},
			where level=2{text width=14.5em,font=\normalsize,}{},
			where level=3{text width=10.5em,font=\normalsize,}{},
			where level=3{text width=10em,font=\normalsize,}{},
			[
			Security/Privacy Threats and Corresponding Solutions in SemComNet, ver
			% 模型相关
			[Semantic \\Model\\-Related 
			[{[T2.2] Semantic Model Poisoning},layer2
			[
			\S \ref{subsec:semantic_model_defense}  Anti-poisoning 
			[ 
			{Modol sanitization \cite{wang2019neural}, Robust aggregation \cite{blanchard2017machine}, \\ Anonmolous updates detection \cite{zhao2021shielding}}
			, leaf, text width=21.3em
			]
			]]
			[{[T2.3] Semantic Adversarial},layer2
			[
			\S \ref{subsec:adversarial_training} Adversarial \\ Training
			[ 
			{R-DeepSC \cite{peng2022robustb}, SemMixed \cite{nan2023physicallayera}, Semantic distance minimization \cite{kang2023adversariala}}
			, leaf, text width=28.8em
			]
			]
			[
			\S \ref{subsec:adversarial_training} Others 
			[   
			{Defensive distillation \cite{zhang2019adversariala}, Weight perturbation \cite{hu2023robust}, Detector \cite{zhang2019adversariala}}
			, leaf, text width=27.8em]
			]]
   			[{[T1.6] IP Infringement},layer1
			[\S \ref{subsec:IP_protection} IP Protection 
			[{DNN-watermarking \cite{zhang2018protectinga, zhang2022deepa}, Blockchain \cite{wang2023surveyc}}, leaf, text width=20.1em]
			]
			]
			]
			% 传输相关 \textcolor{blue}{IV-B3}\,\&\textcolor{blue}{IV-B6}\,\&\textcolor{blue}{IV-C2} 
			[Semantic \\Transmission\\-Related
			[{[T2.4] Semantic Jamming},layer2
			[\S    \ref{subsec:PLST_defense}\&\ref{subsec:emerging_defense}\&\ref{subsec:reliable_transmission_defense} \\ PLS 
			[{Spread spectrum \cite{li2023ubiquitousa}, Anti-jamming \cite{mitev2023whata, tang2023ganinspired}, Jamming recognition \cite{zhao2024generative}, \\ Reliable transmission \cite{zhou2022adaptive, wang2023wireless}}, leaf, text width=31.8em]
			]
			[\S \ref{subsec:covertcomm_defense} Covert Comm. 
			[{PL-LPDC \cite{wang2023multiagent,hu2024covert}, Steganography \cite{xie2022security, yamaguchi2020physicallayer}}, leaf, text width=21.2em]% , GAN-inspired anti-jamming \cite{tang2023ganinspired}, GAN-based jamming recognition \cite{zhao2024generative}, \\Semantic noise suppression \cite{han2023semanticpreserved, xiao2023imitation}
			]
			]	
			[{[T2.5] Semantic Eavesdropping},layer2
			[\S \ref{subsec:data_encryption_defense} Encryption 			[{Classic encryption \cite{chen2023model, qin2023securinga}, MPC \cite{knott2021crypten}, DL-based encryption \cite{xu2023deepa, luo2023encrypted}}, leaf, text width=31.8em]]
			[\S \ref{subsec:PLK_defense}\&\ref{subsec:PLST_defense}\&\ref{subsec:emerging_defense} \\ PLS  
			[{PLKG \cite{zhao2022semkey, mitev2023whata}, Secure transmission \cite{hu2023interference, lin2018physicallayer, yin2023multidomain, li2021exploiting}, RIS \cite{wang2023starrisassisted, wang2023starrisassisted}}, leaf, text width=30.4em]]%    \cite{wang2023starrisassisted, du2023semantica, li2023secureb}, AN \cite{hu2023interference}\\ Beamforming \cite{lin2018physicallayer}, Power allocation \cite{yin2023multidomain}, Friendly jamming \cite{li2021exploiting}
            [\S \ref{subsec:covertcomm_defense} Covert Comm.
			[{PL-LPDC \cite{wang2023multiagent,hu2024covert}, Steganography \cite{xie2022security, yamaguchi2020physicallayer}}, leaf, text width=21.2em]% , GAN-inspired anti-jamming \cite{tang2023ganinspired}, GAN-based jamming recognition \cite{zhao2024generative}, \\Semantic noise suppression \cite{han2023semanticpreserved, xiao2023imitation}
			]			
            [\S \ref{subsec:quantum_defense} Quantum 
			[{Quantum key distribution \cite{kaewpuang2023cooperative}, Quantum semantic encoding \cite{khalid2023quantum}}, leaf, text width = 27.7em]
			]
			]
                % 无解决方案的
                [ \lbrack {T2.6}\rbrack \ Threats from Semantic Relay ,layer2]
                [ \lbrack {T2.7}\rbrack \ Feedback Exploitation,layer2]
			]
			% 可靠性问题
			[Reliability \\Issues
%			[{[T1.1] Denial of Service}
%			[{GAI-aided contextual awareness \cite{yao2024survey}}, leaf, text width=15.3em]
%			]
			% Intrusion detection \cite{Franco2021Survey}, Anomaly detection \cite{jiang2024adgym},\\ Honeynets \cite{Franco2021Survey}, 
   	        [{[T1.1] Sponge Example},layer1
		[\S \ref{subsec:sponge_examples_defense} Anti-sponge
		[{Sponge example defense \cite{shumailov2021sponge,wong2020fast}, Sponge example detection \cite{chen2020stateful}}, leaf, text width=29.5em]
		]
		]
			[{[T1.4] Mismatches of KBs},layer1
			[\S \ref{subsec:relay_defense} Translator 			[{ Semantic-forward relay \cite{lin2024semanticforward, luo2022autoencoderbaseda}, Multi-modal relay \cite{guo2024distributed}}, leaf, text width=24.8em]
			]
			[\S \ref{subsec:GAI_defense}\&\ref{subsec:reasoning_defense}  SI \\ Resilient Understanding 
			[{GAI-empowered semantic noise suppression \cite{erdemir2023generative, chen2023commin}, Reasoning-driven \\ disambiguation \cite{jiang2022reliablea, yao2021typing, thomas2023neurosymbolic, xiao2023imitation}}, leaf, text width=31.2em]
			]
			]
                [ \lbrack {T1.8}\rbrack \  Semantic-aware Resource\\  Allocation Threats,layer1]
			]
			% 信任相关
			[Trust Issues
			[{[T3.2] False Data Injection},layer3
			[{\S \ref{subsec:auth_defense} Authentication}
			[{Passive authentication \cite{gao2023esanet}, Active authentication \cite{jorswieck2015broadcasting, tan2024optimization}}, leaf, text width=26em
			]]
			%[{Reputation\\ system}
			%[{ML-based trust assessment \cite{jayasinghe2019machine}, DRL-driven reputation update \cite{gyawali2021deep}, \\ Reputation management \cite{wang2021survey}, E-R trust mechanism \cite{truong2019trust},}, leaf, text width=29.3em]]
			[{\S \ref{subsec:bockchain_trustless_defense} Provenance}
			[{ProvChain \cite{liang2017provchain}, Auditable data sharing \cite{wang2021spdsa}}, leaf, text width=19.2em
			]]
			]
                % Free-riding
   			[{[T3.3] Free-riding Attack},layer3
			[
			\S \ref{subsec:trust_eval_defense} Trust Evaluation
			[ 
			{ML-based trust assessment \cite{jayasinghe2019machine}, DRL-driven reputation update \cite{gyawali2021deep}, \\ Trust-impact attribute aggregation \cite{wang2021survey, truong2019trust}}
			, leaf, text width=29.2em
			]
			]
			[
			\S \ref{Data_Knowledge_AC} AC  
			[   
			{Role-based AC \cite{sultan2023rolebased}, Attribute-based AC \cite{xue2019attributebased, xu2021integrated}, Purpose-based AC \cite{xue2023sparkac}}
			, leaf, text width=33.2em]
			]]
			[{[T3.4] Impersonation Attack},layer3
			[\S \ref{subsec:auth_defense} Authentication 
			[{Passive authentication \cite{gao2023esanet}, Active authentication \cite{jorswieck2015broadcasting, tan2024optimization}}, leaf, text width=26em]
			]
			]
			[{[T3.5] Trust Management Threats},layer3
			[{\S \ref{subsec:bockchain_trustless_defense} Trust-free}
			[{Identity management \cite{bao2023pbidm,xu2020identity}, Distributed authentication \cite{shen2020blockchainassisted, chen2022xautha}}, leaf, text width=30.4em]
			]
			[{\S \ref{subsec:trust_eval_defense} Trust Evaluation}
			[{ML-based trust assessment \cite{jayasinghe2019machine}, DRL-driven reputation update \cite{gyawali2021deep}, \\ Trustimpact attribute aggregation \cite{wang2021survey, truong2019trust}}, leaf, text width=29.7em]
			]
			]
			]
			% 数据、知识相关
			[Data \& \\ Knowledge\\-Related
   			[{[T2.1] Semantic Data Poisoning},layer2
			[
			\S \ref{subsec:semantic_data_defense}  Anti-poisoning \\  
			[ 
			{Data precessing \cite{borgnia2021strong}, Poisoning filter \cite{li2017learning}, Robustness training \cite{wang2022threats}
			}
			, leaf, text width=28em
			]
			]
			]
			[{[T1.2] KBs Poisoning}\\
			{[T1.5] Unauthorized KBs Access},layer1
			%[AuthN \S IV-D1
			%[{Multi-factor authentication \cite{wang2023understanding}, Cross-domain authentication \cite{shen2020blockchainassisted, chen2022xautha} }, leaf, text width=31.1em]]
			[{\S \ref{Data_Knowledge_AC} AC}
			[{Role-based AC \cite{sultan2023rolebased}, Attribute-based AC \cite{xue2019attributebased, xu2021integrated}, Purpose-based AC \cite{xue2023sparkac}}, leaf, text width=33.2em]
			]
			[{\S \ref{Data_Knowledge_Blockchain} Blockchain}
			[{ProvChain \cite{liang2017provchain},  Blockchain-based secure sharing \cite{wang2022blockchainbaseda, fotiou2016decentralized}, \\ Provable data possession \cite{wang2021blockchainbased}, Trustworthy redact \cite{ateniese2017redactable}}, leaf, text width=26.3em]
			]
			]
			[{[T1.7] Query Misguiding},layer1
			[{Filtering of poisoning facts \cite{xi2023security}, Training with adversarial queries \cite{bahramali2021robust, xi2023security}}, leaf, text width=32.2em]
			]
			[{[T3.1] Cognitive Manipulation},layer3
			[{Adversarial information detection \cite{xia2024mmnet}, GAI-aided detecting \cite{yao2024survey} }, leaf, text width=27.8em]
			]	
			[{[T3.6] Malware},layer3
			[ {Malware detection \cite{oz2022survey, yao2024survey}, Malware defense \cite{mcintosh2021ransomware}, Malware prevention \cite{mcintosh2021ransomware}}, leaf, text width=34.1em]
			]
               [ \lbrack {T1.3}\rbrack \ KBs Desynochonization,layer1]]
			]
		\end{forest}
	}
% 图例
 \begin{tikzpicture}
    % Legend entry for Control Layer
    \draw[fill=layercolor1!20] (0,0) rectangle (0.3,0.3); % Color box for Control Layer
    \node[right] at (0.35,0.15) {\footnotesize Control Layer Related}; % Text label for Control Layer

    % Legend entry for Transmission Layer
    \draw[fill=layercolor2!20] (4.3,0) rectangle (4.6,0.3); % Color box for Transmission Layer
    \node[right] at (4.65,0.15) {\footnotesize Semantic Transmission Layer Related}; % Text label for Transmission Layer

    % Legend entry for Sensing Layer
    \draw[fill=layercolor3!20] (10,0) rectangle (10.3,0.3); % Color box for Sensing Layer
    \node[right] at (10.35,0.15) {\footnotesize Cognitive Sensing Layer Related}; % Text label for Sensing Layer
\end{tikzpicture}

	\caption{{The taxonomy of security/privacy threats to SemComNet from five aspects (i.e., model, transmission, reliability, trust, and data \& knowledge) and corresponding existing/potential defenses in the SemComNet.}}
	\label{threats_solutions}
\end{figure*}

	\begin{itemize}
		\item \emph{Threats to Cognitive Manipulation [T3.1]}.
        These threats in SemComNet involve deliberate interference with agents' cognitive processes to influence their judgment and reasoning. Cognition manipulation can stem from adversarial information dissemination, such as spreading rumors, or deceptive content (e.g., Deepfake images generated by GANs \cite{xia2024mmnet}). Such tactics compromise decision-making and the integrity of SemComNet.
        Additionally, the easy-to-use GAI models enable malicious agents to create biased ``information bombs" \cite{yao2024survey} via malicious prompts, disrupting public cognition and shaping virtual opinion leaders.
        {Defenses against these threats include adversarial information detection techniques, categorized as spatial-based, frequency-based, and data-driven methods \cite{xia2024mmnet}. Spatial-based techniques enhance local forgery detection by focusing on specific spatial regions, frequency-based methods analyze differences in the frequency domain, and data-driven approaches improve detection generalization by training with diverse fake data \cite{yao2024survey}.}

		\item \emph{False Data Injection Attack [T3.2]}.
		{To enrich private KBs and improve environmental understanding, agents in SemComNet autonomously collect environmental semantics via on-body sensors \cite{qin2023generalized}.}
        However, as illustrated in the bottom part of Fig.~\ref{fig:attack_layered}, attackers might attempt to mislead SemComNet by injecting false data. 
        {For instance, GPS spoofing attack \cite{wang2023infrastructure} can override legitimate signals with counterfeit ones, causing GPS receivers to misinterpret geographical information. }
        This deception can introduce flawed environmental semantics into the KBs, leading to incorrect decisions and undermining SemComNet's integrity.  {To effectively mitigate such threat, techniques such as robust authentication and provenance tracking may be beneficial, as discussed in Sects.~\ref{subsec:auth_defense} and ~\ref{subsec:bockchain_trustless_defense}.}

		\item \emph{Free-riding Attack [T3.3]}.
		{In multi-agent SemComNet, collaboration is essential for task completion (e.g., collaborate perception) and ensuring smooth system operations. However, selfish agents may avoid contributions while reaping benefits or accessing its resources \cite{wang2023surveyc}.}
		For instance, selfish agents may deliberately provide low-quality \cite{wang2021spdsa} or semantic-poor sensing data to save their energy and resources.
        {If widespread, such free-riding behavior can deplete system resources, reduce service availability, and degrade the QoE of legitimate agents.} {Mitigation strategies include trust evaluation (detailed in Sect.~\ref{subsec:trust_eval_defense}) and effective AC (discussed in Sect.~\ref{Data_Knowledge_AC}).}

		\item  \emph{Impersonation Attack [T3.4]}.
		{Adversaries can exploit authentication vulnerabilities in SemComNet to impersonate legitimate agents, compromising system confidentiality \cite{mitev2023whata}. This could be achieved by stealing the credentials through password cracking or phishing \cite{oz2022survey}.} Once successful, attackers can misuse compromised identities for malicious activities, such as submitting false sensory data or violating privacy. The authentication mechanism (discussed in Sect.~\ref{subsec:auth_defense}) with information-theoretic security guarantees can help prevent such impersonation threats.

		\item  \emph{Threats to Trust Management [T3.5]}.
        In the dynamic environment of SemComNet, where autonomous agents frequently change their behavior (e.g., uploading sensory data), managing trust is crucial to mitigate interaction risks \cite{wang2022blockchainbaseda}. However, establishing an effective trust management system is challenging due to the need for real-time monitoring of agent behaviors and reputations \cite{wang2021survey}, which may compromise confidentiality. Furthermore, agent interactions often involve limited or incomplete data sharing, increasing the complexity of determining the reputation of other agents and establishing trust relationships. To address these issues, trust-free architectures (discussed in Sect.~\ref{subsec:bockchain_trustless_defense}) and efficient trust evaluation mechanisms (covered in Sect.~\ref{subsec:trust_eval_defense}) offer alternative solutions to enhance trust management in SemComNet.

		\item \emph{Malware Attack [T3.6]}.
		This attack compromises service availability, confidentiality, and integrity of data/knowledge in SemComNet.
		Specifically, malware can swiftly spread between agents, causing deviations in their behavior patterns and roles. For instance, malicious code or instructions \cite{yao2024survey} injected into a benign agent can turn it into a aggressive entity. 
		In addition, certain malware families such as ransomware, spyware, and worms may steal private data, erase sensory data, or delete accumulated knowledge \cite{oz2022survey}.
		{To resist these threats, malware detection, defense, and prevention are crucial \cite{mcintosh2021ransomware}.} 
		Malware detection helps identify threats before damage occurs \cite{oz2022survey}, while defense and prevention focus on mitigating and reversing malware activities before and after infection \cite{gaber2024malware}. In practice, a multilayered approach combining these strategies can enhance resilience against complex malware threats in SemComNet.

	\end{itemize}

	\vspace{-3mm}
	\subsection{Summary and Lessons Learned}

{
SemComNet inherit several security/privacy threats from traditional networks, with varying impacts. These can be classified into: i) enhanced impact of existing attacks. Traditional attacks, such as threats from relay nodes, become more severe in SemComNet, leading to increased risks of privacy breaches and semantic manipulation. ii) Reduced impact of existing attacks. Some existing attacks are easier to defend in SemComNet. For instance, semantic eavesdropping is harder to execute, as decoding intercepted SI without proper codecs and knowledge is highly challenges for attackers. iii) Variant of existing attacks. Traditional attacks such as DoS evolve into sponge example attacks, exploiting SemComNet’s reliance on computational resources to cause excessive energy consumption and delays.
Additionally, some brand-new threats arise from the cutting-edge technologies underpinning SemComNet, such as its reliance on multiple KBs and AI techniques. These introduce risks such as KB poisoning, unauthorized access, and desynchronization attacks.}

Besides, cross-layer attacks present significant challenges in the multi-layered architecture of SemComNet. Such attacks exploit interactions and dependencies among layers, complicating detection and defense. 
Attackers can exploit the dependencies between layers, amplifying the effects of their attacks. For instance, cognitive manipulation in the sensing layer may contaminate agents' private KBs, which can then spread to shared KBs in the control layer. The contaminated shared KBs can adversely affect the semantic transmission layer, leading to incorrect semantic interpretations and reduced communication performance. 
The interconnected nature of SemComNet amplifies potential damage, highlighting the need for robust, multi-layered defense mechanisms to detect and mitigate threats at every level.

In summary, we have categorized SemComNet's security and privacy threats (i.e., from Sect.~\ref{subsec:threat1} to Sect.~\ref{subsec:threat3}) across its three functional layers, i.e., control layer, semantic transmission layer, and cognitive sensing layer. Besides, as depicted in Fig.~\ref{threats_solutions}, we have reviewed existing/potential defense approaches for the above security/privacy issues within SemComNet from five perspectives: semantic model security, semantic transmission security, reliability, trust, and data/knowledge security.

	\section{Security and Privacy Countermeasures in SemComNet}
	In this section, we provide an in-depth and up-to-date discussion of security and privacy defenses tailored for SemComNet across five crucial aspects, i.e., semantic model security, semantic transmission security, reliable SemComNet, trust management in SemComNet, and data \& knowledge security in detail.
	%-----------------------------------------------------------------------
	\vspace{-3.0mm}
	\subsection{Semantic Model Security} \label{subsec:defense1}

\begin{table*}[!t]
	\centering \setlength{\abovecaptionskip}{0cm}
	\caption{Summary of Key Literature on Semantic Model Security in Semantic Communication Networks}\label{Summary1}
	\begin{tabular}{cclc}\hline
		\textbf{Ref.} & \textbf{\begin{tabular}[c]{@{}c@{}}Security\\ Threat\end{tabular}} &  \multicolumn{1}{c}{\textbf{\begin{tabular}[c]{@{}l@{}}$\star$ Purpose\\$\bullet$ Advantages\\$\circ$ Limitations\\ $\dagger$ {Evaluation Metrics}\end{tabular}}} & \textbf{\begin{tabular}[c]{@{}c@{}}Utilized Technology\end{tabular}} \\\hline

		% Anti-poisoning Centeral _backdoor setting
		{\cite{wang2019neural}}& {{\begin{tabular}[l]{@{}c@{}}Semantic \\ backdoor  attack \end{tabular}}}  
		& {\begin{tabular}[l]{@{}l@{}}
				$\star$ Detect whether a model is backdoored exploit activation statistics\\
				%Prevents contamination of semantic models through poisoned data\\
				$\bullet$ Robust and general tools for detecting and mitigating backdoor attack \\
				%Increases security and integrity of model training; adaptable to different poisoning strategies\\
				$\circ$ Unscalable to large-scale model and strong assumption (e.g., clean training dataset) \\
				% Heavily rely on a clean training dataset
				$\dagger$ Attack success rate, classification accuracy, and false positive/negative rate
				%Requires ongoing adjustment to counter novel poisoning techniques; may impact model performance
		\end{tabular}} 
		
		& \begin{tabular}[c]{@{}c@{}} Backdoor detection,\\ optimization\end{tabular} 
		\\\hline
		
		% Anti-poisoning Federated__Poisoning setting
		{\cite{zhao2021shielding}}& {{\begin{tabular}[l]{@{}c@{}}Semantic \\ poisoning  attack \end{tabular}}}  
		& {\begin{tabular}[l]{@{}l@{}}
				$\star$ Leverage clean data through cross-validation for detecting poisoning updates \\
				$\bullet$ High defense strength and robustness without compromising model accuracy\\
				$\circ$ 
				Limited by validation data size and susceptible to malicious clients \\
				%The detection effect may be limited by validation data size and malicious participants \\

				% Requires ongoing adjustment to counter novel poisoning techniques; may impact model performance\\
				$\dagger$ Model accuracy, computation cost, and probability of evading the detection
		\end{tabular}} 
		& \begin{tabular}[c]{@{}c@{}} Model   validation \end{tabular} 
		\\\hline
  
		% Adversarial Training Example
		{\cite{peng2022robustb}}& {{\begin{tabular}[l]{@{}c@{}}Semantic \\ adversarial attack  \end{tabular}}}  
		& {\begin{tabular}[l]{@{}l@{}}
				$\star$ Enhance the resilience of semantic codecs against adversarial examples\\
				$\bullet$ High robustness, semantic fidelity, and compatibility with existing training pipelines\\
				$\circ$ 
				Increased computational cost, difficulty in converging, and low defense generalization\\ 
				% May not generalize well to unseen attack methods; can increase training complexity\\
				$\dagger$ BLEU score and BERT score \\
		\end{tabular}} & Adversarial training \\\hline
  
		% Watermark-based IP Protection Example
		{\cite{zhang2022deepa}}& {{\begin{tabular}[l]{@{}c@{}}
					Semantic model stealing,\\	surrogate model attacks \end{tabular}}} 
		&{{\begin{tabular}[l]{@{}l@{}}
					$\star$ 
					Trace illegal usage and copyright verification of AI models \\ 
					$\bullet$ 
					High watermark capacity and strong generalization ability \\
					$\circ$ 
					Cannot resist ambiguity attack and lack resilience against pre-processing techniques\\
					$\dagger$
					Peak SNR, SSIM, normalized correlation, and success rate of watermark extracted
					% PSNR and SSIM for visual quality, and normalized correlation,
					
		\end{tabular}}} 
		& \begin{tabular}[c]{@{}c@{}} Model   \\watermarking \end{tabular} 
		\\\hline

		% % Blockchain-based IP Protection Example
		% {\cite{somy2019ownership}}& {{\begin{tabular}[l]{@{}c@{}}IP infringement of \\ shared semantic models \end{tabular}}} 
		% & {\begin{tabular}[l]{@{}l@{}}
		% 		$\star$ Create a verifiable, traceable, unalterable record of transactions via blockchain ledger\\
		% 		$\bullet$ Preserved ownership/privacy of semantic assets and enhanced trust among agents\\
		% 		$\circ$ Performance and scalability concerns in large-scale distributed setting\\
		% 	   $\dagger$ Latency, throughput, and transaction efficiency
		% \end{tabular}} 
		% & Blockchain\\\hline
		
	\end{tabular}
\end{table*}

    Advanced AI techniques (e.g., DL and RL) empower semantic codecs to efficiently extract and interpret SI. As the performance of SemComNet is heavily dependent on the capabilities of these AI-driven semantic codecs, ensuring their security becomes crucial \cite{luo2022semantic, yang2023semanticb}. In the following, we discuss countermeasures to protect semantic models from poisoning attacks, AEs, and sponge examples, as well as strategies to safeguard the IP of these models.

	\subsubsection{{Defense Strategies Against Semantic Model Poisoning}} \label{subsec:semantic_model_defense}
        Semantic model poisoning attacks directly manipulate the parameters of semantic codecs, rather than contaminating training data.
        In centralized learning, defenses against model poisoning focus on model sanitization, which detects and mitigates malicious parameter modifications. Techniques such as fine-pruning, which removes dormant or suspicious neurons and retrains the model on clean data, can restore normal model function. For instance, Wang \emph{et al.} \cite{wang2019neural} present a system for detecting and mitigating DNNs backdoor attacks by exploiting the model's sensitivity to input perturbations. However, this method faces scalability issues for large models and assumes the availability of clean training data.
        In distributed learning, robust aggregation methods such as secure aggregation protocols \cite{blanchard2017machine} help prevent malicious updates. For instance, Krum as a Byzantine-resilient aggregation rule, filters out abnormal model weights by sorting and selecting participants’ updates, thus enhancing the robustness of the global model \cite{blanchard2017machine}. However, Krum’s complexity increases with more participants due to pairwise distance calculations. An alternative approach involves detecting malicious clients. For instance, Zhao \emph{et al.} \cite{zhao2021shielding} propose a client-side cross-validation scheme where each update is evaluated using other clients' local data. This method adjusts update weights based on the evaluations during aggregation.

	\subsubsection{Adversarial Training for Semantic Noise Resistant SemComNet}	\label{subsec:adversarial_training}
	SemComNet are susceptible to semantic adversarial attack due to the vulnerability of DNN-based codecs and the broadcast of wireless medium \cite{nan2023physicallayera,du2023rethinking,kang2023adversariala}. These attacks may induce semantic noise that will gradually distort the desired meaning conveyed in SI.
	Current defense schemes span diverse categories \cite{zhang2019adversariala}, i.e., adversarial training, defensive distillation, AEs detector, and methods involving weight concealment and interference \cite{hu2023robust}. 
	Among these approaches, adversarial training stands out as a simple yet highly effective method. It has been extensively studied and proven to enhance the resilience of semantic models to AEs \cite{peng2022robustb, du2023rethinking,nan2023physicallayera}.
	This approach involves incorporating AEs into the training data and continually generating new AEs during each iteration of training process.
	For instance, Peng \emph{et al.} \cite{peng2022robustb} employ adversarial training, specifically the fast gradient method to identify perturbations that significantly disrupt the system and subsequently train the system to withstand these AEs.	
    {In \cite{peng2022robustb}, the \textit{Bilingual Evaluation Understudy} (BLEU) score is utilized to assess the semantic quality of received data, and the \textit{Bidirectional Encoder Representations from Transformers} (BERT) score measures the semantic similarity between text sentences. The results, indicated by high BLEU and BERT scores, demonstrate the proposed scheme's effectiveness in mitigating semantic distortions caused by adversarial noise.}

	However, this training strategy might increase computational costs and may not generalize well to AEs from other adversaries because it trains the model on narrowly crafted AEs, which increases the risk of overfitting.
	Besides, the work \cite{peng2022robustb} does not consider the interference caused by the massive connections between agents, which can introduce various adversarial perturbations in wireless channels.
	To simulate this, Nan \emph{et al.} \cite{nan2023physicallayera} introduce \emph{SemAdv} to create semantic-oriented perturbations for physical-layer adversarial attacks during SI transmission. To defend this, they propose an adversarial training approach \emph{SemMixed} to enhance the resilience of SemComNet against various physical adversarial perturbations. 
    {The experiments not only use standard evaluation metrics (e.g., PSNR and SSIM) but also introduce a new security metric called \textit{misleading rate}, which evaluates the model's tendency to mislead predictions toward specific image categories. Simulation results highlight that the proposed defense strategy significantly reduces the attack success rate.}

\subsubsection{{Safeguarding SemComNet Against Sponge Examples}}\label{subsec:sponge_examples_defense}
To mitigate the threat posed by sponge examples, a straightforward solution is to adopt worst-case performance analysis, as proposed by Shumailov \emph{et al.} in \cite{shumailov2021sponge}. By establishing processing time and energy thresholds from model profiling with natural examples, inputs surpassing these limits will be rejected, thereby helping to ensure system availability.
However, this solution has limitations. Specifically, setting optimal thresholds requires careful calibration tailored to the hardware platform and model performance. 
If thresholds are not correctly set, the defense could yield suboptimal results. Additionally, attackers could manipulate sponge examples to stay within these thresholds while still consuming excessive resources and causing delays. With large volumes of sponge examples, the system's resources may be overwhelmed, potentially leading to a DoS situation.
Thus, relying solely on threshold-based defenses may be insufficient. To bolster the robustness of SemComNet, supplementary defense mechanisms from both defense \cite{wong2020fast} and detection \cite{chen2020stateful} could be considered. For instance, Wong \emph{et al.} \cite{wong2020fast} propose training models with AEs generated using techniques such as FGSM. This approach improves semantic models' resilience to perturbations and reduces their susceptibility to manipulation by sponge examples.

When direct defense strategies are not feasible, detection becomes critical. For instance, Chen \emph{et al.} \cite{chen2020stateful} introduce a novel detection scheme by analyzing the sequence of queries made to the victim model. Unlike traditional defenses that focus on identifying malicious inputs, they leverage the temporal nature of attacks and track the similarity between successive queries. By calculating the k-nearest-neighbor distance between a new query and stored examples, the system can detect suspicious patterns indicative of ongoing attacks. Simulation results show its effectiveness even under black-box attack scenarios, where attackers may use query blinding techniques to obscure the sequence. Additionally, this detection method is compatible with existing defenses against zero-query attacks \cite{chen2020stateful}, offering a comprehensive solution for detecting adversarial examples.

    \subsubsection{{IP Protection for Securing Model in SemComNet}} \label{subsec:IP_protection}
{Powerful semantic codecs are crucial for intelligent communication tasks in SemComNet, but training process is often expensive and time-consuming. To protect the IP of these models, watermarks are vital \cite{zhang2018protectinga, zhang2022deepa}.} In \cite{zhang2018protectinga}, Zhang \emph{et al.} extend digital watermarking from multimedia to DNNs and propose three DNN-compatible watermark generation algorithms to confirm model ownership via detecting preset patterns. However, the work \cite{zhang2018protectinga} fails to resist surrogate model attack\footnote{A surrogate model attack targets the IP of DL models by creating a substitute model that mimics the target model’s behavior. In a black-box scenario, the attacker does not have access to the target model's internal structure or parameters but can observe its input-output pairs. By feeding input samples into the target model to obtain predictions, the attacker constructs a training dataset and trains a surrogate model to approximate the target model. After several iterations, the attacker successfully creates a model that behaves similarly to the original, thus stealing its IP.}. In response, Zhang \emph{et al.} \cite{zhang2022deepa} introduce a task-agnostic method that embeds a spatially invisible watermark within the networks' outputs, maintaining high extraction success rates even against surrogate models. Apart from watermark-based IP protection, blockchain technology offers a trustworthy and traceable mechanism for semantic codecs and knowledge sharing within SemComNet \cite{wang2023surveyc},  providing an additional layer of trustworthiness and transparency.

%For instance, SemComNet has developed DL-based semantic codecs for VR video transmission which is valuable due to the extensive data and computational resources required for training. An attacker inputs query samples $x$ into the victim semantic model to obtain the predicted output $f(x)$.

	\vspace{-3mm}
	%-----------------------------------------------------------------------
	\subsection{Semantic Transmission Security}\label{subsec:defense2}

\begin{table*}[!t]
	\centering \setlength{\abovecaptionskip}{0cm}
	\caption{Summary of Key Literature on Semantic Transmission Security in Semantic Communication Networks}\label{Summary2}
	\begin{tabular}{cclc}\hline
		\textbf{Ref.} & \textbf{\begin{tabular}[c]{@{}c@{}}Security\\ Threat\end{tabular}} &  \multicolumn{1}{c}{\textbf{\begin{tabular}[c]{@{}l@{}}$\star$ Purpose\\$\bullet$ Advantages\\$\circ$ Limitations\\ $\dagger$ {Evaluation Metrics} \end{tabular}}} & \textbf{\begin{tabular}[c]{@{}c@{}}Utilized Technology\end{tabular}} \\\hline
		
		% Semantic Data Encryption Example
		{\cite{luo2023encrypted}}& {{\begin{tabular}[l]{@{}c@{}}Semantic \\ eavesdropping attack \end{tabular}}}  
		& {\begin{tabular}[l]{@{}l@{}}
				$\star$ Encrypts SI to protect against unauthorized access during transmission\\
				$\bullet$ 
				Ensures the confidentiality and integrity of transmitted SI\\
				$\circ$ Low scalability, complicated key management, and high processing delay\\
				$\dagger$ BLEU score and convergence speed
		\end{tabular}} 
		& \begin{tabular}[c]{@{}c@{}}Symmetric encryption, \\ adversarial training \end{tabular}
		\\\hline
		% Physical-Layer Secret Key Generation Example——>窃听攻击
		{\cite{zhao2022semkey}}& 
		{{\begin{tabular}[l]{@{}c@{}}Semantic \\ eavesdropping, \\ spoofing attacks\end{tabular}}}  
		& {\begin{tabular}[l]{@{}l@{}}
				$\star$ Generate secret key by leveraging semantic drifts and update under RIS assistance\\
				$\bullet$ High-security guarantee, fast key generation, and enhanced channel randomness\\
				$\circ$ Idealized spatial constraint and simplified channel model\\
				$\dagger$ Key generation rate, P-value, and randomness pass rate
		\end{tabular}} 
		& \begin{tabular}[c]{@{}c@{}}Physical layer key \\ generation, RIS \end{tabular}
		\\\hline
		
		% Secure Transmission for Anti-eavesdrop
		{\cite{hu2023interference}}& {{\begin{tabular}[l]{@{}c@{}}Passive eavesdropping, \\interference attacks\end{tabular}}}  
		& {\begin{tabular}[l]{@{}l@{}}
				$\star$ Artificial noise-assisted interference alignment for mitigating eavesdropping\\
				$\bullet$ Practical assumption, enhanced security, and mitigated interference\\
				$\circ$ Lack cross-layer optimization and may impact transmission efficiency\\
				$\dagger$ Secrecy outage probability and power allocation ratio
		\end{tabular}} 
		& \begin{tabular}[c]{@{}c@{}}Interference alignment, \\ artificial noise \end{tabular}
		
		\\\hline
		
		% Covert Communication Example
		{\cite{wang2023multiagent}}& {{\begin{tabular}[l]{@{}c@{}}Semantic \\ eavesdropping,\\ jamming attacks\end{tabular}}}  
		& {\begin{tabular}[l]{@{}l@{}}
				$\star$ Covert and secure SI transmission assisted with friendly jammer\\
				$\bullet$ Low probability of detection and suitable for highly confidential scenarios\\
				$\circ$ Lack complexity and scalability analysis, as well as complex training procedures\\
				$\dagger$ Average accuracy of answering the questions
		\end{tabular}} 
		& \begin{tabular}[c]{@{}c@{}}Covert comm.,\\ multi-agent RL \end{tabular}
		\\\hline
		
		% Quantum Technology for Security-enhanced  but the reliance on fragile and expensive photonic quantum resources renders QCN resource optimization challenging
		{\cite{khalid2023quantum}}& {{\begin{tabular}[l]{@{}c@{}} Semantic poisoning, \\ adversarial attacks\end{tabular}}}  
		& {\begin{tabular}[l]{@{}l@{}}
				$\star$ Quantum SemCom for reliable and secure interaction\\
				$\bullet$ Strong security guarantees and future-proof against quantum attacks\\
				$\circ$ Challenging in fragile/expensive photonic quantum resources and  scalability issues\\
				$\dagger$ Distance of semantic Hilbert space, semantic decoding error, and fidelity
		\end{tabular}} 
		& \begin{tabular}[c]{@{}c@{}}Quantum embedding, \\ quantum ML \end{tabular}
		\\\hline				

		% RIS-enhanced 		
		\cite{du2023semantica} & {{\begin{tabular}[l]{@{}c@{}}
		Unauthorized access, \\eavesdropping attack\end{tabular}}}  
		& {\begin{tabular}[l]{@{}l@{}}
		$\star$ RIS-assisted secure and inverse semantic-aware wireless sensing \\
		$\bullet$ Efficiency in transmission, enhanced security, cross-layer  compatibility\\	
		$\circ$  Hardware cost and implementation complexity, and productive attenuation		
		\end{tabular}} 
		& \begin{tabular}[c]{@{}c@{}}RIS \end{tabular}
		\\\hline
		
	\end{tabular}
\end{table*}

 	\begin{figure*}[!t]\setlength{\abovecaptionskip}{-0.1cm}
		\centering
		%\hspace{-0.75cm}
    \includegraphics[width=0.7\textwidth]{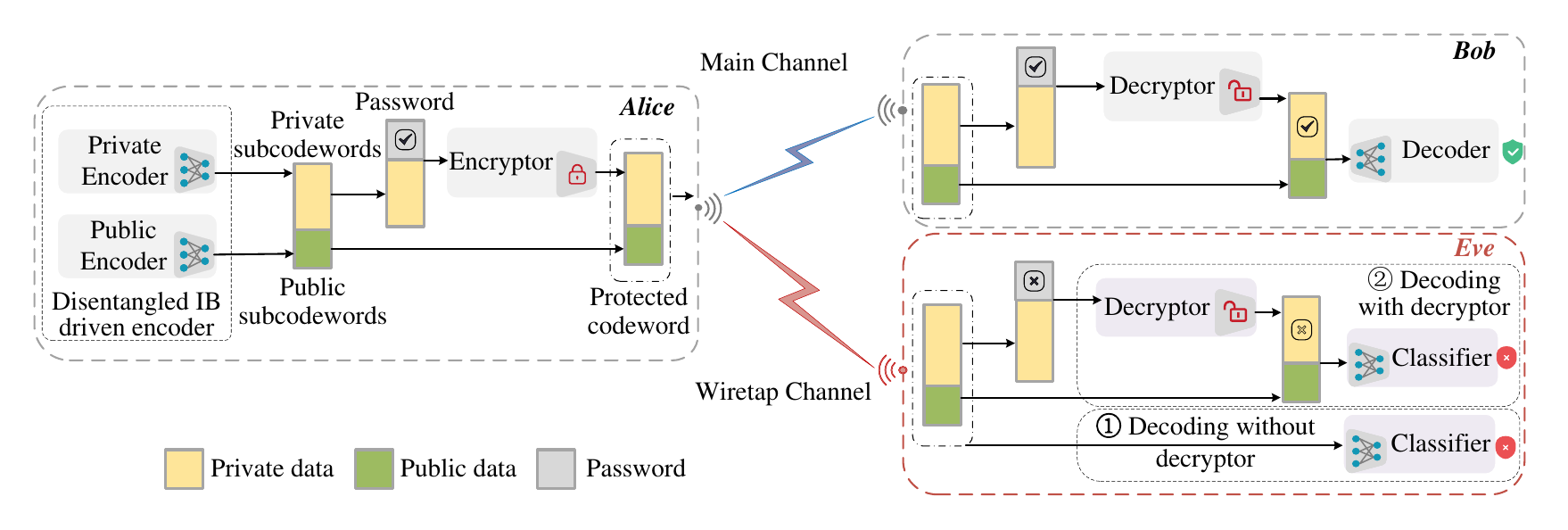}
		\caption{{Illustration of the DPJSCC Scheme \cite{sun2024disentangled}, which depicts the disentangled information bottleneck (IB) guided privacy-protective JSCC framework for image transmission.}}\label{fig: IB-Encry}\vspace{-2mm}
	\end{figure*}
 
		Existing and potential solutions to ensure transmission security in SemComNet include cryptography, physical layer security (PLS), covert communication, quantum technology, and other emerging communication techniques, as detailed below.
	
    \subsubsection{Semantic Data Encryption for Secure Transmission} \label{subsec:data_encryption_defense}
	Encryption schemes play a crucial role in safeguarding the privacy, security, and integrity of semantic data, thereby enhancing the overall reliability of SemComNet, which has received increasing attention \cite{chen2023model,chen2024lightweight,knott2021crypten}.
	In \cite{chen2023model}, Chen \emph{et al.} propose a cryptography method using random permutation and substitution to counter model inversion eavesdropping attack \cite{wang2024privacy}, where attackers intercept and reconstruct the original message. Experiments show this approach is effective in both white-box and black-box scenarios. Similarly, Chen \emph{et al.} \cite{chen2024lightweight} employ RSA and AES algorithms in SemCom to prevent eavesdroppers from accessing sensitive information, reducing eavesdropping accuracy (i.e., eavesdropping ACC) to approximately 10\% on CIFAR10 and ImageNet datasets.

	% HE、MPC
    However, this approach \cite{chen2023model} relies heavily on key management and distribution, posing risks associated with key exposure. To address these challenges, advanced cryptographic techniques such as secure multi-party computation (MPC) \cite{knott2021crypten} and homomorphic encryption (HE) \cite{knott2021crypten} eliminate the need for key sharing. 
    MPC enables agents in SemComNet to perform collaborative computations (e.g., addition, comparisons, and searches) without revealing raw data. To promote its integration into ML, Knott \emph{et al.} \cite{knott2021crypten} propose a user-friendly framework named \textit{CRYPTEN} by offering MPC services through familiar ML abstractions. Meanwhile, HE enhances data security by allowing computations directly on encrypted data, ensuring privacy during transmission and processing without requiring decryption.

	However, the above solutions \cite{knott2021crypten} may impose excessive communication and computing burdens on agents. Moreover, traditional encryption \cite{chen2023model} such as permutation and substitution disrupt intrinsic SI correlations, reducing reconstruction quality. To solve this, Xu \emph{et al.} \cite{xu2023deepa} propose a DL-based joint encryption and source-channel coding (DJESCC) for SemComNet. This approach secures visual content by transforming images into visually protected forms while preserving reconstruction quality. However, encrypting and decrypting entire information may result in reconstruction errors. In response, Sun \emph{et al.} \cite{sun2024disentangled} introduce a privacy-preserving JSCC scheme named DPJSCC, as shown in Fig.~\ref{fig: IB-Encry}. It separates private and public information into distinct subcodewords using disentangled IB. The private subcodewords are encrypted with a password-based algorithm, while public ones are directly transmitted. DPJSCC is effective even against advanced eavesdroppers, achieving eavesdropping accuracy near random guessing while maintaining strong reconstruction quality. Experiments show DPJSCC reduces eavesdropping accuracy by up to 18\% compared to DeepJSCC \cite{eirina2019deep} and adversarial JSCC \cite{thomas2022adversarial}.

	\subsubsection{Physical-Layer Lightweight Secret Key Generation for SemComNet}\label{subsec:PLK_defense}
	Physical layer key generation (PLKG) emerges as a quantum-resistant and key transmission-free solution, offering a promising avenue for securing SemComNet. By leveraging the unique properties of the transmission medium, such as wireless channel fading, PLKG enables secret key generation directly between communicating parties.
	For instance, Zhao \emph{et al.} \cite{zhao2022semkey} present a PLKG scheme using a reconfigurable intelligent surface (RIS) \cite{li2022robust} to enhance the key generation rate by exploiting the randomness of semantic drifts between the transmitter and receiver.
    However, the idealized assumption of spatial constraints (e.g., eavesdroppers being half a wavelength away) in \cite{zhao2022semkey} may not hold in real-world scenarios, where eavesdroppers could occupy various positions, potentially impacting system vulnerability differently. Additionally, PLKG can integrate seamlessly with hybrid cryptographic schemes, as demonstrated in \cite{mitev2023whata}, combining traditional cryptographic strengths with PLKG's unique advantages. This integration bolsters security and efficiency, making PLKG an advanced solution for safeguarding SemComNet.

	\subsubsection{Physical Layer Secure Transmission for  SemComNet}\label{subsec:PLST_defense}
	Since its inception \cite{wyner1975wiretap}, physical layer secure transmission technology has attracted substantial attention \cite{liu2017physical}. This technology seamlessly integrates the aspects of data transmission and encryption, offering a promising approach for securing SemComNet.
	Currently, there are several key technical approaches \cite{liu2017physical}: beamforming \cite{lin2018physicallayer}, power allocation \cite{du2023rethinking}, cooperative interference \cite{li2021exploiting}, and artificial noise (AN) injection \cite{goel2008guaranteeing,hu2023interference}. 
	These methods intentionally amplify the gap between the legitimate and eavesdropping communication channels. This allows authorized recipients to decode confidential SI successfully, while semantic signals at eavesdroppers are deliberately scrambled and irreversibly corrupted.
	
	\textit{(i) Beamforming.} Implementing the beamforming technique in SemComNet makes the SI transmission in specific directions which ensures only intended receivers can capture it. For instance, Lin \emph{et al.} \cite{lin2018physicallayer} propose a frequency diverse array beamforming approach that optimizes frequency offsets and transmits beamforming to maximize secrecy rates, particularly in close legitimate user and eavesdropper scenarios.
	\textit{(ii) Power allocation.}
	A well-designed power allocation scheme enhances the security of the transmission significantly \cite{du2023rethinking}. For instance, Yin \emph{et al.} \cite{yin2023multidomain} introduce a multi-domain resource multiplexing scheme leveraging co-channel interference among IoT nodes. Their alternating optimization with successive convex approximation methods enhances physical layer security.
        \textit{(iii) Cooperative interference.}
	To effectively disrupt unauthorized channels, cooperative relay, and friendly jamming techniques can be introduced.
	In \cite{li2021exploiting}, Li \emph{et al.} introduce two innovative cooperative interference alignment (IA) schemes, the former adjusts the spatial signature of one interference and strength of all interference, ensuring orthogonality with the desired transmission, and the other modifies all interfering signal strength, preserving orthogonality.
	\textit{(iv) AN injection.}
	As traditional IA methods are prone to secret signal cancellation, Hu \emph{et al.} \cite{hu2023interference} propose an AN-assisted IA scheme that minimizes secrecy outage probability through optimized power allocation. 
    Although these techniques are effective at the physical layer, achieving holistic security across SemComNet requires exploring cross-layer optimization, such as optimizing and integrating physical layer security with other layers (e.g., the control layer) within SemComNet.

	\subsubsection{Covert Communication for Secure Transmission}   \label{subsec:covertcomm_defense}
\begin{figure}[!t]\setlength{\abovecaptionskip}{-0.0cm}%\vspace{-3mm}
		\centering
		\includegraphics[width=9.5cm]{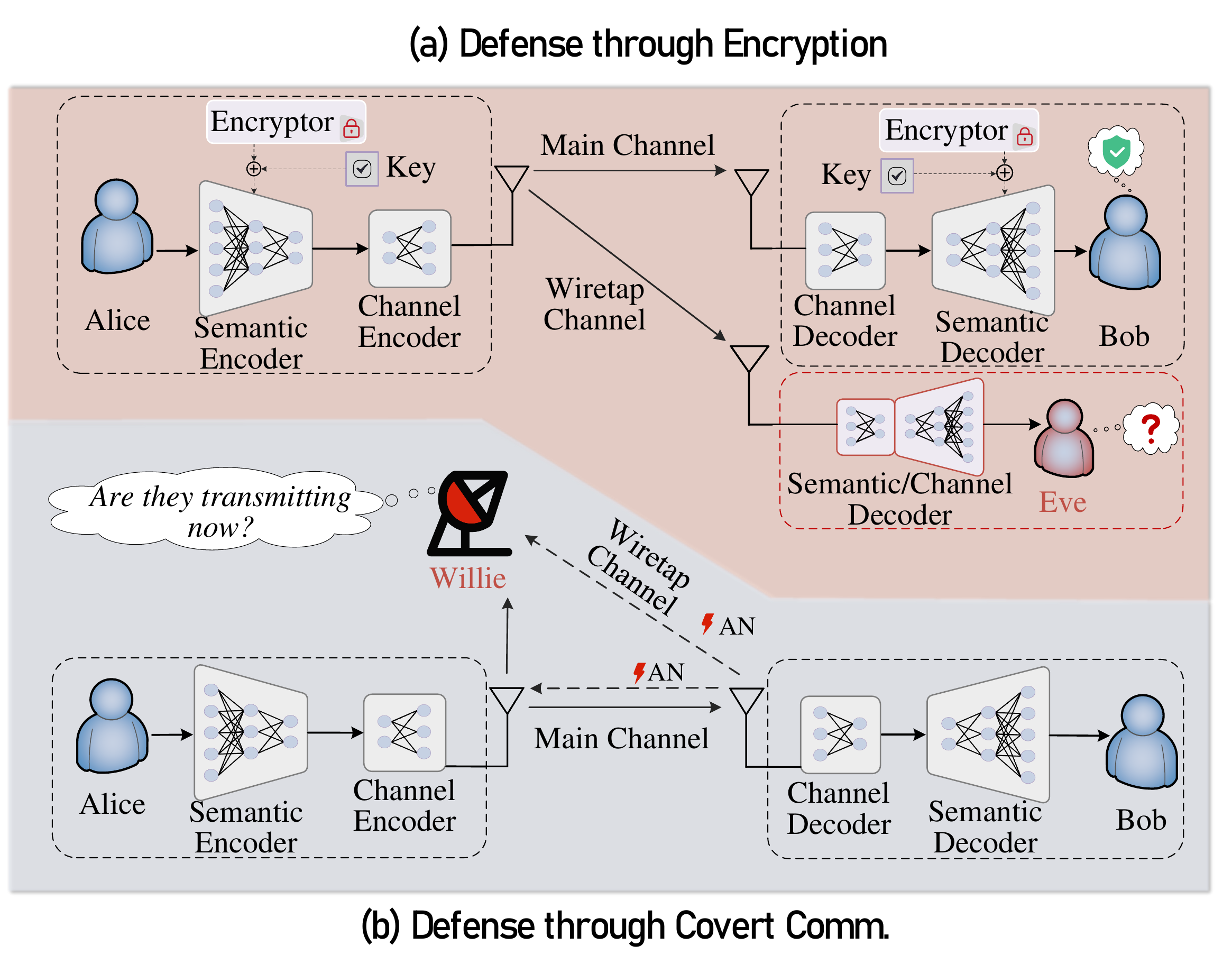}
		\caption{{An illustrative comparison of defense mechanisms through semantic encryption \cite{luo2023encrypted} and covert communication \cite{hu2024covert}.}}\label{fig:Defense_Comparsion_Covert_Encryption}
		\vspace{-2mm}
		\end{figure}

        In SemComNet, as shown in Fig.~\ref{fig:Defense_Comparsion_Covert_Encryption}, semantic data encryption ensures content security through advanced encryption \cite{sun2024disentangled,luo2023encrypted,xu2023deepa}, but it incurs higher computational complexity and key management overhead. PLS protects the content by ensuring the message remains inaccessible to eavesdroppers, even if communication is detected. Key performance indicators for PLS include \textit{secrecy outage probability} (SOP) and \textit{average secrecy capacity} (ASC) \cite{liu2017physical,du2023rethinking}. In contrast, as shown in Fig.~\ref{fig:Defense_Comparsion_Covert_Encryption}, covert communication focuses on undetectable transmissions. It conceals the transmission itself, preventing detection by a warden, and is measured by metrics such as \textit{covert communication capacity} \cite{xu2024covert}, \textit{detection error probability} (DEP), and \textit{covert rate} \cite{chen2023coverta, hu2024covert,du2023rethinking}.
  	
    One paradigm of covert communication is physical-layer low probability of detection communication (PL-LPDC). For text semantic transmission in wireless networks, as shown in Fig.~\ref{fig:Defense_Comparsion_Covert_Encryption}, Xu \emph{et al.} \cite{hu2024covert} propose a covert communication scheme using a full-duplex receiver to generate AN, which interferes with Willie’s detection. By jointly optimizing transmit power, AN power, and the number of semantic symbols per word, they maximize semantic spectral efficiency while ensuring minimum semantic similarity and maintaining covertness. Detection performance is evaluated using the false alarm rate and missed detection rate. Their analysis shows that a fixed AN power can enhance covert transmission. For image transmission, Wang \emph{et al.} \cite{wang2023multiagent} present a PL-LPDC framework, where friendly jammers deploy jamming signals to defend against eavesdroppers. A multi-agent policy gradient algorithm is proposed to improve system performance, allowing devices and jammers to identify vulnerable devices and optimize transmission parameters in SemComNet. However, the lack of complexity and scalability analysis, along with the intricate training procedures, may limit its practical deployment in SemComNet.

    Unlike PL-LPDC, which focuses on concealment and low detection probability, physical-layer steganography \cite{xie2022security} aims to hide and encrypt the content of communication, offering a higher level of confidentiality. It embeds secret SI within the physical properties of the communication channel (e.g., signal power, phase, or frequency), making it imperceptible to unauthorized users. For instance, Yamaguchi \emph{et al.} \cite{yamaguchi2020physicallayer} propose a steganography security approach that conceals the secret signal by embedding it into the cover data, making detection difficult without prior knowledge. Future research efforts are required in designing intelligent covert communication in SemComNet, integrating adaptive optimization, theoretical completeness, and lightweight implementation to accommodate diverse environmental conditions.

	\subsubsection{Quantum Technology for Secure-enhanced SemComNet}\label{subsec:quantum_defense}
	Quantum technology, based on the principles of quantum physics, can significantly enhance security within SemComNet \cite{khalid2023quantum, chehimi2024quantum}. Quantum key distribution (QKD) plays a pivotal role in establishing secure key exchanges in SemComNet, providing a tamper-evident framework for SemComNet.  Unlike classical cryptographic methods, QKD can detect eavesdropping attempts. Any interference is rapidly identified by the communicating parties, alerting them to potential intruders and safeguarding the confidentiality of semantic data.

	In \cite{kaewpuang2023cooperative}, Kaewpuang \emph{et al.} focus on QKD-aided secure SI transmission within SemComNet. Furthermore, they address resource allocation challenges in QKD deployment for SI transmission by proposing a two-stage stochastic optimization model that optimizes QKD resource deployment, with Shapley values ensuring fair cost allocation among cooperative QKD service providers.
	Apart from QKD, quantum semantic encoding can further protect the privacy of semantic content, offering enhanced security and efficient data transmission. For instance, Khalid \emph{et al.} \cite{khalid2023quantum} explore quantum SemCom employing quantum embedding and quantum ML to encode data into quantum states, which are securely teleported using quantum principles. 
	However, such quantum solutions face challenges in resource optimization and scalability, primarily due to the reliance on fragile and expensive photonic quantum resources.

	\subsubsection{Emerging Communication Technology Assisting Secure SemComNet}\label{subsec:emerging_defense}
	Reconfigurable intelligent surface (RIS), an emerging wireless technology \cite{li2022robust}, can enhance the security and efficiency of SI transmission in SemComNet. 
	For instance, Wang \emph{et al.} \cite{wang2023starrisassisted} propose using a simultaneous transmitting and reflecting RIS to prevent eavesdropping during SI transmission. By optimizing transmission and reflection coefficients, they enhance legitimate semantic signal transmission and create interference for eavesdroppers.
	In \cite{du2023semantica}, Du \emph{et al.} present an inverse semantic-aware wireless sensing framework for SemComNet, which uses the RIS amplitude response matrix for secure data encryption and semantic hash sampling for efficient self-supervised decoding.
	Experimental results show a 95\% reduction in data volume without affecting sensing tasks, offering a resource-saving solution for the secure SemComNet.  
	
	Furthermore, the above passive RIS schemes  
\cite{wang2023starrisassisted,du2023semantica} face challenges, such as productive attenuation where reflection link fading is proportional to distances. To tackle this issue, Li \emph{et al.} \cite{li2023secureb} introduce active RIS to reconfigure the propagation within the wireless environment, while utilizing the on-off control scheme for active RIS phase shifts. Apart from RIS, spread spectrum techniques such as frequency hopping spread spectrum and direct sequence spread spectrum can also be utilized in SemComNet \cite{li2023ubiquitousa}. As such, the semantic signal is spread across a wide spectrum of frequencies, or pseudo-random sequences are employed. This helps mitigate the disruptive effects of interference on the channel, ensuring a robust and secure communication environment.

\begin{table*}[!t]
	\centering \setlength{\abovecaptionskip}{0cm}
	\caption{Summary of Key Literature on Ensuring Reliable SemComNet}\label{Summary3}
	\begin{tabular}{cclc}\hline
		\textbf{Ref.} & \textbf{\begin{tabular}[c]{@{}c@{}}Security\\ Threat\end{tabular}} &  \multicolumn{1}{c}{\textbf{\begin{tabular}[c]{@{}l@{}}$\star$ Purpose\\$\bullet$ Advantages\\$\circ$ Limitations\\ $\dagger$ {Evaluation Metrics} \end{tabular}}} & \textbf{\begin{tabular}[c]{@{}c@{}}Utilized Technology\end{tabular}} \\\hline
		
		% Misunderstanding Resilient Communication Example
		% {\cite{Misunderstanding Resilient}} 
		\cite{erdemir2023generative}
		& {{\begin{tabular}[l]{@{}c@{}} Semantic noise, \\ distortion \end{tabular}}} 
		& {\begin{tabular}[l]{@{}l@{}}
				$\star$ Leverage GAI model to enhance perceptual quality and semantic reliability \\
				$\bullet$ Improved communication efficiency and reduced misunderstanding errors\\
				$\circ$ 
				High hallucination risk and heavy computational burden on agents\\
				$\dagger$ Peak SNR, MS-SSIM, and LPIPS
				% peak signal-to-noise-ratio (PSNR), multiscale structural similarity index measure (MS-SSIM), and LPIPS.
		\end{tabular}} & StyleGAN 
		\\\hline

		\cite{du2023generative}
		& {{\begin{tabular}[l]{@{}c@{}} Semantic noise, \\eavesdrop and \\ interference attack \end{tabular}}} 
		& {\begin{tabular}[l]{@{}l@{}}
				$\star$ Covert SI transmission and accurate SI decoding via multi-modal GAI models\\
				$\bullet$ 
				Strong secure guarantee and accurate content generation capability\\
				$\circ$ 
				Complex training procedures and costly resource consumption\\
				$\dagger$ SSIM, detection error probability, and bit error probability
		\end{tabular}} 
		& \begin{tabular}[c]{@{}c@{}}GAI, multi-modal \\ prompts, covert comm. \end{tabular}
		\\\hline

		% ML-driven Resilient Interpretation Example
		% \cite{ML-driven Resilience} 
		% \cite{han2023semanticpreserved}
		% & {{\begin{tabular}[l]{@{}c@{}} Physical and \\ semantic interference \end{tabular}}} 
		% & {\begin{tabular}[l]{@{}l@{}}
		% 		$\star$ Pre-trained semantic corrector module to refine and rectify SI\\
		% 		$\bullet$ Effective in suppressing both physical and semantic noise\\
		% 		$\circ$ Challenging in model adaptability and response latency \\
		% 		$\dagger$ Melcepstral distortion, mean opinion score, word-error-rate, and 
		% \end{tabular}} 
		% & \begin{tabular}[c]{@{}c@{}}Pre-trained  \\ strategy, GAN \end{tabular}
		% \\\hline

		% Reliable Adaptive Transmission Example
		% \cite{Adaptive Transmission} 
		\cite{zhou2022adaptive}
		& {{\begin{tabular}[l]{@{}c@{}}Physical and \\ semantic interference  \end{tabular}}} 
		& {\begin{tabular}[l]{@{}l@{}}
				$\star$ 
				Adaptive bit rate control mechanism for SI transmission under harsh conditions\\
				$\bullet$ 
				High efficiency, strong fault tolerance, and cost-saving\\
				$\circ$ 
				Lack of large-scale and real-world test\\
				$\dagger$ BLEU score
		\end{tabular}} 
		& \begin{tabular}[c]{@{}c@{}}Semantic HARQ, \\ adaptive transmission \end{tabular}
		\\\hline
		
		% Reasoning-driven Reliable Example
		% \cite{Reasoning-driven Reliability} 
		\cite{xiao2023imitation}
		& {{\begin{tabular}[l]{@{}c@{}}	Semantic interference, \\ and privacy \\  exposure of KBs \end{tabular}}} 
		& {\begin{tabular}[l]{@{}l@{}}
				$\star$ Leverage RL for reasoning implicit SI while protecting privacy\\
				$\bullet$ 
				Automatic reasoning and robust semantic error correction 
				\\
				$\circ$ Low sample efficiency and lack generalization to unseen data \\
				$\dagger$ Semantic reasoning accuracy, average symbol recovery accuracy, and symbol error rate
		\end{tabular}} & Inverse RL \\\hline
		
		% Collaborative Relay Strategy Example
		% \cite{Collaborative Relay} 
		\cite{luo2022autoencoderbaseda}
		& {{\begin{tabular}[l]{@{}c@{}} Mismatch of KBs \\ between transceiver \end{tabular}}} 
		& {\begin{tabular}[l]{@{}l@{}}
				$\star$ Assist in forwarding and interpreting SI while minimizing semantic noise\\
				$\bullet$ Extended communication range and reliable translation capability\\
				$\circ$ High computational complexity and long transmission delay\\
				$\dagger$ BLEU score and sentence similarity
		\end{tabular}} 
		& 
		\begin{tabular}[c]{@{}c@{}}Semantic \\ relay \end{tabular}
		\\\hline
		
	\end{tabular}
\end{table*}
	
%	\vspace{-3mm}
	%-----------------------------------------------------------------------
	\subsection{Reliable SemComNet}\label{subsec:defense3}
	In SemComNet, reliability stands as a vital property facilitating semantic-oriented service provisioning and effective agent interaction across diverse environments. Specifically, ensuring a reliable SemComNet involves considerations spanning two dimensions, i.e., resilient semantic interpretation and reliable SI transmission.
	The former dimension guarantees accurate understanding and reconstruction of semantic content, thereby enhancing tolerance to semantic ambiguity. On the other hand, the latter dimension focuses on maintaining stable SI transmission performance even in harsh conditions, such as extra-low SNR, strong perturbation \cite{zeng2024USVFleet}, and long transmission distances.
	We delve into discussions regarding enhancing SemComNet's reliability through various approaches, as below.

	\subsubsection{GAI for Misunderstanding Mitigation in SemComNet}\label{subsec:GAI_defense}
\begin{figure}[!t]\setlength{\abovecaptionskip}{-0.0cm}
		\centering
		\includegraphics[width=9.5cm]{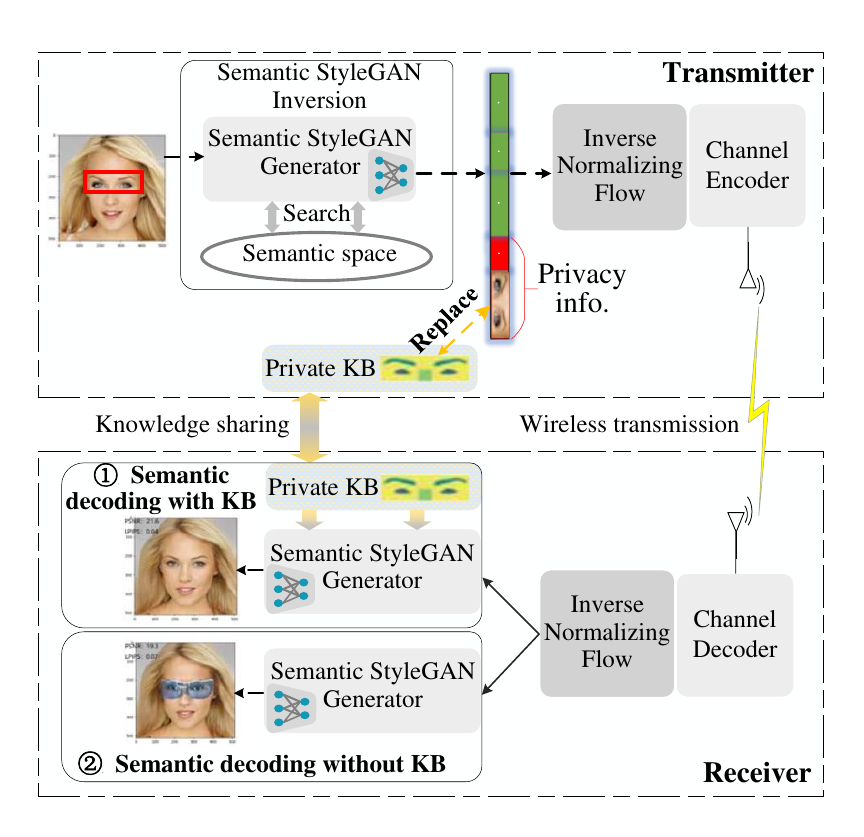}
		\caption{Illustration of a reliable and privacy-preserving SemCom system for image transmission employing GAI and privacy filters in \cite{han2023generative}. This system accurately reconstructs semantically consistent details from SI at the destination, effectively mitigating semantic noise risks. Besides, it utilizes a privacy filter and KBs to replace privacy information with corresponding natural features, ensuring privacy protection.} \label{fig:defense}\vspace{-2mm}
	\end{figure}

	GAI models such as ChatGPT can enhance the resilience of semantic interpretation (e.g., suppressing semantic noise) within SemComNet \cite{zhao2024generative}. Specifically, these models aid in reducing transmission traffic and latency \cite{xia2023generative}. They also excel in reconstructing semantic-consistent details from SI at the destination side, even when encountering challenges such as semantic noise and mismatches in KBs between the transmitter and receiver.
	For instance, to enhance the perceptual quality of reconstructed data, Erdemir \emph{et al.} \cite{erdemir2023generative} introduce two innovative GAI-based JSCC frameworks: InverseJSCC and GenerativeJSCC. {Unlike traditional approaches focusing solely on distortion metrics, the schemes proposed in \cite{erdemir2023generative} optimize a combination of \textit{mean squared error} (MSE) and \textit{learned perceptual image patch similarity} (LPIPS) losses to ensure semantic fidelity. Simulations show that the proposed scheme exceeds DeepJSCC not only in distortion metrics such as \textit{peak signal-to-noise-ratio} (PSNR) and \textit{multiscale structural similarity index measure} (MS-SSIM), but also in perceptual quality measured by LPIPS, even with KBs mismatches.}
	To achieve superior perceptual quality in limited bandwidth and low SNR scenarios, Chen \emph{et al.} \cite{chen2023commin} treat image recovery as an inverse problem. They employ invertible neural networks with the diffusion model to aid the reconstruction process.
	However, the above works \cite{erdemir2023generative,chen2023commin} neglect the aspect of privacy protection during transmission in GAI-driven SemComNet. To address this, as shown in Fig.~\ref{fig:defense}, Han \emph{et al.} \cite{han2023generative} leverage the StyleGAN inversion method to extract disentangled SI from the original image. Meanwhile, they employ privacy filters to replace private SI with natural features guided by KBs, thereby safeguarding sensitive SI. The results have proven successful in enhancing communication efficiency and reducing misunderstanding errors.

	However, a notable challenge arises from the inherent instability generation capabilities of GAI models, such as the issue of content ``hallucination" \cite{du2023generative, tonmoy2024comprehensive}. This instability poses challenges, particularly in applications requiring accurate information transfer (e.g., medical imaging diagnosis). 
    To address this issue, Du \emph{et al.} \cite{du2023generative} propose a GAI-aided approach that employs multi-modal prompts for accurate content decoding. By leveraging generative diffusion models and covert communication, the approach facilitates the transmission of multi-modal prompts and ensures precise image regeneration under energy constraints.
	Nonetheless, the integration of large-scale GAI models into SemComNet presents challenges, demanding substantial computing power for training. This strains the deployment of GAI models in resource-limited agents within SemComNet. 
	In summary, achieving trustworthy and resource-efficient integration of GAI into SemComNet necessitates further investigation.

	\subsubsection{Reliable Adaptive Transmission in SemComNet}\label{subsec:reliable_transmission_defense}
In dynamic network conditions and deteriorating channel quality, ensuring reliable SI transmission is crucial. Adaptive transmission offers a promising solution to enhance SemComNet reliability \cite{lu2023semanticsempowered, yang2023semanticb}.
For instance, to improve noise resilience in ultra-low SNR scenarios, Zhou \emph{et al.} \cite{zhou2022adaptive} propose a multi-bit length selection strategy with a policy network to dynamically adjust coding rates. They also introduce progressive semantic hybrid automatic repeat request (HARQ) schemes, incorporating incremental knowledge to reduce semantic errors. Results, evaluated using the BLEU metric, show improved efficiency, robustness, and cost-effectiveness. However, the absence of extensive, real-world testing may limit the validation of these findings.
Moreover, Wang \emph{et al.} \cite{wang2023wireless} explore an adaptive bitrate transmission strategy for video data. Using nonlinear transformations and conditional coding, they extract SI from video frames and allocate channel bandwidth dynamically for optimal performance under challenging wireless conditions.

	\subsubsection{Reasoning-driven Reliable SemComNet}\label{subsec:reasoning_defense}
	Agents endowed with reasoning capabilities significantly enhance the reliability of SemComNet by mitigating semantic ambiguity and enhancing communication reliability.
	For instance, Jiang \emph{et al.} \cite{jiang2022reliablea} introduced a KG-driven SemComNet which could significantly enhance reliability against semantic noise.
	However, the above work \cite{jiang2022reliablea} does not consider the issue of typing errors in large-scale factual KGs within SemComNet.
	These errors, especially in entity-type pairs, pose a significant obstacle to reliable knowledge extraction and utilization in SemComNet.
	To address this challenge, Yao \emph{et al.} \cite{yao2021typing} propose an innovative active typing error detection algorithm that effectively incorporates both gold and noisy labels. Furthermore, they emphasize the semi-supervised noise models as a feasible solution, offering the possibility to enhance the utilization of varied information for error detection in SemComNet.
	
	RL excels at reasoning implicit SI from transmitted data, which could seamlessly adapt to the dynamic conditions within SemComNet.
	For instance, an implicit semantic-aware communication architecture is proposed in \cite{xiao2023imitation}, focusing on both explicit and implicit message semantics reasoning. The proposed generative imitation-based reasoning mechanism could guide destination users to automatically interpret implicit semantics without requiring access to the source data, significantly safeguarding user privacy and reducing semantic noise.
    However, this method has drawbacks, such as low sample efficiency and limited generalization to unseen data.	In response, Thomas \emph{et al.} \cite{thomas2023neurosymbolic} integrate a signaling game and Neuro-Symbolic (NeSy) AI \cite{wang2022dataand} into semantic transmission. The game-theoretical approach creates a compositional language, aiding in generalization and semantic awareness.
	The NeSy AI integrates experiential learning (neural component) with knowledge-based reasoning (symbolic component), empowering the SemComNet to acquire intricate signaling strategies with limited training samples and minimal data transmission.
    Currently, GAI models \cite{min2023recent, tonmoy2024comprehensive, wang2023surveya} serve as powerful tools with semantic understanding and reasoning capabilities. Leveraging these models could significantly enhance reasoning accuracy in SemComNet \cite{xia2023generative, xu2024unleashing, wang2023surveya}.

	\subsubsection{Collaborative Relay for Reliable SemComNet}\label{subsec:relay_defense}
	In traditional communication systems, collaborative relays ensure reliable information delivery, particularly in long-distance and low-SNR scenarios \cite{guo2024distributed}. Similarly, semantic relay nodes in SemComNet not only assist in reliable transmission but also help translate and interpret SI from source to destination \cite{lin2024semanticforward}.
	For instance, Luo \emph{et al.} \cite{luo2022autoencoderbaseda} introduce the intelligent relay-assisted SemCom with amplify-and-forward (AF) and decode-and-forward (DF) modes. This system assists in forwarding and interpreting SI while minimizing semantic noise. However, it faces challenges with high computational complexity and long transmission delay.

	\vspace{-2mm}
	%-----------------------------------------------------------------------
	\subsection{Trust Management in SemComNet}\label{subsec:defense4}
	Establishing and managing trust among collaborative agents in SemComNet is crucial for building a trustworthy environment.
	In response, this section will introduce three approaches for creating a trustworthy SemComNet, i.e., intrinsic trust via physical-layer authentication mechanism, trustless SemComNet through blockchain, and measurable trust evaluation.
	\begin{table*}[!t]
	\centering \setlength{\abovecaptionskip}{0cm}
	\caption{Summary of Key Literature on Trust Management in Semantic Communication Networks}\label{Summary4}
	\begin{tabular}{cclc}\hline
		\textbf{Ref.} & \textbf{\begin{tabular}[c]{@{}c@{}}Security\\ Threat\end{tabular}} &  \multicolumn{1}{c}{\textbf{\begin{tabular}[c]{@{}l@{}}$\star$ Purpose\\$\bullet$ Advantages\\$\circ$ Limitations\\ $\dagger$ {Evaluation Metrics} \end{tabular}}} & \textbf{\begin{tabular}[c]{@{}c@{}}Utilized Technology\end{tabular}} \\\hline
		
		% Physical-Layer Authentication Example
		% \cite{Physical-Layer Auth.} 
		\cite{tan2024optimization}
		& {{\begin{tabular}[l]{@{}c@{}} Impersonation attack, \\ eavesdropping attack \end{tabular}}}  
		& {\begin{tabular}[l]{@{}l@{}}
				$\star$ 
				Leverages unique physical layer properties for secure authentication\\
				$\bullet$ 
				Information-theoretic guarantee, lightweight processing speed, and high compatibility\\
				$\circ$ 
				Challenging in imperfect channel conditions and large-scale environments\\
				$\dagger$ 
				Secure authentication efficiency
				% , probability of message transmission, message outage, and secure authentication
		\end{tabular}} 
		& \begin{tabular}[c]{@{}c@{}}Physical layer \\ authentication\end{tabular} 
		\\\hline
		
		% Blockchain for Trustless Environments Example
		% \cite{Blockchain Trustlessness} 
		\cite{bao2023pbidm}
		& {{\begin{tabular}[l]{@{}c@{}} False data injection, \\impersonation attack, \\ malware attack \end{tabular}}}  
		% This architecture is vulnerable to various attacks, such as a single point of failure, phishing attacks, malicious tampering attacks, and internal attacks.
		& {\begin{tabular}[l]{@{}l@{}}
				$\star$ 
				Provide robust privacy-preserving identity management for industrial SemComNet\\
				$\bullet$ 
				Guarantee unforgeability, traceability, revocability, and public verifiability\\
				$\circ$ Without consideration of privacy-preserving and processing speed \\
				$\dagger$
				Computation cost, communication cost, and gas cost 
		\end{tabular}} 
		& \begin{tabular}[c]{@{}c@{}} Cryptographic \\tool, blockchain\end{tabular}  \\\hline
		
		\cite{wang2021spdsa}
		& {{\begin{tabular}[l]{@{}c@{}} Free-riding attack, \\ misuse of shared \\ knowledge/data \end{tabular}}}  
		& {\begin{tabular}[l]{@{}l@{}}
				$\star$ Trust-free environment for fine-grained data authorization and traceable audit\\
				$\bullet$ Low computation overhead in shared knowledge/data audit\\ 
				$\circ$ Lack large-scale and real-world performance test \\
				%$\dagger$ 
				%Secure authentication efficiency    
		\end{tabular}} 
		& \begin{tabular}[c]{@{}c@{}} Smart contract, \\ trusted computing\end{tabular}  \\\hline
		
		% Trust Evaluation Techniques Example
		% \cite{Trust Evaluation} 
		\cite{gyawali2021deep}
		& {{\begin{tabular}[l]{@{}c@{}} False feedback attack, \\  internal attack \end{tabular}}}  
		& {\begin{tabular}[l]{@{}l@{}}
				$\star$ Dynamic reputation mechanism for limiting wrong feedback from malicious agents\\
				$\bullet$ Enhanced system reliability and adaptability to agent behavior\\
				$\circ$ Overheads grow exponentially as the number of agents rises\\
				$\dagger$ 	Reputation and average reward
		\end{tabular}} 
		&  \begin{tabular}[c]{@{}c@{}} Dempster-Shafer\\ theory, DRL\end{tabular}
		\\\hline

		\cite{truong2019trust}
		& {{\begin{tabular}[l]{@{}c@{}} Semantic-poor \\ data quality \end{tabular}}}  
		& {\begin{tabular}[l]{@{}l@{}}
				$\star$ Reputation evaluation model to filter out trusted agents with semantic-rich data\\
				$\bullet$ Strong reliability and practicality via real-world validation \\
				$\circ$ Difficulty in obtaining trust indicators\\
				$\dagger$ 	Reputation and QoS score
		\end{tabular}} 
		&  \begin{tabular}[c]{@{}c@{}} Experience-Reputation \\ trust evaluation\end{tabular}
		\\\hline

	\end{tabular}
\end{table*}
	
	\subsubsection{Authentication Mechanism in SemComNet} \label{subsec:auth_defense}    
	Authentication serves as a fundamental component of trust management by providing a preliminary verification of agents' identities, forming the basis for trust establishment and subsequent evaluations. Conventional cryptography-based authentication methods face scalability and complexity challenges in large-scale SemComNet. In contrast, physical layer authentication (PLA) is gaining considerable attention due to its information-theoretic security guarantee, lightweight processing speed, and high compatibility \cite{xie2021survey}. This approach is increasingly recognized as a viable implementation in heterogeneous and decentralized SemComNet.

	PLA schemes can be categorized as passive and active \cite{xie2021survey}. Passive PLA relies on physical-layer features for transmitter authentication without modifying the source message. 
	For instance, Gao \emph{et al.} \cite{gao2023esanet} introduce \textit{EsaNet}, a DL-based passive authentication network that extracts a wireless channel fingerprint from environmental semantics (i.e., CSI) to distinguish legitimate users.
	In contrast, active PLA modifies the source message by embedding a physical-layer tag generated from a secret key, enhancing information-theoretic security according to Shannon's secrecy analysis.
	In \cite{jorswieck2015broadcasting}, a generalized model for achieving data confidentiality and active wireless PLA is proposed. It highlights the role of channel uncertainty and various design dimensions, such as time, frequency, and space, in enhancing security for SemComNet. 
	To evaluate the transmission efficiency of active PLA schemes, Tan \emph{et al.} \cite{tan2024optimization} introduces a metric named \textit{secure authentication efficiency} (SAE). By tuning three key parameters that impact SAE (i.e., the probability of message transmission, the probability of message outage, and the probability of secure authentication), they provide a systematic optimization framework and analyze feasibility constraints and optimal solutions.
	
	However, the aforementioned works \cite{jorswieck2015broadcasting, tan2024optimization} assume perfect channel conditions, which is unrealistic in real-world scenarios. To address this, Perazzone \emph{et al.} \cite{perazzone2021artificial} explore the use of AN in fingerprint embedding for wireless security, focusing on scenarios with imperfect CSI. They examine the impact of AN leakage on security and compare detectors for imperfect CSI. The findings reveal that while AN improves security, its effectiveness diminishes under poor channel conditions.

	% \cite{wang2022blockchainempowered}
	\subsubsection{Blockchain for Trustless SemComNet}\label{subsec:bockchain_trustless_defense}
    Blockchain integration in SemComNet is gaining attention for its transparency, decentralization, and tamper-proof features \cite{xie2019survey}. By enabling decentralized identity management \cite{bao2023pbidm,xu2020identity}, cross-domain authentication \cite{chen2022xautha, shen2020blockchainassisted}, and smart contract-driven authorization \cite{wang2021spdsa}, blockchain provides a robust trust management solution. It eliminates the need for a central certification authority, facilitating decentralized, self-sovereign identity management, where agents can independently publish and query their identities.

For instance, Bao \emph{et al.} \cite{bao2023pbidm} propose a robust privacy-preserving identity management system for industrial IoT, utilizing blockchain and cryptographic tools to ensure unforgeability, traceability, revocability, and verifiability. However, in the open and resource-limited SemComNet scenarios, the need for lightweight, privacy-preserving identity management is crucial, which is neglected in \cite{bao2023pbidm}. To tackle this, Xu \emph{et al.} \cite{xu2020identity} introduce a blockchain-based system for mobile SemComNet, empowering users with self-sovereign identities (SSIs). Blockchain records legitimate user SSIs and public keys on the blockchain for decentralized authentication, while Chameleon hash ensures efficient revocation of unauthorized users to reduce overhead.
	
In the fast-paced SemComNet, cross-domain authentication is crucial for rapid identity verification across diverse agents. Chen \emph{et al.} \cite{chen2022xautha} propose a privacy-preserving cross-domain authentication solution for public key infrastructure. It ensures cross-domain compatibility and rapid response via multiple Merkle hash trees. For industrial networks, Shen \emph{et al.} \cite{shen2020blockchainassisted} employ a consortium blockchain and identity-based signatures for establishing trust among different domains. For authorization, smart contracts running on the blockchain enable automatic authorization policies in SemComNet. For instance, Wang \emph{et al.} \cite{wang2021spdsa} apply smart contracts for fine-grained data access and traceable usage management, with off-chain execution to reduce computational overhead. However, real-world performance in large-scale SemComNet with heterogeneous components remains underexplored.

	\subsubsection{Trust Evaluation in SemComNet}\label{subsec:trust_eval_defense}
	Trust evaluation is vital for ensuring security and reliability during cooperation within large-scale SemComNet.
	Current methods predominantly focus on using Dempster-Shafer theory, fuzzy logic, and Bayes theorem to analyze and combine trust-impact attributes (such as direct/indirect, subjective/objection, local/global, and historical/present) during evaluation \cite{wang2021survey}.
    % wang2023surveyd
	For instance, Parhizkar \emph{et al.} \cite{parhizkar2020combining} propose a trust evaluation mechanism based on both direct trust (from historical transactions) and indirect trust (from social interactions). However, the above work \cite{parhizkar2020combining} approach relies on subjective influence and simple attribute combinations, leading to practical challenges, such as cold start and sparse history data \cite{wang2022surveyb}.

	The integration of DL methods \cite{wang2022surveyb} addresses the challenges of trust evaluation and establishes an automated framework for SemComNet. 
	For instance, Jayasinghe \emph{et al.} \cite{jayasinghe2019machine} propose an ML-based trust assessment model for IoT services. Utilizing unsupervised learning techniques and support vector machines, this model accurately classifies trust features and calculates trust values.
	Besides, deep RL (DRL) which combines DL's perceptual capabilities and RL's decision-making abilities, may be well applied in trust or reputation models within SemComNet. 
    In the context of misbehavior detection, Gyawali \emph{et al.} \cite{gyawali2021deep} use DRL to dynamically update vehicle reputation based on Dempster-Shafer theory, improving system resilience against internal attacks. However, as the number of agents increases, communication and computation overheads may become prohibitive. The above works \cite{parhizkar2020combining,jayasinghe2019machine,gyawali2021deep} mainly focus on entity-based trust models (i.e., rely on agents' credibility), while neglecting the authenticity of SI. To fill this gap, Truong \emph{et al.} \cite{truong2019trust} use virtual interactions and quality assessments to calculate trust indicators (experience and reputation). However, acquiring these indicators may be challenging, and a typically sparse trust matrix may compromise the accuracy of trust assessments in SemComNet.

	%-----------------------------------------------------------------------
	\vspace{-3mm}
	\subsection{Data \& Knowledge Security in SemComNet} \label{subsec:defense5}
     SemComNet operates as a data-knowledge dual-driven paradigm. On the one hand, it relies on training semantic models with large-scale data to provide semantic-oriented transmission services. 
	On the other hand, the accumulated knowledge in KBs improves the agents' abilities in understanding and reasoning, thereby facilitating the extraction and reconstruction of SI.
	Within this paradigm, information security becomes a crucial prerequisite for the development and prosperity of the SemComNet.
	Next, we delve into discussions regarding data \& knowledge security in SemComNet, focusing on aspects such as anti-poisoning, tamper-proofing, access control, and privacy preservation.
	
\begin{table*}[!t]
	\centering \setlength{\abovecaptionskip}{0cm}
	\caption{Summary of Key Literature on Data \& Knowledge Security in Semantic Communication Networks}\label{Summary5}
	\begin{tabular}{cclc}\hline
		\textbf{Ref.} & \textbf{\begin{tabular}[c]{@{}c@{}}Security\\ Threat\end{tabular}} &  \multicolumn{1}{c}{\textbf{\begin{tabular}[c]{@{}l@{}}$\star$ Purpose\\$\bullet$ Advantages\\$\circ$ Limitations\\ $\dagger$ {Evaluation Metrics} \end{tabular}}} & \textbf{\begin{tabular}[c]{@{}c@{}}Utilized Technology\end{tabular}} \\\hline
		
		% Blockchain for Tamper-proof Data Example
		% \cite{Blockchain Tamper-proof} 
		\cite{wang2022blockchainbaseda}
		& {{\begin{tabular}[l]{@{}c@{}}  Desynchronization of \\ KBs, data tampering \end{tabular}}}  
		& {\begin{tabular}[l]{@{}l@{}}
				$\star$ Safeguard data/knowledge integrity and confidentiality during sharing\\
				$\bullet$ High agents' QoE, enhanced efficiency, and strengthened security\\
				$\circ$ Potential scalability and latency issues in large-scale deployments \\
				$\dagger$ Detection ratio of malicious full nodes and ratio of stakes of legitimate nodes  \\
		\end{tabular}} &
		\begin{tabular}[c]{@{}c@{}} Coalition-matching\\ game, blockchain \end{tabular} 
		\\\hline
		
		% Access Control Mechanisms Example
		% \cite{Access Control} 
		\cite{xue2023sparkac}			
		& {{\begin{tabular}[l]{@{}c@{}}  Unauthorized KBs \\ access, poisoning attack\end{tabular}}}  
		& {\begin{tabular}[l]{@{}l@{}}
				$\star$	Flexible access control policies for big data/knowledge sharing and analysis\\
				$\bullet$ Fine-grained, purpose-aware, and low-performance overhead\\
				$\circ$ Complexity of configuration and difficulty in implementing cross-domain access\\
				$\dagger$ Set sum of squared error and query processing efficiency
		\end{tabular}} &  Purpose-based AC 
		\\\hline
		
		% Machine Unlearning for Privacy Example
		% \cite{Machine Unlearning} 
		\cite{sekhari2021remembera}
		& {{\begin{tabular}[l]{@{}c@{}}  Privacy exposure from \\ erased data \end{tabular}}}  
		& {\begin{tabular}[l]{@{}l@{}}
				$\star$ Endow trained data/knowledge with deletion capacity in efficient ways\\
				$\bullet$ Reduce storage and computational overhead without access to the training data\\
				$\circ$ Lack information-theoretic guarantee and adaptability to non-convex loss\\
				$\dagger$ Computational and storage complexity
		\end{tabular}} 
		&\begin{tabular}[c]{@{}c@{}} DP, machine\\ unlearning \end{tabular} \\\hline
		
		% Federated Learning for Privacy Preservation Example
		% \cite{Federated Learning Privacy} 
		\cite{wei2023federateda}
		& {{\begin{tabular}[l]{@{}c@{}}   Privacy exposure from  \\ raw data/KBs \end{tabular}}}  
		& {\begin{tabular}[l]{@{}l@{}}
				$\star$ 
				Federated semantic codecs training with IB for balancing accuracy and privacy\\
				$\bullet$ 
				High rate-distortion performance and convergence in harsh channel conditions\\
				$\circ$ 
				Limited applicability to non-IID data and high communication overhead\\
				$\dagger$ Recall accuracy, latency, and error rate
		\end{tabular}} 
		&\begin{tabular}[c]{@{}c@{}} FL, IB \end{tabular} 
		\\\hline
		
		% Potential Privacy Protection Approaches Example
		% \cite{Privacy Protection Approaches} 
		\cite{zhao2023data}
		& {{\begin{tabular}[l]{@{}c@{}}  Gradient leakage attack \end{tabular}}}  
		& {\begin{tabular}[l]{@{}l@{}}
				$\star$ 
				Offer customized and controllable data utilization for agents\\
				$\bullet$ High personalization, adaptability, and customization\\
				$\circ$ Degradation of model accuracy\\
				$\dagger$ Peak SNR
		\end{tabular}} & DP 
		\\\hline
		
	\end{tabular}
\end{table*}

	\subsubsection{{Defense Strategies Against Semantic Data Poisoning}} \label{subsec:semantic_data_defense}
    {The semantic data poisoning attacker manipulates the training data used by semantic codecs to degrade their performance. 
    Several effective defense strategies have been proposed, such as data preprocessing \cite{borgnia2021strong}, poison filtering \cite{li2017learning}, and robustness training \cite{wang2022threats}. Data preprocessing techniques, such as data augmentation and transformation, can strengthen the model against such attacks. For instance, Borgnia \emph{et al.} \cite{borgnia2021strong} demonstrate that strong data augmentations (e.g., Mixup and CutMix), significantly improve resistance to poisoning attacks without trading off performance. 
    Apart from data preprocessing, the poison-filter-based defenses \cite{wang2022threats} focus on detecting and removing corrupted data before training semantic models. Methods such as outlier detection, ensemble-based filtering, validation-based filtering, and k-nearest neighbors (k-NN) filtering \cite{wang2022threats} can identify and eliminate anomalous data.  For instance, Li \emph{et al.} \cite{li2017learning} propose a defense mechanism based on knowledge distillation to counteract data poisoning. By training an auxiliary model on a validation set and distilling its knowledge to create pseudo-labels, they sanitize the dataset and enhance the model’s robustness. Additionally, robust training techniques \cite{wang2022threats} such as robust regression enhance the semantic models' resistance by optimizing the training process to mitigate the impact of malicious data.}

	\subsubsection{Blockchain for Tamper-proof SemComNet}\label{Data_Knowledge_Blockchain}
	The distributed nature and immutability of blockchain make the SemComNet under the premise of ensuring data availability \cite{liang2017provchain} and sharing \cite{wang2022blockchainbaseda}. It provides SemComNet with integrity \cite{wang2021blockchainbased}, confidentiality \cite{fotiou2016decentralized} of semantic data \& knowledge while maintaining it can be redactable \cite{ateniese2017redactable}.
	Specifically, 
	Liang \emph{et al.} in \cite{liang2017provchain} present a blockchain-based data provenance system named \textit{ProvChain} to ensure semantic data source security from collection, storage, and verification in three stages. 
	For secure decentralized knowledge sharing, Wang \emph{et al.} \cite{wang2022blockchainbaseda} introduce a blockchain-based framework to reduce malicious activities effectively. That holds promise for supporting secure knowledge sharing in SemComNet. However, scalability and latency concerns in extensive SemComNet deployments require careful consideration.

	To ensure remote knowledge integrity in cloud storage services, Wang \emph{et al.} \cite{wang2021blockchainbased} introduce a blockchain-based private provable data possession scheme, to ensure remote data integrity in cloud storage. Compared with existing cryptography schemes, they offer enhanced security, efficiency, and practicality while ensuring agent anonymity.
	As for confidentiality of data and knowledge, Fotiou \emph{et al.} \cite{fotiou2016decentralized} introduce a decentralized security approach for content distribution utilizing blockchain in a fully distributed manner, ensuring security without relying on central authorities.
	Lastly, blockchain technology plays a crucial role in ensuring trustworthy data deletion. For instance, Ateniese \emph{et al.} \cite{ateniese2017redactable} propose a framework for redacting and compressing the content of blocks in blockchain-based systems. Experiment results show the overhead imposed by having a mutable blockchain is negligible.

	\subsubsection{Data \& Knowledge Access Control in SemComNet}\label{Data_Knowledge_AC}
	Various access control (AC) policies, including role-based \cite{sultan2023rolebased}, attribute-based \cite{xue2019attributebased,xu2021integrated}, purpose-based \cite{xue2023sparkac}, can be employed to safeguard knowledge and semantic data in SemComNet, tailored to specific requirements.
	For instance, Sultan \emph{et al.} \cite{sultan2023rolebased} introduce a cryptographic role-based encryption for AC, where only authorized users decrypt data. It includes efficient user revocation and outsourced decryption, reducing computational load. Theoretical analysis proves its security against chosen-plaintext attacks, making it ideal for practical SemComNet applications.
	For fine-grained AC in collaborative scenarios, 
	Xue \emph{et al.} \cite{xue2019attributebased} employs attribute-based encryption to facilitate collaborative access based on owner-defined policies, which may be applied to SemComNet to ensure the clustered and networked agents collaboration while preventing unauthorized collusion attempts.

	However, the above role and attribute-based AC models usually fall short in allowing agents to perform statistical analysis on sensitive semantic data and knowledge without direct access.
	To tackle this challenge, on the one hand, Xu \emph{et al.} \cite{xu2021integrated} propose a privacy-preserving, revocable attribute-based AC using a linear secret sharing scheme and an extended path oblivious random access memory protocol. This ensures privacy and allows for fine-grained access control, such as write access and policy updates.
    On the other hand, the purpose-aware AC model allows knowledge owners to define intended usage purposes, ensuring privacy-preserving access control. For instance, Xue \emph{et al.} \cite{xue2023sparkac} propose an automatic purpose-aware AC model that distinguishes data processing purposes and enforces access using two mechanisms for structured and unstructured data. While experiments show efficiency in large-scale platforms, the model's complexity and challenges in cross-domain access hinder its broader adoption in SemComNet. Further research is needed to develop context-aware AC that adapts permissions based on specific communication environments, enhancing SemComNet's adaptability and intelligence.

	\subsubsection{Machine Unlearning for Privacy Preservation in SemComNet}\label{subsec:machine_unlearning}
	In SemComNet, agents may need to erase their communication history and private knowledge for privacy and security reasons. {Traditional methods, such as retraining DNN-based codecs from scratch, are resource-intensive and disrupt communication services. 
	The emerging concept of machine unlearning \cite{xu2024machine} offers an efficient solution for data removal and model adaptation.  It provides a lightweight approach to safeguard agents' right to be forgotten while maintaining seamless communication, effectively balancing privacy and efficiency.}

    {In \cite{golatkar2020eternal}, Golatkar \emph{et al.} propose a method called scrubbing to selectively remove any information about a dataset from the network's weights without requiring access to the original data or retraining. 
    While effective, its implementation in SemComNet is challenging due to its reliance on simple and well-structured models.
    To simplify unlearning, Sekhari \emph{et al.} \cite{sekhari2021remembera} present an approximate unlearning algorithm that reduces computational and storage complexities. By applying disturbance updates during training to subtract the influence of specific samples, an unlearned model was produced. However, the work \cite{sekhari2021remembera} has practical limitations, including reliance on convex loss functions and the lack of an information-theoretic guarantee for unlearning.}

    The above centralized unlearning methods \cite{golatkar2020eternal, sekhari2021remembera} require access to all training data, while the federated unlearning, which does not, is more practical in the distributed training setting of SemComNet. {It aims to erase a client’s contributions from the trained models and can be categorized into server-side and client-side approaches \cite{wang2024machine}.
    Server-side unlearning is more efficient as it allows clients' contributions to be erased without their participation. For instance, Wu \emph{et al.} \cite{wu2022federated} propose a server-based approach that subtracts a client's accumulated updates from the global model and uses knowledge distillation to restore its performance. In contrast, client-side federated unlearning involves clients erasing the influence of specific data samples. Liu \emph{et al.} \cite{liu2022right} propose a Newton-type retraining algorithm that removes the influence of specific samples, using the Fisher information matrix to minimize retraining costs. Their approach achieves efficient data erasure while maintaining model accuracy.}

	\subsubsection{FL for Privacy Preservation in SemComNet}	\label{subsec:FL_privacy_preservation}
     Centralized learning in SemComNet requires aggregating raw data from users, which leads to traffic congestion and privacy concerns \cite{xie2022taskoriented,zhao2023data,wei2023federateda}. FL offers a more privacy-preserving and efficient alternative by keeping user data on devices and reducing the training burden through shared efforts. Researchers have explored integrating FL with SemComNet for tasks such as training semantic codecs and constructing KBs \cite{zhao2023data, wei2023federateda, lu2024efficient}. For semantic codecs training, Zhao \emph{et al.} \cite{zhao2023data} propose an online inference and offline FL framework that balances privacy with data utilization by combining privacy-protected model training and personalized deployment. For shared KB training, Wei \emph{et al.} \cite{wei2023federateda} introduce the federated semantic learning (FedSem) framework. This approach allows agents to train local KBs and transmit learned knowledge (as model parameters) to the server. The IB mechanism is also employed to limit the amount of shared knowledge, ensuring privacy protection and improving rate-distortion and convergence performance.

	\subsubsection{Potential Privacy Protection Approaches for SemComNet}\label{subsec:potential_privacy_protection}
	Differential privacy (DP), another widely used technique in SemComNet, safeguards semantic data and knowledge by adding noise. For instance, 
    Min \emph{et al.} \cite{min2021reinforcement} apply the DP mechanism to randomize semantic locations, using RL to adaptively select the perturbation policies based on location sensitivity and attack history. Experimental results show effectiveness in balancing privacy and QoE loss. Additionally, to resist the gradient leakage attacks where attackers can infer private training data from shared semantic model parameters or gradients, Zhao \emph{et al.} \cite{zhao2023data} employ edge-side model aggregation with DP. This approach allows agents to add random noise to parameter sharing based on their diverse privacy needs, ensuring personalized privacy protection. 
	However, adding noise may markedly degrade the accuracy of semantic codecs, presenting a trade-off between model accuracy and privacy protection that needs careful consideration. In response, Liu \emph{et al.} \cite{liu2024adaptive} propose APB-DP, an adaptive privacy budget-based DP scheme that balances model performance and privacy in collaborative training. Simulations show that APB-DP reduces the \textit{privacy leakage rate} (PRL) by 13\% and \textit{performance loss rate} (PLoss) by 71\% compared to the standard FL scheme. Additionally, techniques such as knowledge distillation \cite{chen2023trustworthya}, secure multi-party computation \cite{knott2021crypten}, and trusted execution environment \cite{jauernig2020trusted} also offer potential solutions safeguarding privacy in SemComNet.

	\vspace{-3mm}
	\subsection{Summary and Lessons Learned}
	{This section has discussed advanced security/privacy countermeasures for SemComNet}, encompassing semantic model security, transmission security, robustness, trust mechanisms, and data \& knowledge security (i.e., from Sect.~\ref{subsec:defense1} to Sect.~\ref{subsec:defense5}).
	The insights from the literature review highlight potential pathways to building a secure, privacy-preserving, robust, and trustworthy SemComNet.
	The summary and key lessons learned from this section are listed as follows.
	\begin{itemize}
		\item \emph{Semantic model security}. 
        {Semantic models, foundational to SemComNet, leverage AI techniques for semantic understanding tasks. However, they face threats like AEs, poisoning examples, sponge examples, and model theft. In Sect.~\ref{subsec:defense1}, we have explored key strategies to mitigate these risks. Specifically, adversarial training, defensive distillation, and AE detectors improve model robustness against AEs, while detection and adversarial training methods mitigate sponge examples. For poisoning threats, anomaly detection, data sanitization, and robust aggregation prove effective. Besides, watermarking and blockchain-based techniques are promising for IP protection. Table~\ref{Summary1} summarizes these defenses for semantic model-related threats.}

		\item \emph{Semantic transmission security}. 
        The broadcast nature of communication channels allows any nearby user to intercept transmitted SI, creating opportunities for adversaries to launch attacks such as semantic eavesdropping and jamming \cite{du2023rethinking}.
		For secure SI transmission in SemComNet, we have learned that existing cryptographic schemes, physical layer security techniques (including secret key generation and secure transmission approaches), covert communication, quantum technology, and other emerging communication technologies (e.g., RIS and spread spectrum techniques) can offer some insights for protecting SI transmission.
        {Table~\ref{Summary2} compares existing/potential countermeasures for addressing semantic transmission threats.}
		
		\item \emph{Reliable SemComNet}. 
		Reliability in SemComNet is essential to tolerate perturbations during transmission and maintain resilience in SI interpretation (e.g., reducing semantic noise). We have learned that AI techniques including ML, GAI, and RL could enhance the resilience of semantic interpretation while optimizing bandwidth usage.
		{Besides, adaptive transmission strategies could offer some insights for enhancing reliability.	Moreover, incorporating reasoning abilities and collaborative relays within SemComNet further enhances robustness by improving communication reliability and ensuring precise SI interpretation.    Table~\ref{Summary3} summarizes existing/potential countermeasures for addressing reliability risks in SemComNet.}

        \item \emph{Trust management in SemComNet}. 
            {
            To build a trustworthy SemComNet, trust solutions can be characterized as intrinsic, trust-free, and quantifiable approaches. Specifically, PLA mechanisms establish intrinsic trust by verifying agents' identities through inherent properties (e.g., RF fingerprints), forming the foundation for trust evaluation. Moreover, blockchain creates a trustless environment by eliminating reliance on central authorities, making it ideal for large-scale identity management. Besides, measurable trust evaluation mechanisms assess the reliability of agents or semantic content. Table~\ref{Summary4} summarizes existing and potential countermeasures for addressing trust management challenges in SemComNet.}

% Traditional methods rely on evaluating specific influencing attributes, while advanced approaches leverage AI techniques (e.g., DL and DRL) for intelligent and automated trust assessments.
		
		\item \emph{Data \& knowledge security in SemComNet}. % 数据\知识安全  from Summary-4 avail
		{For data \& knowledge security in SemComNet, the distributed nature and expanded attack surface intensify existing threats.
        In response, Blockchain emerges as a promising solution, ensuring availability, integrity, and confidentiality throughout their life cycle (e.g., collection, storage, sharing, and destruction). Moreover, various AC policies (e.g., role-based, attribute-based, and purpose-based) allow for tailored restrictions on agents' access to critical information.
		Lastly, for privacy protection, advanced approaches such as machine unlearning, FL, and DP offer effective solutions. However, further advancements tailored to SemComNet's unique characteristics are needed.}
	A comparison of existing/potential defenses tailored for data/knowledge security and privacy issues is presented in Table~\ref{Summary5}.
	\end{itemize}

	\section{Future Research Directions}
	Recent SemComNet research has shown significant advancements. However, several critical issues remain unexplored. This section explores key challenges in SemComNet research and outlines potential future directions.
	
	\subsection{Green SemComNet Architecture}
	In large-scale SemComNet, frequent knowledge/model sharing significantly increases resource and energy consumption. {To address this, SemComNet should adopt green and eco-friendly designs to minimize resource waste and prolong agents' battery life.}
	On the one hand, various communication technologies (e.g., ultra-massive MIMO and RIS) are increasingly deployed to meet the growing demand for high data throughput \cite{zhang2022wisdomevolutionary}. 
    A potential research direction is ensuring seamless compatibility between SemComNet and existing communication infrastructure. It ensures a smooth transition to a more intelligent and green paradigm without wasting current infrastructure investments.
	On the other hand, the growing complexity of communication tasks imposes a significant burden on resource-constrained agents. For instance, the reliability of data-driven codecs requires rich high-quality datasets for training and sufficient background knowledge for support. It is necessary to explore eco-friendly strategies that efficiently construct self-learning \cite{xiao2020selflearninga} and self-updating datasets \& KBs, reducing the maintenance cost. Additionally, from an economic perspective, exploring the utilization of low-power hardware, such as neuromorphic chips \cite{davies2021advancing} and quantum photonic chips \cite{luo2023recent}, offers a promising avenue for energy-efficient and secure task execution.

	\subsection{Explainable Semantic Model}
        The mathematical representation of semantics is inherently challenging, but the powerful semantic extraction capabilities of DL have made SemCom a reality \cite{xie2021deep, qin2022semantic}. 
        However, the opaque, black-box nature of DL-based semantic codecs results in a lack of explainability, raising concerns about reliability and social acceptance, particularly in safety-sensitive domains such as transportation and healthcare.
        {Explainable AI (XAI) offers a promising avenue by making decision-making processes transparent and traceable. In SemComNet, XAI can be applied across its three layers, such as improving resource allocation interpretation in the control layer, clarifying SI extraction and reconstruction in the transmission layer, and enhancing knowledge acquisition clarity in the cognitive sensing layer.
        Additionally, integrating domain-specific knowledge to enhance contextual explainability remains an open challenge. For instance, KGs offer structured and interpretable representations of knowledge \cite{yang2023taskdrivena}, enabling contextual understanding and the incorporation of domain expertise. However, adapting KGs to SemComNet and ensuring their dynamic updates in response to evolving contexts pose critical challenges. }

	\subsection{SemComNet Orchestrated with Generative AI}
        The integration of GAI models (e.g., DALL-E and GPT-4) with SemComNet holds transformative potential, enabling applications such as KBs creation and content refinement. These models can generate semantically consistent, context-aware content to enrich KBs and enhance semantic reasoning \cite{xia2023generative}. However, their incorporation into SemComNet faces challenges in trustworthiness, sustainability, and personalization, which require further research. Firstly, GAI models are prone to \textit{hallucination} that generates outputs that appear plausible but are nonsensical or adversarial, thereby compromising services reliability \cite{tonmoy2024comprehensive}.
        Secondly, the resource-intensive and time-consuming nature of GAI models, both for training and inference, pose challenges for resource-constrained agents \cite{xia2023generative}. Techniques such as quantization and model compression can enhance efficiency in large-scale SemComNet deployments while maintaining performance \cite{pan2024cloudedge}.
        Thirdly, while GAI models offer general-purpose solutions, adapting them to personalized scenarios through fine-tuning remains an open research area requiring further attention.

\subsection{Endogenous Secure SemComNet}
    
    As security threats in SemComNet become more complex and sophisticated, traditional patch-like protection solutions are increasingly ineffective. There is an urgent need to develop a SemComNet framework with endogenous security \cite{wu2022development}, where security mechanisms are integrated by design into the system architecture from the outset. This proactive approach, which incorporates self-protection, self-evolution, rapid response, and autoimmunity capabilities, is aimed at adapting to dynamic environments and resisting both known and unknown threats.
    For instance, physical-layer steganography technology \cite{xie2022security} can encrypt and conceal SI by exploiting the unique properties of the physical channel, providing quantum-resistant security for data transmission within SemComNet.
    {However, developing a comprehensive and systematically integrated endogenous security framework for SemComNet remains a significant challenge. Moreover, given the collaborative nature of SemComNet, where multiple agents work together to achieve common objectives, further research is needed to design privacy-preserving and endogenous security mechanisms for collaborative decision-making and conflict resolution.} Such mechanisms would ensure secure cooperation among agents, especially in knowledge sharing and SI transmission, while safeguarding confidentiality and integrity in multi-agent interactions.

	\subsection{Adaptive SemComNet Design}
{As a future direction, adaptive SemComNet should address two critical challenges to support flexible collaboration and diverse demands in practical scenarios.
Firstly, achieving effective SemCom among multiple users in dynamic and complex networks requires overcoming challenges such as knowledge synchronization and adaptation to evolving network topologies. Synchronizing shared KBs is crucial for maintaining consistent semantic understanding, yet it becomes challenging in networked SemCom with changing topologies or intermittent connections \cite{shi2021semanticd}. Despite its significance, the implementation of adaptive networked SemCom remains underexplored and warrants further investigation.
Secondly, efficiently handling multiple semantic tasks simultaneously is another pressing issue \cite{zhang2024unified}. The SI extracted by different DL-based codecs for semantic tasks (e.g., behavior identification and anomaly detection) is often incompatible and non-interchangeable \cite{tian2023asynchronous}, leading to inefficiencies when facing diverse tasks. A promising solution lies in developing unified semantic models capable of supporting multiple tasks. Such models could leverage advanced techniques such as continual learning, meta-learning, and multi-task learning to dynamically adapt to various tasks without requiring extensive retraining or storing multiple models. }

	\section{Conclusion}
	In this work, a thorough survey of SemComNet on fundamental concepts, security/privacy concerns, and countermeasures aspects has been presented.
	Firstly, we have introduced a novel three-layered architecture of SemComNet, consisting of the control layer, semantic transmission layer, and cognitive sensing layer.
	Afterward, we have discussed three working modes (i.e., paired, clustered, and networked) of SemComNet, along with its supporting technologies, use cases, and evaluation metrics.
	Next, our survey has revealed critical security and privacy threats of SemComNet from the three functional layers, which have not been comprehensively investigated in existing research.
	Then, to build a secure and robust SemComNet, the security and privacy countermeasures have been reviewed and examined from both academic and industrial perspectives, and the key challenges have been discussed to build tailored defenses in SemComNet.
	Finally, we have outlined the future research directions for SemComNet.
    We hope this work serves as a valuable guideline for security and privacy in SemComNet and encourages further research in this emerging area.

%\vspace{-0.2cm}
\section*{Acknowledgment} %\vspace{-0.05cm}
This work was supported in part by the 
NSFC (Nos. U20A20175, U22A2029, 62302387, and U23A20276), Key Research and Development Program of Shaanxi (No. 2022GXLH-01-25).
	
    \bibliographystyle{IEEEtran}
    \bibliography{SemCom.bib}

\begin{thebibliography}{100}
\providecommand{\url}[1]{#1}
\csname url@samestyle\endcsname
\providecommand{\newblock}{\relax}
\providecommand{\bibinfo}[2]{#2}
\providecommand{\BIBentrySTDinterwordspacing}{\spaceskip=0pt\relax}
\providecommand{\BIBentryALTinterwordstretchfactor}{4}
\providecommand{\BIBentryALTinterwordspacing}{\spaceskip=\fontdimen2\font plus
\BIBentryALTinterwordstretchfactor\fontdimen3\font minus \fontdimen4\font\relax}
\providecommand{\BIBforeignlanguage}[2]{{%
\expandafter\ifx\csname l@#1\endcsname\relax
\typeout{** WARNING: IEEEtran.bst: No hyphenation pattern has been}%
\typeout{** loaded for the language `#1'. Using the pattern for}%
\typeout{** the default language instead.}%
\else
\language=\csname l@#1\endcsname
\fi
#2}}
\providecommand{\BIBdecl}{\relax}
\BIBdecl

\bibitem{zhang2022toward}
P.~Zhang, H.~Yang, Z.~Feng, Y.~Cui, J.~Dai, X.~Qin, J.~Li, and Q.~Zhang, ``Toward intelligent and efficient 6g networks: Jcsc enabled on-purpose machine communications,'' \emph{IEEE Wireless Communications}, vol.~30, no.~1, pp. 150--157, 2022.

\bibitem{luo2022semantic}
X.~Luo, H.-H. Chen, and Q.~Guo, ``Semantic {{Communications}}: {{Overview}}, {{Open Issues}}, and {{Future Research Directions}},'' \emph{IEEE Wireless Communications}, vol.~29, no.~1, pp. 210--219, 2022.

\bibitem{xie2021deep}
H.~Xie, Z.~Qin, G.~Y. Li, and B.-H. Juang, ``Deep {{Learning Enabled Semantic Communication Systems}},'' \emph{IEEE Transactions on Signal Processing}, vol.~69, pp. 2663--2675, 2021.

\bibitem{han2023semanticpreserved}
T.~Han, Q.~Yang, Z.~Shi, S.~He, and Z.~Zhang, ``Semantic-{{Preserved Communication System}} for {{Highly Efficient Speech Transmission}},'' \emph{IEEE Journal on Selected Areas in Communications}, vol.~41, no.~1, pp. 245--259, 2023.

\bibitem{wang2022taskoffloading}
Y.~Wang, W.~Chen, T.~H. Luan, Z.~Su, Q.~Xu, R.~Li, and N.~Chen, ``Task offloading for post-disaster rescue in unmanned aerial vehicles networks,'' \emph{IEEE/ACM Transactions on Networking}, vol.~30, no.~4, pp. 1525--1539, 2022.

\bibitem{zeng2024USV}
H.~Zeng, Z.~Su, Q.~Xu, R.~Li, Y.~Wang, M.~Dai, T.~H. Luan, X.~Sun, and D.~Liu, ``Usv fleet-assisted collaborative computation offloading for smart maritime services: An energy-efficient design,'' \emph{IEEE Transactions on Vehicular Technology}, vol.~73, no.~10, pp. 14\,718--14\,733, 2024.

\bibitem{huang2023semantica}
K.~Huang, Q.~Lan, Z.~Liu, and L.~Yang, ``Semantic {{Data Sourcing}} for {{6G Edge Intelligence}},'' \emph{IEEE Communications Magazine}, pp. 1--7, 2023.

\bibitem{peng2022robustb}
X.~Peng, Z.~Qin, D.~Huang, X.~Tao, J.~Lu, G.~Liu, and C.~Pan, ``A {{Robust Deep Learning Enabled Semantic Communication System}} for {{Text}},'' in \emph{Proceedings of {{IEEE Global Communications Conference}} ({{GLOBECOM}})}, 2022, pp. 2704--2709.

\bibitem{zhao2022semkey}
R.~Zhao, Q.~Qin, N.~Xu, G.~Nan, Q.~Cui, and X.~Tao, ``{{SemKey}}: {{Boosting Secret Key Generation}} for {{RIS-assisted Semantic Communication Systems}},'' in \emph{Proceedings of {{IEEE Vehicular Technology Conference}} ({{VTC2022-Fall}})}, 2022, pp. 1--5.

\bibitem{nan2023physicallayera}
G.~Nan, Z.~Li, J.~Zhai, Q.~Cui, G.~Chen, X.~Du, X.~Zhang, X.~Tao, Z.~Han, and T.~Q.~S. Quek, ``Physical-{{Layer Adversarial Robustness}} for {{Deep Learning-Based Semantic Communications}},'' \emph{IEEE Journal on Selected Areas in Communications}, vol.~41, no.~8, pp. 2592--2608, 2023.

\bibitem{ChineseEngineers6G}
\BIBentryALTinterwordspacing
C.~T. Blog. (2024) Chinese engineers field test a 6g network with semantic communications on 4g infrastructure. Accessed: 2024-09-22. [Online]. Available: \url{https://techblog.comsoc.org/2024/07/15/chinese-engineers-field-test-a-6g-network-with-semantic-communications-on-4g-infrastructure/}
\BIBentrySTDinterwordspacing

\bibitem{zhang2022wisdomevolutionary}
P.~Zhang, W.~Xu, H.~Gao, K.~Niu, X.~Xu, X.~Qin, C.~Yuan, Z.~Qin, H.~Zhao, J.~Wei, and F.~Zhang, ``Toward {{Wisdom-Evolutionary}} and {{Primitive-Concise 6G}}: {{A New Paradigm}} of {{Semantic Communication Networks}},'' \emph{Engineering}, vol.~8, pp. 60--73, 2022.

\bibitem{shi2021semanticd}
G.~Shi, Y.~Xiao, Y.~Li, and X.~Xie, ``From {{Semantic Communication}} to {{Semantic-Aware Networking}}: {{Model}}, {{Architecture}}, and {{Open Problems}},'' \emph{IEEE Communications Magazine}, vol.~59, no.~8, pp. 44--50, 2021.

\bibitem{qin2024ai}
Z.~Qin, L.~Liang, Z.~Wang, S.~Jin, X.~Tao, W.~Tong, and G.~Y. Li, ``Ai empowered wireless communications: From bits to semantics,'' \emph{Proceedings of the IEEE}, pp. 1--32, 2024, doi:10.1109/JPROC.2024.3437730.

\bibitem{tian2023asynchronous}
Z.~Tian, H.~Vo, C.~Zhang, G.~Min, and S.~Yu, ``An asynchronous multi-task semantic communication method,'' \emph{IEEE Network}, vol.~38, no.~4, pp. 275--283, 2024.

\bibitem{shen2023secure}
M.~Shen, J.~Wang, H.~Du, D.~Niyato, X.~Tang, J.~Kang, Y.~Ding, and L.~Zhu, ``Secure semantic communications: Challenges, approaches, and opportunities,'' \emph{IEEE Network}, vol.~38, no.~4, pp. 197--206, 2024.

\bibitem{peng2024semantic}
H.~Peng, Z.~Zhang, Y.~Liu, Z.~Su, T.~H. Luan, and N.~Cheng, ``Semantic communication in non-terrestrial networks: A future-ready paradigm,'' \emph{IEEE Network}, vol.~38, no.~4, pp. 119--127, 2024.

\bibitem{shumailov2021sponge}
I.~Shumailov, Y.~Zhao, D.~Bates, N.~Papernot, R.~Mullins, and R.~Anderson, ``Sponge {{Examples}}: {{Energy-Latency Attacks}} on {{Neural Networks}},'' in \emph{Proceedings of {{IEEE European Symposium}} on {{Security}} and {{Privacy}} ({{EuroS}}\&{{P}})}, 2021, pp. 212--231.

\bibitem{du2023rethinking}
H.~Du, J.~Wang, D.~Niyato, J.~Kang, Z.~Xiong, M.~Guizani, and D.~I. Kim, ``Rethinking {{Wireless Communication Security}} in {{Semantic Internet}} of {{Things}},'' \emph{IEEE Wireless Communications}, vol.~30, no.~3, pp. 36--43, 2023.

\bibitem{li2022crossmodal}
A.~Li, X.~Wei, D.~Wu, and L.~Zhou, ``Cross-{{Modal Semantic Communications}},'' \emph{IEEE Wireless Communications}, vol.~29, no.~6, pp. 1--8, 2022.

\bibitem{wang2024privacy}
Y.~Wang, S.~Guo, Y.~Deng, H.~Zhang, and Y.~Fang, ``Privacy-preserving task-oriented semantic communications against model inversion attacks,'' \emph{IEEE Transactions on Wireless Communications}, vol.~23, no.~8, pp. 10\,150--10\,165, 2024.

\bibitem{kang2023adversariala}
J.~Kang, J.~He, H.~Du, Z.~Xiong, Z.~Yang, X.~Huang, and S.~Xie, ``Adversarial {{Attacks}} and {{Defenses}} for {{Semantic Communication}} in {{Vehicular Metaverses}},'' \emph{IEEE Wireless Communications}, vol.~30, no.~4, pp. 48--55, 2023.

\bibitem{wang2023surveyc}
Y.~Wang, Z.~Su, S.~Guo, M.~Dai, T.~H. Luan, and Y.~Liu, ``A {{Survey}} on {{Digital Twins}}: {{Architecture}}, {{Enabling Technologies}}, {{Security}} and {{Privacy}}, and {{Future Prospects}},'' \emph{IEEE Internet of Things Journal}, vol.~10, no.~17, pp. 14\,965--14\,987, 2023.

\bibitem{cheng2024knowledge}
S.~Cheng, X.~Zhang, Y.~Sun, Q.~Cui, and X.~Tao, ``Knowledge discrepancy oriented privacy preserving for semantic communication,'' \emph{IEEE Transactions on Vehicular Technology}, vol.~73, no.~8, pp. 11\,637--11\,646, 2024.

\bibitem{li2023secure}
C.~Li, L.~Zeng, X.~Huang, X.~Miao, and S.~Wang, ``Secure {{Semantic Communication Model}} for {{Black-Box Attack Challenge Under Metaverse}},'' \emph{IEEE Wireless Communications}, vol.~30, no.~4, pp. 56--62, 2023.

\bibitem{qiu2018how}
T.~Qiu, N.~Chen, K.~Li, M.~Atiquzzaman, and W.~Zhao, ``How {{Can Heterogeneous Internet}} of {{Things Build Our Future}}: {{A Survey}},'' \emph{IEEE Communications Surveys \& Tutorials}, vol.~20, no.~3, pp. 2011--2027, 2018.

\bibitem{lan2021whatd}
Q.~Lan, D.~Wen, Z.~Zhang, Q.~Zeng, X.~Chen, P.~Popovski, and K.~Huang, ``What is {{Semantic Communication}}? {{A View}} on {{Conveying Meaning}} in the {{Era}} of {{Machine Intelligence}},'' \emph{Journal of Communications and Information Networks}, vol.~6, no.~4, pp. 336--371, 2021.

\bibitem{qin2022semantic}
Z.~Qin, X.~Tao, J.~Lu, W.~Tong, and G.~Y. Li, ``Semantic {{Communications}}: {{Principles}} and {{Challenges}},'' \emph{arXiv preprint arXiv:2201.01389}, 2022.

\bibitem{uysal2022semantica}
E.~Uysal, O.~Kaya, A.~Ephremides, J.~Gross, M.~Codreanu, P.~Popovski, M.~Assaad, G.~Liva, A.~Munari, B.~Soret, T.~Soleymani, and K.~H. Johansson, ``Semantic communications in networked systems: {{A}} data significance perspective,'' \emph{IEEE Network}, vol.~36, no.~4, pp. 233--240, 2022.

\bibitem{gunduz2023transmitting}
D.~G{\"u}nd{\"u}z, Z.~Qin, I.~E. Aguerri, H.~S. Dhillon, Z.~Yang, A.~Yener, K.~K. Wong, and C.-B. Chae, ``Beyond {{Transmitting Bits}}: {{Context}}, {{Semantics}}, and {{Task-Oriented Communications}},'' \emph{IEEE Journal on Selected Areas in Communications}, vol.~41, no.~1, pp. 5--41, 2023.

\bibitem{yang2023semanticb}
W.~Yang, H.~Du, Z.~Q. Liew, W.~Y.~B. Lim, Z.~Xiong, D.~Niyato, X.~Chi, X.~Shen, and C.~Miao, ``Semantic {{Communications}} for {{Future Internet}}: {{Fundamentals}}, {{Applications}}, and {{Challenges}},'' \emph{IEEE Communications Surveys \& Tutorials}, vol.~25, no.~1, pp. 213--250, 2023.

\bibitem{lu2023semanticsempowered}
Z.~Lu, R.~Li, K.~Lu, X.~Chen, E.~Hossain, Z.~Zhao, and H.~Zhang, ``Semantics-{{Empowered Communications}}: {{A Tutorial-cum-Survey}},'' \emph{IEEE Communications Surveys \& Tutorials}, vol.~26, no.~1, pp. 41--79, 2024.

\bibitem{trevlakis2024natively}
S.~E. Trevlakis, N.~Pappas, and A.-A.~A. Boulogeorgos, ``Toward natively intelligent semantic communications and networking,'' \emph{IEEE Open Journal of the Communications Society}, vol.~5, pp. 1486--1503, 2024.

\bibitem{zhang2024unified}
G.~Zhang, Q.~Hu, Z.~Qin, Y.~Cai, G.~Yu, and X.~Tao, ``A unified multi-task semantic communication system for multimodal data,'' \emph{IEEE Transactions on Communications}, 2024, doi: 10.1109/TCOMM.2024.3364990.

\bibitem{qin2023generalized}
Z.~Qin, F.~Gao, B.~Lin, X.~Tao, G.~Liu, and C.~Pan, ``A {{Generalized Semantic Communication System}}: {{From Sources}} to {{Channels}},'' \emph{IEEE Wireless Communications}, vol.~30, no.~3, pp. 18--26, 2023.

\bibitem{xie2022taskoriented}
H.~Xie, Z.~Qin, X.~Tao, and K.~B. Letaief, ``Task-{{Oriented Multi-User Semantic Communications}},'' \emph{IEEE Journal on Selected Areas in Communications}, vol.~40, no.~9, pp. 2584--2597, 2022.

\bibitem{zhang2023deep}
H.~Zhang, S.~Shao, M.~Tao, X.~Bi, and K.~B. Letaief, ``Deep {{Learning-Enabled Semantic Communication Systems With Task-Unaware Transmitter}} and {{Dynamic Data}},'' \emph{IEEE Journal on Selected Areas in Communications}, vol.~41, no.~1, pp. 170--185, 2023.

\bibitem{weng2023deep}
Z.~Weng, Z.~Qin, X.~Tao, C.~Pan, G.~Liu, and G.~Y. Li, ``Deep learning enabled semantic communications with speech recognition and synthesis,'' \emph{IEEE Transactions on Wireless Communications}, vol.~22, no.~9, pp. 6227--6240, 2023.

\bibitem{gong2023adaptive}
W.~Gong, H.~Tong, S.~Wang, Z.~Yang, X.~He, and C.~Yin, ``Adaptive bitrate video semantic communication over wireless networks,'' in \emph{Proceedings of International Conference on Wireless Communications and Signal Processing (WCSP)}, 2023, pp. 122--127.

\bibitem{wang2023wireless}
S.~Wang, J.~Dai, Z.~Liang, K.~Niu, Z.~Si, C.~Dong, X.~Qin, and P.~Zhang, ``Wireless {{Deep Video Semantic Transmission}},'' \emph{IEEE Journal on Selected Areas in Communications}, vol.~41, no.~1, pp. 214--229, 2023.

\bibitem{al-fuqaha2015internet}
A.~{Al-Fuqaha}, M.~Guizani, M.~Mohammadi, M.~Aledhari, and M.~Ayyash, ``Internet of {{Things}}: {{A Survey}} on {{Enabling Technologies}}, {{Protocols}}, and {{Applications}},'' \emph{IEEE Communications Surveys \& Tutorials}, vol.~17, no.~4, pp. 2347--2376, 2015.

\bibitem{xu2024unleashing}
M.~Xu, H.~Du, D.~Niyato, J.~Kang, Z.~Xiong, S.~Mao, Z.~Han, A.~Jamalipour, D.~I. Kim, X.~Shen, V.~C.~M. Leung, and H.~V. Poor, ``Unleashing the power of edge-cloud generative ai in mobile networks: A survey of aigc services,'' \emph{IEEE Communications Surveys \& Tutorials}, vol.~26, no.~2, pp. 1127--1170, 2024.

\bibitem{min2023recent}
B.~Min, H.~Ross, E.~Sulem, A.~P.~B. Veyseh, T.~H. Nguyen, O.~Sainz, E.~Agirre, I.~Heintz, and D.~Roth, ``Recent {{Advances}} in {{Natural Language Processing}} via {{Large Pre-trained Language Models}}: {{A Survey}},'' \emph{ACM Computing Surveys}, vol.~56, no.~2, pp. 1--40, 2023.

\bibitem{xiao2023imitation}
Y.~Xiao, Z.~Sun, G.~Shi, and D.~Niyato, ``Imitation {{Learning-Based Implicit Semantic-Aware Communication Networks}}: {{Multi-Layer Representation}} and {{Collaborative Reasoning}},'' \emph{IEEE Journal on Selected Areas in Communications}, vol.~41, no.~3, pp. 639--658, 2023.

\bibitem{pan2024cloudedge}
Y.~Pan, Z.~Su, Y.~Wang, S.~Guo, H.~Liu, R.~Li, and Y.~Wu, ``Cloud-edge collaborative large model services: Challenges and solutions,'' \emph{IEEE Network}, 2024, doi:10.1109/MNET.2024.3442880.

\bibitem{zhao2023data}
L.~Zhao, D.~Wu, and L.~Zhou, ``Data {{Utilization Versus Privacy Protection}} in {{Semantic Communications}},'' \emph{IEEE Wireless Communications}, vol.~30, no.~3, pp. 44--50, 2023.

\bibitem{meng2020survey}
T.~Meng, X.~Jing, Z.~Yan, and W.~Pedrycz, ``A survey on machine learning for data fusion,'' \emph{Information Fusion}, vol.~57, pp. 115--129, 2020.

\bibitem{luo2022multimodal}
X.~Luo, R.~Gao, H.-H. Chen, S.~Chen, Q.~Guo, and P.~N. Suganthan, ``Multi-{{Modal}} and {{Multi-User Semantic Communications}} for {{Channel-Level Information Fusion}},'' \emph{IEEE Wireless Communications}, pp. 1--18, 2022.

\bibitem{bourtsoulatze2019deep}
E.~Bourtsoulatze, D.~B. Kurka, and D.~Gunduz, ``Deep {{Joint Source-Channel Coding}} for {{Wireless Image Transmission}},'' \emph{IEEE Transactions on Cognitive Communications and Networking}, vol.~5, no.~3, pp. 567--579, 2019.

\bibitem{xu2023deepa}
J.~Xu, B.~Ai, W.~Chen, N.~Wang, and M.~Rodrigues, ``Deep joint source-channel coding for image transmission with visual protection,'' \emph{IEEE Transactions on Cognitive Communications and Networking}, vol.~9, no.~6, pp. 1399--1411, 2023.

\bibitem{xia2023generative}
L.~Xia, Y.~Sun, C.~Liang, L.~Zhang, M.~A. Imran, and D.~Niyato, ``Generative {{AI}} for {{Semantic Communication}}: {{Architecture}}, {{Challenges}}, and {{Outlook}},'' \emph{arXiv preprint arXiv:2308.15483}, 2023.

\bibitem{wu2014cognitive}
Q.~Wu, G.~Ding, Y.~Xu, S.~Feng, Z.~Du, J.~Wang, and K.~Long, ``Cognitive {{Internet}} of {{Things}}: {{A New Paradigm Beyond Connection}},'' \emph{IEEE Internet of Things Journal}, vol.~1, no.~2, pp. 129--143, 2014.

\bibitem{seo2023semanticsnative}
H.~Seo, J.~Park, M.~Bennis, and M.~Debbah, ``Semantics-{{Native Communication}} via {{Contextual Reasoning}},'' \emph{IEEE Transactions on Cognitive Communications and Networking}, vol.~9, no.~3, pp. 604--617, 2023.

\bibitem{zhao2023tutorial}
Q.~Zhao, G.~Li, J.~Cai, M.~Zhou, and L.~Feng, ``A {{Tutorial}} on {{Internet}} of {{Behaviors}}: {{Concept}}, {{Architecture}}, {{Technology}}, {{Applications}}, and {{Challenges}},'' \emph{IEEE Communications Surveys \& Tutorials}, vol.~25, no.~2, pp. 1227--1260, 2023.

\bibitem{bourgin2019cognitive}
D.~D. Bourgin, J.~C. Peterson, D.~Reichman, S.~J. Russell, and T.~L. Griffiths, ``Cognitive model priors for predicting human decisions,'' in \emph{Proceedings of {{International Conference}} on {{Machine Learning}} ({{ICML}})}, 2019, pp. 5133--5141.

\bibitem{wu2021unified}
Q.~Wu, T.~Ruan, F.~Zhou, Y.~Huang, F.~Xu, S.~Zhao, Y.~Liu, and X.~Huang, ``A {{Unified Cognitive Learning Framework}} for {{Adapting}} to {{Dynamic Environments}} and {{Tasks}},'' \emph{IEEE Wireless Communications}, vol.~28, no.~6, pp. 208--216, 2021.

\bibitem{rabinowitz2018machinea}
N.~Rabinowitz, F.~Perbet, F.~Song, C.~Zhang, S.~M.~A. Eslami, and M.~Botvinick, ``Machine {{Theory}} of {{Mind}},'' in \emph{Proceedings of {{International Conference}} on {{Machine Learning}} ({{ICML}})}, 2018, pp. 4218--4227.

\bibitem{wang2022social}
Y.~Wang, Z.~Su, Y.~Pan, T.~H. Luan, R.~Li, and S.~Yu, ``Social-aware clustered federated learning with customized privacy preservation,'' \emph{IEEE/ACM Transactions on Networking}, vol.~32, no.~5, pp. 3654--3668, 2024.

\bibitem{guo2024distributed}
J.~Guo, H.~Chen, B.~Song, Y.~Chi, C.~Yuen, F.~R. Yu, G.~Y. Li, and D.~Niyato, ``Distributed task-oriented communication networks with multimodal semantic relay and edge intelligence,'' \emph{IEEE Communications Magazine}, vol.~62, no.~6, pp. 82--89, 2024.

\bibitem{tang2024cooperative}
B.~Tang, L.~Huang, Q.~Li, A.~Pandharipande, and X.~Ge, ``Cooperative {{Semantic Communication With On-Demand Semantic Forwarding}},'' \emph{IEEE Open Journal of the Communications Society}, vol.~5, pp. 349--363, 2024.

\bibitem{luo2022autoencoderbaseda}
X.~Luo, B.~Yin, Z.~Chen, B.~Xia, and J.~Wang, ``Autoencoder-based {{Semantic Communication Systems}} with {{Relay Channels}},'' in \emph{Proceedings of {{IEEE International Conference}} on {{Communications Workshops}} ({{ICC WKSHPS}})}, 2022, pp. 711--716.

\bibitem{lin2024semanticforward}
W.~Lin, Y.~Yan, L.~Li, Z.~Han, and T.~Matsumoto, ``Semantic-forward relaying: A novel framework toward 6g cooperative communications,'' \emph{IEEE Communications Letters}, vol.~28, no.~3, pp. 518--522, 2024.

\bibitem{abbasi2007survey}
A.~A. Abbasi and M.~Younis, ``A survey on clustering algorithms for wireless sensor networks,'' \emph{Computer Communications}, vol.~30, no.~14, pp. 2826--2841, 2007.

\bibitem{delange2022continual}
M.~De~Lange, R.~Aljundi, M.~Masana, S.~Parisot, X.~Jia, A.~Leonardis, G.~Slabaugh, and T.~Tuytelaars, ``A {{Continual Learning Survey}}: {{Defying Forgetting}} in {{Classification Tasks}},'' \emph{IEEE Transactions on Pattern Analysis and Machine Intelligence}, vol.~44, no.~7, pp. 3366--3385, 2022.

\bibitem{yin2023multidomain}
Z.~Yin, N.~Cheng, Y.~Hui, W.~Wang, L.~Zhao, K.~Aldubaikhy, and A.~Alqasir, ``Multi-domain {{Resource Multiplexing Based Secure Transmission}} for {{Satellite-Assisted IoT}}: {{AO-SCA Approach}},'' \emph{IEEE Transactions on Wireless Communications}, vol.~22, no.~11, pp. 7319--7330, 2023.

\bibitem{leonardi2022lora}
L.~Leonardi, L.~Lo~Bello, and G.~Patti, ``{{LoRa}} support for long-range real-time inter-cluster communications over {{Bluetooth Low Energy}} industrial networks,'' \emph{Computer Communications}, vol. 192, pp. 57--65, 2022.

\bibitem{wang2023knowledgeb}
B.~Wang, R.~Li, J.~Zhu, Z.~Zhao, and H.~Zhang, ``Knowledge {{Enhanced Semantic Communication Receiver}},'' \emph{IEEE Communications Letters}, vol.~27, no.~7, pp. 1794--1798, 2023.

\bibitem{qiu2020survey}
J.~Qiu, Z.~Tian, C.~Du, Q.~Zuo, S.~Su, and B.~Fang, ``A survey on access control in the age of internet of things,'' \emph{IEEE Internet of Things Journal}, vol.~7, no.~6, pp. 4682--4696, 2020.

\bibitem{zhang2023optimization}
W.~Zhang, Y.~Wang, M.~Chen, T.~Luo, and D.~Niyato, ``Optimization of image transmission in cooperative semantic communication networks,'' \emph{IEEE Transactions on Wireless Communications}, vol.~23, no.~2, pp. 861--873, 2024.

\bibitem{zhang2024intellicise}
P.~Zhang, W.~Xu, Y.~Liu, X.~Qin, K.~Niu, S.~Cui, G.~Shi, Z.~Qin, X.~Xu, F.~Wang, Y.~Meng, C.~Dong, J.~Dai, Q.~Yang, Y.~Sun, D.~Gao, H.~Gao, S.~Han, and X.~Song, ``Intellicise wireless networks from semantic communications: A survey, research issues, and challenges,'' \emph{IEEE Communications Surveys \& Tutorials}, 2024, doi:10.1109/COMST.2024.3443193.

\bibitem{feng2024semantic}
Y.~Feng, H.~Shen, Z.~Shan, Q.~Yang, and X.~Shi, ``Semantic communication for edge intelligence enabled autonomous driving system,'' \emph{IEEE Network}, 2024, doi:10.1109/MNET.2024.3468328.

\bibitem{akyildiz2022holographic}
I.~F. Akyildiz and H.~Guo, ``Holographic-type communication: A new challenge for the next decade,'' \emph{ITU Journal on Future and Evolving Technologies}, vol.~3, no.~2, pp. 421--442, 2022.

\bibitem{cheng2023enriching}
R.~Cheng, K.~Liu, N.~Wu, and B.~Han, ``Enriching telepresence with semantic-driven holographic communication,'' in \emph{Proceedings of the 22nd ACM Workshop on Hot Topics in Networks}, 2023, pp. 147--156.

\bibitem{wang2022survey}
Z.~Wang, D.~Liu, Y.~Sun, X.~Pang, P.~Sun, F.~Lin, J.~C.~S. Lui, and K.~Ren, ``A {{Survey}} on {{IoT-Enabled Home Automation Systems}}: {{Attacks}} and {{Defenses}},'' \emph{IEEE Communications Surveys \& Tutorials}, vol.~24, no.~4, pp. 2292--2328, 2022.

\bibitem{hsu2024socialaware}
B.-Y. Hsu, C.-Y. Shen, H.~S. Yuan, W.-C. Lee, and D.-N. Yang, ``Social-{{Aware Group Display Configuration}} in {{VR Conference}},'' vol.~38, no.~8, 2024, pp. 8517--8525.

\bibitem{chen2023trustworthya}
J.~Chen, J.~Wang, C.~Jiang, Y.~Ren, and L.~Hanzo, ``Trustworthy {{Semantic Communications}} for the {{Metaverse Relying}} on {{Federated Learning}},'' \emph{IEEE Wireless Communications}, vol.~30, no.~4, pp. 18--25, 2023.

\bibitem{popovski2022perspective}
P.~Popovski, F.~Chiariotti, K.~Huang, A.~E. Kalør, M.~Kountouris, N.~Pappas, and B.~Soret, ``A perspective on time toward wireless 6g,'' \emph{Proceedings of the IEEE}, vol. 110, no.~8, pp. 1116--1146, 2022.

\bibitem{kadam2024semantic}
S.~Kadam and D.~I. Kim, ``Semantic communication-empowered vehicle count prediction for traffic management,'' in \emph{Proceeding of IEEE Wireless Communications and Networking Conference (WCNC)}, 2024, doi:10.1109/WCNC57260.2024.10571211.

\bibitem{zhang2023intelligent}
H.~Zhang, H.~Wang, Y.~Li, K.~Long, and V.~C.~M. Leung, ``Toward {{Intelligent Resource Allocation}} on {{Task-Oriented Semantic Communication}},'' \emph{IEEE Wireless Communications}, vol.~30, no.~3, pp. 70--77, 2023.

\bibitem{mitev2023whata}
M.~Mitev, A.~Chorti, H.~V. Poor, and G.~P. Fettweis, ``What {{Physical Layer Security Can Do}} for {{6G Security}},'' \emph{IEEE Open Journal of Vehicular Technology}, vol.~4, pp. 375--388, 2023.

\bibitem{yao2024survey}
Y.~Yao, J.~Duan, K.~Xu, Y.~Cai, Z.~Sun, and Y.~Zhang, ``A survey on large language model (llm) security and privacy: The good, the bad, and the ugly,'' \emph{High-Confidence Computing}, p. 100211, 2024.

\bibitem{tian2022comprehensive}
Z.~Tian, L.~Cui, J.~Liang, and S.~Yu, ``A {{Comprehensive Survey}} on {{Poisoning Attacks}} and {{Countermeasures}} in {{Machine Learning}},'' \emph{ACM Computing Surveys}, vol.~55, no.~8, pp. 1--35, 2022.

\bibitem{zhang2018protectinga}
J.~Zhang, Z.~Gu, J.~Jang, H.~Wu, M.~P. Stoecklin, H.~Huang, and I.~Molloy, ``Protecting {{Intellectual Property}} of {{Deep Neural Networks}} with {{Watermarking}},'' in \emph{Proceedings of {{Asia Conference}} on {{Computer}} and {{Communications Security}} ({{AsiaCCS}})}, 2018, pp. 159--172.

\bibitem{nie2024deep}
H.~Nie, S.~Lu, J.~Wu, and J.~Zhu, ``Deep model intellectual property protection with compression-resistant model watermarking,'' \emph{IEEE Transactions on Artificial Intelligence}, 2024, doi: 10.1109/TAI.2024.3351116.

\bibitem{xi2023security}
Z.~Xi, T.~Du, C.~Li, R.~Pang, S.~Ji, X.~Luo, X.~Xiao, F.~Ma, and T.~Wang, ``On the security risks of knowledge graph reasoning,'' in \emph{Proceeding of 32nd USENIX Security Symposium (USENIX Security)}, 2023, pp. 3259--3276.

\bibitem{zhang2023drldrivena}
H.~Zhang, H.~Wang, Y.~Li, K.~Long, and A.~Nallanathan, ``{{DRL-Driven Dynamic Resource Allocation}} for {{Task-Oriented Semantic Communication}},'' \emph{IEEE Transactions on Communications}, vol.~71, no.~7, pp. 3992--4004, 2023.

\bibitem{wang2021feature}
Z.~Wang, H.~Guo, Z.~Zhang, W.~Liu, Z.~Qin, and K.~Ren, ``Feature {{Importance-aware Transferable Adversarial Attacks}},'' in \emph{Proceedings of {{IEEE}}/{{CVF International Conference}} on {{Computer Vision}} ({{ICCV}})}, 2021, pp. 7619--7628.

\bibitem{wang2022threats}
Z.~Wang, J.~Ma, X.~Wang, J.~Hu, Z.~Qin, and K.~Ren, ``Threats to training: A survey of poisoning attacks and defenses on machine learning systems,'' \emph{ACM Computing Surveys}, vol.~55, no.~7, pp. 1--36, 2022.

\bibitem{borgnia2021strong}
E.~Borgnia, V.~Cherepanova, L.~Fowl, A.~Ghiasi, J.~Geiping, M.~Goldblum, T.~Goldstein, and A.~Gupta, ``Strong {{Data Augmentation Sanitizes Poisoning}} and {{Backdoor Attacks Without}} an {{Accuracy Tradeoff}},'' in \emph{Proceeding of IEEE International Conference on Acoustics, Speech and Signal Processing (ICASSP)}.\hskip 1em plus 0.5em minus 0.4em\relax IEEE, 2021, pp. 3855--3859.

\bibitem{li2017learning}
Y.~Li, J.~Yang, Y.~Song, L.~Cao, J.~Luo, and L.-J. Li, ``Learning from {{Noisy Labels}} with {{Distillation}},'' in \emph{Proceedings of {{IEEE International Conference}} on {{Computer Vision}} ({{ICCV}})}, 2017, pp. 1928--1936.

\bibitem{wang2019neural}
B.~Wang, Y.~Yao, S.~Shan, H.~Li, B.~Viswanath, H.~Zheng, and B.~Y. Zhao, ``Neural {{Cleanse}}: {{Identifying}} and {{Mitigating Backdoor Attacks}} in {{Neural Networks}},'' in \emph{Proceedings of {{IEEE Symposium}} on {{Security}} and {{Privacy}} ({{SP}})}, 2019, pp. 707--723.

\bibitem{blanchard2017machine}
P.~Blanchard, E.~M. El~Mhamdi, R.~Guerraoui, and J.~Stainer, ``Machine {{Learning}} with {{Adversaries}}: {{Byzantine Tolerant Gradient Descent}},'' in \emph{Proceedings of {{Advances}} in {{Neural Information Processing Systems}} ({{NeurIPS}})}, vol.~30, 2017.

\bibitem{zhao2021shielding}
L.~Zhao, S.~Hu, Q.~Wang, J.~Jiang, C.~Shen, X.~Luo, and P.~Hu, ``Shielding {{Collaborative Learning}}: {{Mitigating Poisoning Attacks Through Client-Side Detection}},'' \emph{IEEE Transactions on Dependable and Secure Computing}, vol.~18, no.~5, pp. 2029--2041, 2021.

\bibitem{zhang2019adversariala}
J.~Zhang and C.~Li, ``Adversarial {{Examples}}: {{Opportunities}} and {{Challenges}},'' \emph{IEEE Transactions on Neural Networks and Learning Systems}, vol.~31, no.~7, pp. 2578--2593, 2019.

\bibitem{hu2023robust}
Q.~Hu, G.~Zhang, Z.~Qin, Y.~Cai, G.~Yu, and G.~Y. Li, ``Robust {{Semantic Communications}} with {{Masked VQ-VAE Enabled Codebook}},'' \emph{IEEE Transactions on Wireless Communications}, pp. 8707--8722, 2023.

\bibitem{liu2017physical}
Y.~Liu, H.-H. Chen, and L.~Wang, ``Physical {{Layer Security}} for {{Next Generation Wireless Networks}}: {{Theories}}, {{Technologies}}, and {{Challenges}},'' \emph{IEEE Communications Surveys \& Tutorials}, vol.~19, no.~1, pp. 347--376, 2017.

\bibitem{zhao2024generative}
C.~Zhao, H.~Du, D.~Niyato, J.~Kang, Z.~Xiong, D.~I. Kim, X.~Shen, and K.~B. Letaief, ``Generative ai for secure physical layer communications: A survey,'' \emph{IEEE Transactions on Cognitive Communications and Networking}, 2024, doi:10.1109/TCCN.2024.3438379.

\bibitem{yang2023secure}
Z.~Yang, M.~Chen, G.~Li, Y.~Yang, and Z.~Zhang, ``Secure semantic communications: Fundamentals and challenges,'' \emph{IEEE Network}, 2024, doi:10.1109/MNET.2024.3411027.

\bibitem{tang2023ganinspired}
R.~Tang, D.~Gao, M.~Yang, T.~Guo, H.~Wu, and G.~Shi, ``Gan-inspired intelligent jamming and anti-jamming strategy for semantic communication systems,'' in \emph{Proceeding of IEEE International Conference on Communications Workshops (ICC WKSHPS)}, 2023, pp. 1623--1628.

\bibitem{li2023ubiquitousa}
K.~Li, B.~P.~L. Lau, X.~Yuan, W.~Ni, M.~Guizani, and C.~Yuen, ``Toward {{Ubiquitous Semantic Metaverse}}: {{Challenges}}, {{Approaches}}, and {{Opportunities}},'' \emph{IEEE Internet of Things Journal}, vol.~10, no.~24, pp. 21\,855--21\,872, 2023.

\bibitem{chen2023coverta}
X.~Chen, J.~An, Z.~Xiong, C.~Xing, N.~Zhao, F.~R. Yu, and A.~Nallanathan, ``Covert {{Communications}}: {{A Comprehensive Survey}},'' \emph{IEEE Communications Surveys \& Tutorials}, vol.~25, no.~2, pp. 1173--1198, 2023.

\bibitem{hu2024covert}
J.~Hu, L.~Ye, Y.~Chen, X.~Zhang, J.~Wang, and Z.~Chen, ``Covert communications for text semantic with finite blocklength,'' \emph{IEEE Wireless Communications Letters}, 2024, doi:10.1109/LWC.2024.3448614.

\bibitem{chen2023model}
Y.~Chen, Q.~Yang, Z.~Shi, and J.~Chen, ``The model inversion eavesdropping attack in semantic communication systems,'' in \emph{Proceedings of IEEE Global Communications Conference (GLOBECOM)}.\hskip 1em plus 0.5em minus 0.4em\relax IEEE, 2023, pp. 5171--5177.

\bibitem{xie2024towards}
H.~Xie, Z.~Qin, X.~Tao, and Z.~Han, ``Towards intelligent communications: Large model empowered semantic communications,'' \emph{arXiv preprint arXiv:2402.13073}, 2024.

\bibitem{zhu2019deep}
L.~Zhu, Z.~Liu, and S.~Han, ``Deep leakage from gradients,'' in \emph{Proceeding of Advances in neural information processing systems (NeurIPS)}, 2019, pp. 14\,747--14\,756.

\bibitem{gaber2024malware}
M.~G. Gaber, M.~Ahmed, and H.~Janicke, ``Malware detection with artificial intelligence: A systematic literature review,'' \emph{ACM Computing Surveys}, vol.~56, no.~6, pp. 1--33, 2024.

\bibitem{zhou2022adaptive}
Q.~Zhou, R.~Li, Z.~Zhao, Y.~Xiao, and H.~Zhang, ``Adaptive {{Bit Rate Control}} in {{Semantic Communication With Incremental Knowledge-Based HARQ}},'' \emph{IEEE Open Journal of the Communications Society}, vol.~3, pp. 1076--1089, 2022.

\bibitem{zhang2022deepa}
J.~Zhang, D.~Chen, J.~Liao, W.~Zhang, H.~Feng, G.~Hua, and N.~Yu, ``Deep {{Model Intellectual Property Protection}} via {{Deep Watermarking}},'' \emph{IEEE Transactions on Pattern Analysis and Machine Intelligence}, vol.~44, no.~8, pp. 4005--4020, 2022.

\bibitem{wang2023multiagent}
Y.~Wang, Y.~Hu, H.~Du, T.~Luo, and D.~Niyato, ``Multi-{{Agent Reinforcement Learning}} for {{Covert Semantic Communications}} over {{Wireless Networks}},'' in \emph{Proceedings of {{IEEE International Conference}} on {{Acoustics}}, {{Speech}} and {{Signal Processing}} ({{ICASSP}})}, 2023, pp. 1--5.

\bibitem{xie2022security}
N.~Xie, J.~Zhang, and Q.~Zhang, ``Security {Provided} by the {Physical Layer} in {Wireless Communications},'' \emph{IEEE Network}, vol.~37, no.~5, pp. 42--48, 2023.

\bibitem{yamaguchi2020physicallayer}
R.~Yamaguchi, H.~Ochiai, and J.~Shikata, ``A {{Physical-Layer Security Based}} on {{Wireless Steganography Through OFDM}} and {{DFT-Precoded OFDM Signals}},'' in \emph{Proceedings of {{IEEE}} {{Vehicular Technology Conference}} ({{VTC2020-Spring}})}, 2020, pp. 1--5.

\bibitem{qin2023securinga}
Q.~Qin, Y.~Rong, G.~Nan, S.~Wu, X.~Zhang, Q.~Cui, and X.~Tao, ``Securing {{Semantic Communications}} with {{Physical-Layer Semantic Encryption}} and {{Obfuscation}},'' in \emph{Proceedings of {{IEEE International Conference}} on {{Communications}} ({{ICC}})}, 2023, pp. 5608--5613.

\bibitem{knott2021crypten}
B.~Knott, S.~Venkataraman, A.~Hannun, S.~Sengupta, M.~Ibrahim, and L.~{van der Maaten}, ``{{CrypTen}}: {{Secure Multi-Party Computation Meets Machine Learning}},'' in \emph{Proceedings of {{Advances}} in {{Neural Information Processing Systems}} ({{NeurIPS}})}, vol.~34, 2021, pp. 4961--4973.

\bibitem{luo2023encrypted}
X.~Luo, Z.~Chen, M.~Tao, and F.~Yang, ``Encrypted {{Semantic Communication Using Adversarial Training}} for {{Privacy Preserving}},'' \emph{IEEE Communications Letters}, vol.~27, no.~6, pp. 1486--1490, 2023.

\bibitem{hu2023interference}
L.~Hu, S.~Tan, H.~Wen, J.~Wu, J.~Fan, S.~Chen, and J.~Tang, ``Interference {{Alignment}} for {{Physical Layer Security}} in {{Multi-User Networks With Passive Eavesdroppers}},'' \emph{IEEE Transactions on Information Forensics and Security}, vol.~18, pp. 3692--3705, 2023.

\bibitem{lin2018physicallayer}
J.~Lin, Q.~Li, J.~Yang, H.~Shao, and W.-Q. Wang, ``Physical-{{Layer Security}} for {{Proximal Legitimate User}} and {{Eavesdropper}}: {{A Frequency Diverse Array Beamforming Approach}},'' \emph{IEEE Transactions on Information Forensics and Security}, vol.~13, no.~3, pp. 671--684, 2018.

\bibitem{li2021exploiting}
Z.~Li, J.~Li, Y.~Liu, X.~Liang, K.~G. Shin, Z.~Yan, and H.~Li, ``Exploiting {{Interactions}} of {{Multiple Interference}} for {{Cooperative Interference Alignment}},'' \emph{IEEE Transactions on Wireless Communications}, vol.~20, no.~11, pp. 7072--7085, 2021.

\bibitem{wang2023starrisassisted}
Y.~Wang, W.~Yang, P.~Guan, Y.~Zhao, and Z.~Xiong, ``Star-ris-assisted privacy protection in semantic communication system,'' \emph{IEEE Transactions on Vehicular Technology}, vol.~73, no.~9, pp. 13\,915--13\,920, 2024.

\bibitem{kaewpuang2023cooperative}
R.~Kaewpuang, M.~Xu, W.~Y.~B. Lim, D.~Niyato, H.~Yu, J.~Kang, and X.~S. Shen, ``Cooperative {{Resource Management}} in {{Quantum Key Distribution}} ({{QKD}}) {{Networks}} for {{Semantic Communication}},'' \emph{IEEE Internet of Things Journal}, vol.~11, no.~3, pp. 4454--4469, 2024.

\bibitem{khalid2023quantum}
U.~Khalid, M.~S. Ulum, A.~Farooq, T.~Q. Duong, O.~A. Dobre, and H.~Shin, ``Quantum {{Semantic Communications}} for {{Metaverse}}: {{Principles}} and {{Challenges}},'' \emph{IEEE Wireless Communications}, vol.~30, no.~4, pp. 26--36, 2023.

\bibitem{wong2020fast}
E.~Wong, L.~Rice, and J.~Z. Kolter, ``Fast is better than free: Revisiting adversarial training,'' \emph{arXiv preprint arXiv:2001.03994}, 2020.

\bibitem{chen2020stateful}
S.~Chen, N.~Carlini, and D.~Wagner, ``Stateful detection of black-box adversarial attacks,'' in \emph{Proceedings of the 1st ACM Workshop on Security and Privacy on Artificial Intelligence}, 2020, pp. 30--39.

\bibitem{erdemir2023generative}
E.~Erdemir, T.-Y. Tung, P.~L. Dragotti, and D.~G{\"u}nd{\"u}z, ``Generative {{Joint Source-Channel Coding}} for {{Semantic Image Transmission}},'' \emph{IEEE Journal on Selected Areas in Communications}, pp. 1--1, 2023.

\bibitem{chen2023commin}
J.~Chen, D.~You, D.~Gündüz, and P.~L. Dragotti, ``Commin: Semantic image communications as an inverse problem with inn-guided diffusion models,'' in \emph{Proceeding of IEEE International Conference on Acoustics, Speech and Signal Processing (ICASSP)}, 2024, pp. 6675--6679.

\bibitem{jiang2022reliablea}
S.~Jiang, Y.~Liu, Y.~Zhang, P.~Luo, K.~Cao, J.~Xiong, H.~Zhao, and J.~Wei, ``Reliable {{Semantic Communication System Enabled}} by {{Knowledge Graph}},'' \emph{Entropy}, vol.~24, no.~6, p. 846, 2022.

\bibitem{yao2021typing}
P.~Yao and D.~Barbosa, ``Typing {{Errors}} in {{Factual Knowledge Graphs}}: {{Severity}} and {{Possible Ways Out}},'' in \emph{Proceedings of the {{Web Conference}} ({{WWW}})}, 2021, pp. 3305--3313.

\bibitem{thomas2023neurosymbolic}
C.~K. Thomas and W.~Saad, ``Neuro-{{Symbolic Causal Reasoning Meets Signaling Game}} for {{Emergent Semantic Communications}},'' \emph{IEEE Transactions on Wireless Communications}, 2023, doi: 10.1109/TWC.2023.3319981.

\bibitem{gao2023esanet}
N.~Gao, Q.~Huang, C.~Li, S.~Jin, and M.~Matthaiou, ``Esanet: Environment semantics enabled physical layer authentication,'' \emph{IEEE Wireless Communications Letters}, vol.~13, no.~1, pp. 178--182, 2024.

\bibitem{jorswieck2015broadcasting}
E.~Jorswieck, S.~Tomasin, and A.~Sezgin, ``Broadcasting {{Into}} the {{Uncertainty}}: {{Authentication}} and {{Confidentiality}} by {{Physical-Layer Processing}},'' \emph{Proceedings of the IEEE}, vol. 103, no.~10, pp. 1702--1724, 2015.

\bibitem{tan2024optimization}
H.~Tan, N.~Xie, and A.~X. Liu, ``An {{Optimization Framework}} for {{Active Physical-Layer Authentication}},'' \emph{IEEE Transactions on Mobile Computing}, vol.~23, no.~1, pp. 164--179, 2024.

\bibitem{liang2017provchain}
X.~Liang, S.~Shetty, D.~Tosh, C.~Kamhoua, K.~Kwiat, and L.~Njilla, ``{{ProvChain}}: {{A Blockchain-Based Data Provenance Architecture}} in {{Cloud Environment}} with {{Enhanced Privacy}} and {{Availability}},'' in \emph{Proceedings of 17th {{IEEE}}/{{ACM International Symposium}} on {{Cluster}}, {{Cloud}} and {{Grid Computing}} ({{CCGRID}})}, 2017, pp. 468--477.

\bibitem{wang2021spdsa}
Y.~Wang, Z.~Su, N.~Zhang, J.~Chen, X.~Sun, Z.~Ye, and Z.~Zhou, ``{{SPDS}}: {{A Secure}} and {{Auditable Private Data Sharing Scheme}} for {{Smart Grid Based}} on {{Blockchain}},'' \emph{IEEE Transactions on Industrial Informatics}, vol.~17, no.~11, pp. 7688--7699, 2021.

\bibitem{jayasinghe2019machine}
U.~Jayasinghe, G.~M. Lee, T.-W. Um, and Q.~Shi, ``Machine {{Learning Based Trust Computational Model}} for {{IoT Services}},'' \emph{IEEE Transactions on Sustainable Computing}, vol.~4, no.~1, pp. 39--52, 2019.

\bibitem{gyawali2021deep}
S.~Gyawali, Y.~Qian, and R.~Q. Hu, ``Deep {{Reinforcement Learning Based Dynamic Reputation Policy}} in {{5G Based Vehicular Communication Networks}},'' \emph{IEEE Transactions on Vehicular Technology}, vol.~70, no.~6, pp. 6136--6146, 2021.

\bibitem{wang2021survey}
J.~Wang, X.~Jing, Z.~Yan, Y.~Fu, W.~Pedrycz, and L.~T. Yang, ``A {{Survey}} on {{Trust Evaluation Based}} on {{Machine Learning}},'' \emph{ACM Computing Surveys}, vol.~53, no.~5, pp. 1--36, 2021.

\bibitem{truong2019trust}
N.~B. Truong, G.~M. Lee, T.-W. Um, and M.~Mackay, ``Trust {{Evaluation Mechanism}} for {{User Recruitment}} in {{Mobile Crowd-Sensing}} in the {{Internet}} of {{Things}},'' \emph{IEEE Transactions on Information Forensics and Security}, vol.~14, no.~10, pp. 2705--2719, 2019.

\bibitem{sultan2023rolebased}
N.~H. Sultan, V.~Varadharajan, L.~Zhou, and F.~A. Barbhuiya, ``A {{Role-Based Encryption}} ({{RBE}}) {{Scheme}} for {{Securing Outsourced Cloud Data}} in a {{Multi-Organization Context}},'' \emph{IEEE Transactions on Services Computing}, vol.~16, no.~3, pp. 1647--1661, 2023.

\bibitem{xue2019attributebased}
Y.~Xue, K.~Xue, N.~Gai, J.~Hong, D.~S.~L. Wei, and P.~Hong, ``An {{Attribute-Based Controlled Collaborative Access Control Scheme}} for {{Public Cloud Storage}},'' \emph{IEEE Transactions on Information Forensics and Security}, vol.~14, no.~11, pp. 2927--2942, 2019.

\bibitem{xu2021integrated}
R.~Xu, J.~Joshi, and P.~Krishnamurthy, ``An {{Integrated Privacy Preserving Attribute-Based Access Control Framework Supporting Secure Deduplication}},'' \emph{IEEE Transactions on Dependable and Secure Computing}, vol.~18, no.~2, pp. 706--721, 2021.

\bibitem{xue2023sparkac}
T.~Xue, Y.~Wen, B.~Luo, G.~Li, Y.~Li, B.~Zhang, Y.~Zheng, Y.~Hu, and D.~Meng, ``{{SparkAC}}: {{Fine-Grained Access Control}} in {{Spark}} for {{Secure Data Sharing}} and {{Analytics}},'' \emph{IEEE Transactions on Dependable and Secure Computing}, vol.~20, no.~2, pp. 1104--1123, 2023.

\bibitem{bao2023pbidm}
Z.~Bao, D.~He, M.~K. Khan, M.~Luo, and Q.~Xie, ``{{PBidm}}: {{Privacy-Preserving Blockchain-Based Identity Management System}} for {{Industrial Internet}} of {{Things}},'' \emph{IEEE Transactions on Industrial Informatics}, vol.~19, no.~2, pp. 1524--1534, 2023.

\bibitem{xu2020identity}
J.~Xu, K.~Xue, H.~Tian, J.~Hong, D.~S.~L. Wei, and P.~Hong, ``An {{Identity Management}} and {{Authentication Scheme Based}} on {{Redactable Blockchain}} for {{Mobile Networks}},'' \emph{IEEE Transactions on Vehicular Technology}, vol.~69, no.~6, pp. 6688--6698, 2020.

\bibitem{shen2020blockchainassisted}
M.~Shen, H.~Liu, L.~Zhu, K.~Xu, H.~Yu, X.~Du, and M.~Guizani, ``Blockchain-{{Assisted Secure Device Authentication}} for {{Cross-Domain Industrial IoT}},'' \emph{IEEE Journal on Selected Areas in Communications}, vol.~38, no.~5, pp. 942--954, 2020.

\bibitem{chen2022xautha}
J.~Chen, Z.~Zhan, K.~He, R.~Du, D.~Wang, and F.~Liu, ``{{XAuth}}: {{Efficient Privacy-Preserving Cross-Domain Authentication}},'' \emph{IEEE Transactions on Dependable and Secure Computing}, vol.~19, no.~5, pp. 3301--3311, 2022.

\bibitem{wang2022blockchainbaseda}
Y.~Wang, Z.~Su, J.~Li, N.~Zhang, K.~Zhang, K.-K.~R. Choo, and Y.~Liu, ``Blockchain-{{Based Secure}} and {{Cooperative Private Charging Pile Sharing Services}} for {{Vehicular Networks}},'' \emph{IEEE Transactions on Vehicular Technology}, vol.~71, no.~2, pp. 1857--1874, 2022.

\bibitem{fotiou2016decentralized}
N.~Fotiou and G.~C. Polyzos, ``Decentralized name-based security for content distribution using blockchains,'' in \emph{Proceedings of {{IEEE Conference}} on {{Computer Communications Workshops}} ({{INFOCOM WKSHPS}})}, 2016, pp. 415--420.

\bibitem{wang2021blockchainbased}
H.~Wang, Q.~Wang, and D.~He, ``Blockchain-{{Based Private Provable Data Possession}},'' \emph{IEEE Transactions on Dependable and Secure Computing}, vol.~18, no.~5, pp. 2379--2389, 2021.

\bibitem{ateniese2017redactable}
G.~Ateniese, B.~Magri, D.~Venturi, and E.~Andrade, ``Redactable {{Blockchain}} {\textendash} or {\textendash} {{Rewriting History}} in {{Bitcoin}} and {{Friends}},'' in \emph{Proceedings of {{IEEE European Symposium}} on {{Security}} and {{Privacy}} ({{EuroS}}\&{{P}})}, 2017, pp. 111--126.

\bibitem{bahramali2021robust}
A.~Bahramali, M.~Nasr, A.~Houmansadr, D.~Goeckel, and D.~Towsley, ``Robust {{Adversarial Attacks Against DNN-Based Wireless Communication Systems}},'' in \emph{Proceedings of the {{ACM SIGSAC Conference}} on {{Computer}} and {{Communications Security}} ({{CCS}})}, 2021, pp. 126--140.

\bibitem{xia2024mmnet}
R.~Xia, D.~Liu, J.~Li, L.~Yuan, N.~Wang, and X.~Gao, ``Mmnet: Multi-collaboration and multi-supervision network for sequential deepfake detection,'' \emph{IEEE Transactions on Information Forensics and Security}, vol.~19, pp. 3409--3422, 2024.

\bibitem{oz2022survey}
H.~Oz, A.~Aris, A.~Levi, and A.~S. Uluagac, ``A {{Survey}} on {{Ransomware}}: {{Evolution}}, {{Taxonomy}}, and {{Defense Solutions}},'' \emph{ACM Computing Surveys}, vol.~54, no. 11s, pp. 1--37, 2022.

\bibitem{mcintosh2021ransomware}
T.~McIntosh, A.~Kayes, Y.-P.~P. Chen, A.~Ng, and P.~Watters, ``Ransomware mitigation in the modern era: A comprehensive review, research challenges, and future directions,'' \emph{ACM Computing Surveys}, vol.~54, no.~9, pp. 1--36, 2021.

\bibitem{wang2023infrastructure}
F.~Wang, Y.~Hong, and X.~Ban, ``Infrastructure-enabled gps spoofing detection and correction,'' \emph{IEEE Transactions on Intelligent Transportation Systems}, vol.~24, no.~9, pp. 9462--9475, 2023.

\bibitem{du2023semantica}
H.~Du, J.~Wang, D.~Niyato, J.~Kang, Z.~Xiong, J.~Zhang, and X.~Shen, ``Semantic {{Communications}} for {{Wireless Sensing}}: {{RIS-Aided Encoding}} and {{Self-Supervised Decoding}},'' \emph{IEEE Journal on Selected Areas in Communications}, vol.~41, no.~8, pp. 2547--2562, 2023.

\bibitem{sun2024disentangled}
L.~Sun, Y.~Yang, M.~Chen, and C.~Guo, ``Disentangled information bottleneck guided privacy-protective joint source and channel coding for image transmission,'' \emph{IEEE Transactions on Communications}, 2024, doi:10.1109/TCOMM.2024.3406381.

\bibitem{chen2024lightweight}
G.~Chen, G.~Nan, Z.~Jiang, H.~Du, R.~Shi, Q.~Cui, and X.~Tao, ``Lightweight and robust wireless semantic communications,'' \emph{IEEE Communications Letters}, 2024.

\bibitem{eirina2019deep}
E.~Bourtsoulatze, D.~B. Kurka, and D.~Gündüz, ``Deep joint source-channel coding for wireless image transmission,'' in \emph{Proceeding of IEEE International Conference on Acoustics, Speech and Signal Processing (ICASSP)}, 2019, pp. 4774--4778.

\bibitem{thomas2022adversarial}
T.~Marchioro, N.~Laurenti, and D.~Gündüz, ``Adversarial networks for secure wireless communications,'' in \emph{Proceeding of IEEE International Conference on Acoustics, Speech and Signal Processing (ICASSP)}, 2020, pp. 8748--8752.

\bibitem{li2022robust}
Z.~Li, W.~Chen, Q.~Wu, H.~Cao, K.~Wang, and J.~Li, ``Robust beamforming design and time allocation for irs-assisted wireless powered communication networks,'' \emph{IEEE Transactions on Communications}, vol.~70, no.~4, pp. 2838--2852, 2022.

\bibitem{wyner1975wiretap}
A.~D. Wyner, ``The wire-tap channel,'' \emph{The Bell System Technical Journal}, vol.~54, no.~8, pp. 1355--1387, 1975.

\bibitem{goel2008guaranteeing}
S.~Goel and R.~Negi, ``Guaranteeing {{Secrecy}} using {{Artificial Noise}},'' \emph{IEEE Transactions on Wireless Communications}, vol.~7, no.~6, pp. 2180--2189, 2008.

\bibitem{xu2024covert}
R.~Xu, G.~Li, Z.~Yang, M.~Chen, Y.~Liu, and J.~Li, ``Covert and reliable semantic communication against cross-layer privacy inference over wireless edge networks,'' in \emph{Proceeding of IEEE Wireless Communications and Networking Conference (WCNC)}, 2024, doi:10.1109/WCNC57260.2024.10570800.

\bibitem{chehimi2024quantum}
M.~Chehimi, C.~Chaccour, C.~K. Thomas, and W.~Saad, ``Quantum semantic communications for resource-efficient quantum networking,'' \emph{IEEE Communications Letters}, vol.~28, no.~4, pp. 803--807, 2024.

\bibitem{li2023secureb}
X.~Li, Y.~Pei, X.~Yue, Y.~Liu, and Z.~Ding, ``Secure communication of active ris assisted noma networks,'' \emph{IEEE Transactions on Wireless Communications}, vol.~23, no.~5, pp. 4489--4503, 2024.

\bibitem{du2023generative}
H.~Du, G.~Liu, D.~Niyato, J.~Zhang, J.~Kang, Z.~Xiong, B.~Ai, and D.~I. Kim, ``Generative al-aided joint training-free secure semantic communications via multi-modal prompts,'' pp. 12\,896--12\,900, 2024.

\bibitem{zeng2024USVFleet}
H.~Zeng, Z.~Su, Q.~Xu, and D.~Fang, ``Usv fleet-assisted collaborative data backup in marine internet of things,'' \emph{IEEE Internet of Things Journal}, vol.~11, no.~22, pp. 36\,308--36\,321, 2024.

\bibitem{han2023generative}
T.~Han, J.~Tang, Q.~Yang, Y.~Duan, Z.~Zhang, and Z.~Shi, ``Generative {{Model}} based {{Highly Efficient Semantic Communication Approach}} for {{Image Transmission}},'' in \emph{Proceedings of {{IEEE International Conference}} on {{Acoustics}}, {{Speech}} and {{Signal Processing}} ({{ICASSP}})}, 2023, pp. 1--5.

\bibitem{tonmoy2024comprehensive}
S.~Tonmoy, S.~Zaman, V.~Jain, A.~Rani, V.~Rawte, A.~Chadha, and A.~Das, ``A comprehensive survey of hallucination mitigation techniques in large language models,'' \emph{arXiv preprint arXiv:2401.01313}, 2024.

\bibitem{wang2022dataand}
W.~Wang, Y.~Yang, and F.~Wu, ``Towards data-and knowledge-driven ai: A survey on neuro-symbolic computing,'' \emph{IEEE Transactions on Pattern Analysis and Machine Intelligence}, 2024, doi:10.1109/TPAMI.2024.3483273.

\bibitem{wang2023surveya}
Y.~Wang, Y.~Pan, M.~Yan, Z.~Su, and T.~H. Luan, ``A survey on {ChatGPT}: {AI}-generated contents, challenges, and solutions,'' \emph{IEEE Open Journal of the Computer Society}, vol.~4, pp. 280--302, 2023.

\bibitem{xie2021survey}
N.~Xie, Z.~Li, and H.~Tan, ``A {{Survey}} of {{Physical-Layer Authentication}} in {{Wireless Communications}},'' \emph{IEEE Communications Surveys \& Tutorials}, vol.~23, no.~1, pp. 282--310, 2021.

\bibitem{perazzone2021artificial}
J.~B. Perazzone, P.~L. Yu, B.~M. Sadler, and R.~S. Blum, ``Artificial {{Noise-Aided MIMO Physical Layer Authentication With Imperfect CSI}},'' \emph{IEEE Transactions on Information Forensics and Security}, vol.~16, pp. 2173--2185, 2021.

\bibitem{xie2019survey}
J.~Xie, H.~Tang, T.~Huang, F.~R. Yu, R.~Xie, J.~Liu, and Y.~Liu, ``A {{Survey}} of {{Blockchain Technology Applied}} to {{Smart Cities}}: {{Research Issues}} and {{Challenges}},'' \emph{IEEE Communications Surveys \& Tutorials}, vol.~21, no.~3, pp. 2794--2830, 2019.

\bibitem{parhizkar2020combining}
E.~Parhizkar, M.~H. Nikravan, R.~C. Holte, and S.~Zilles, ``Combining direct trust and indirect trust in multi-agent systems.'' in \emph{Proceeding of International Joint Conference on Artificial Intelligence (IJCAI)}, 2020, pp. 311--317.

\bibitem{wang2022surveyb}
J.~Wang, Z.~Yan, H.~Wang, T.~Li, and W.~Pedrycz, ``A {{Survey}} on {{Trust Models}} in {{Heterogeneous Networks}},'' \emph{IEEE Communications Surveys \& Tutorials}, vol.~24, no.~4, pp. 2127--2162, 2022.

\bibitem{sekhari2021remembera}
A.~Sekhari, J.~Acharya, G.~Kamath, and A.~T. Suresh, ``Remember {{What You Want}} to {{Forget}}: {{Algorithms}} for {{Machine Unlearning}},'' in \emph{Proceedings of {{Advances}} in {{Neural Information Processing Systems}} ({{NeurIPS}})}, vol.~34, 2021, pp. 18\,075--18\,086.

\bibitem{wei2023federateda}
H.~Wei, W.~Ni, W.~Xu, F.~Wang, D.~Niyato, and P.~Zhang, ``Federated {{Semantic Learning Driven}} by {{Information Bottleneck}} for {{Task-Oriented Communications}},'' \emph{IEEE Communications Letters}, vol.~27, no.~10, pp. 2652--2656, 2023.

\bibitem{xu2024machine}
H.~Xu, T.~Zhu, L.~Zhang, W.~Zhou, and P.~S. Yu, ``Machine {{Unlearning}}: {{A Survey}},'' \emph{ACM Computing Surveys}, vol.~56, no.~1, pp. 1--36, 2024.

\bibitem{golatkar2020eternal}
A.~Golatkar, A.~Achille, and S.~Soatto, ``Eternal {{Sunshine}} of the {{Spotless Net}}: {{Selective Forgetting}} in {{Deep Networks}},'' in \emph{Proceedings of the {{IEEE}}/{{CVF Conference}} on {{Computer Vision}} and {{Pattern Recognition}} ({{CVPR}})}, 2020, pp. 9301--9309.

\bibitem{wang2024machine}
W.~Wang, Z.~Tian, and S.~Yu, ``Machine unlearning: A comprehensive survey,'' \emph{arXiv preprint arXiv:2405.07406}, 2024.

\bibitem{wu2022federated}
C.~Wu, S.~Zhu, and P.~Mitra, ``Federated unlearning with knowledge distillation,'' \emph{arXiv preprint arXiv:2201.09441}, 2022.

\bibitem{liu2022right}
Y.~Liu, L.~Xu, X.~Yuan, C.~Wang, and B.~Li, ``The right to be forgotten in federated learning: An efficient realization with rapid retraining,'' in \emph{Proceeding of IEEE Conference on Computer Communications (INFOCOM)}.\hskip 1em plus 0.5em minus 0.4em\relax IEEE, 2022, pp. 1749--1758.

\bibitem{lu2024efficient}
X.~Lu, K.~Zhu, J.~Li, and Y.~Zhang, ``Efficient knowledge base synchronization in semantic communication network: A federated distillation approach,'' in \emph{Proceeding of IEEE Wireless Communications and Networking Conference (WCNC)}, 2024, pp. 1--6.

\bibitem{min2021reinforcement}
M.~Min, W.~Wang, L.~Xiao, Y.~Xiao, and Z.~Han, ``Reinforcement learning-based sensitive semantic location privacy protection for {{VANETs}},'' \emph{China Communications}, vol.~18, no.~6, pp. 244--260, 2021.

\bibitem{liu2024adaptive}
H.~Liu, B.~Wang, R.~Meng, S.~Han, and X.~Xu, ``Adaptive privacy budget-based differential privacy co-training for wireless semantic communication,'' in \emph{Proceeding of IEEE Wireless Communications and Networking Conference (WCNC)}, 2024, pp. 1--6.

\bibitem{jauernig2020trusted}
P.~Jauernig, A.-R. Sadeghi, and E.~Stapf, ``Trusted {{Execution Environments}}: {{Properties}}, {{Applications}}, and {{Challenges}},'' \emph{IEEE Security \& Privacy}, vol.~18, no.~2, pp. 56--60, 2020.

\bibitem{xiao2020selflearninga}
Y.~Xiao, G.~Shi, Y.~Li, W.~Saad, and H.~V. Poor, ``Toward {{Self-Learning Edge Intelligence}} in {{6G}},'' \emph{IEEE Communications Magazine}, vol.~58, no.~12, pp. 34--40, 2020.

\bibitem{davies2021advancing}
M.~Davies, A.~Wild, G.~Orchard, Y.~Sandamirskaya, G.~A.~F. Guerra, P.~Joshi, P.~Plank, and S.~R. Risbud, ``Advancing {{Neuromorphic Computing With Loihi}}: {{A Survey}} of {{Results}} and {{Outlook}},'' \emph{Proceedings of the IEEE}, vol. 109, no.~5, pp. 911--934, 2021.

\bibitem{luo2023recent}
W.~Luo, L.~Cao, Y.~Shi, L.~Wan, H.~Zhang, S.~Li, G.~Chen, Y.~Li, S.~Li, Y.~Wang, S.~Sun, M.~F. Karim, H.~Cai, L.~C. Kwek, and A.~Q. Liu, ``Recent progress in quantum photonic chips for quantum communication and internet,'' \emph{Light: Science \& Applications}, vol.~12, no.~1, p. 175, 2023.

\bibitem{yang2023taskdrivena}
W.~Yang, X.~Chi, L.~Zhao, Z.~Xiong, and W.~Jiang, ``Task-driven {{Semantic-aware Green Cooperative Transmission Strategy}} for {{Vehicular Networks}},'' \emph{IEEE Transactions on Communications}, pp. 5783--5798, 2023.

\bibitem{wu2022development}
J.~Wu, ``Development paradigms of cyberspace endogenous safety and security,'' \emph{Science China Information Sciences}, vol.~65, no.~5, p. 156301, 2022.

\end{thebibliography}

\begin{IEEEbiography}[{\includegraphics[width=1in,height=1.25in,clip,keepaspectratio]{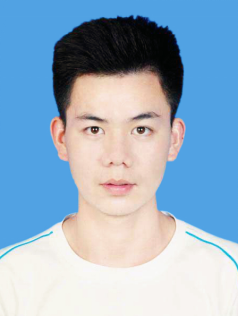}}]{Shaolong Guo}
is working on his Ph.D degree with the School of Cyber Science and Engineering of Xi'an Jiaotong University, Xi'an, China. His research interests include security and privacy protection in deep learning, semantic communications, and network games.
\end{IEEEbiography}

\begin{IEEEbiography}[{\includegraphics[width=1in,height=1.25in,clip,keepaspectratio]{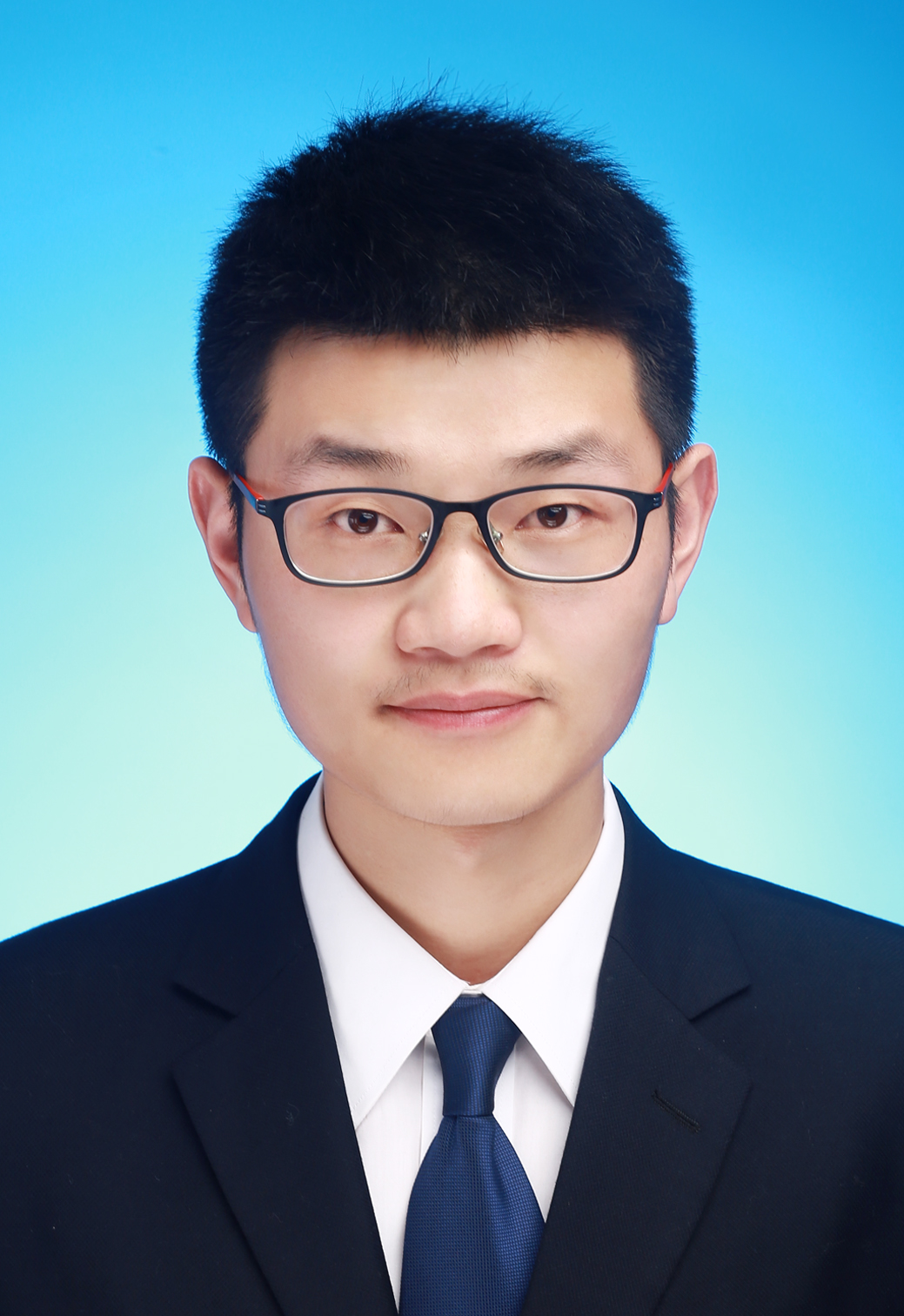}}]{Yuntao Wang} (Member, IEEE) received the Ph.D degree in Cyberspace Security from Xi'an Jiaotong University, Xi'an, China, in 2022, where he is currently an Assistant Professor with the School of Cyber Science and Engineering. His research interests include security and privacy in intelligent IoT, network games, and blockchain.
\end{IEEEbiography}

\begin{IEEEbiography}[{\includegraphics[width=1in,height=1.25in,clip,keepaspectratio]{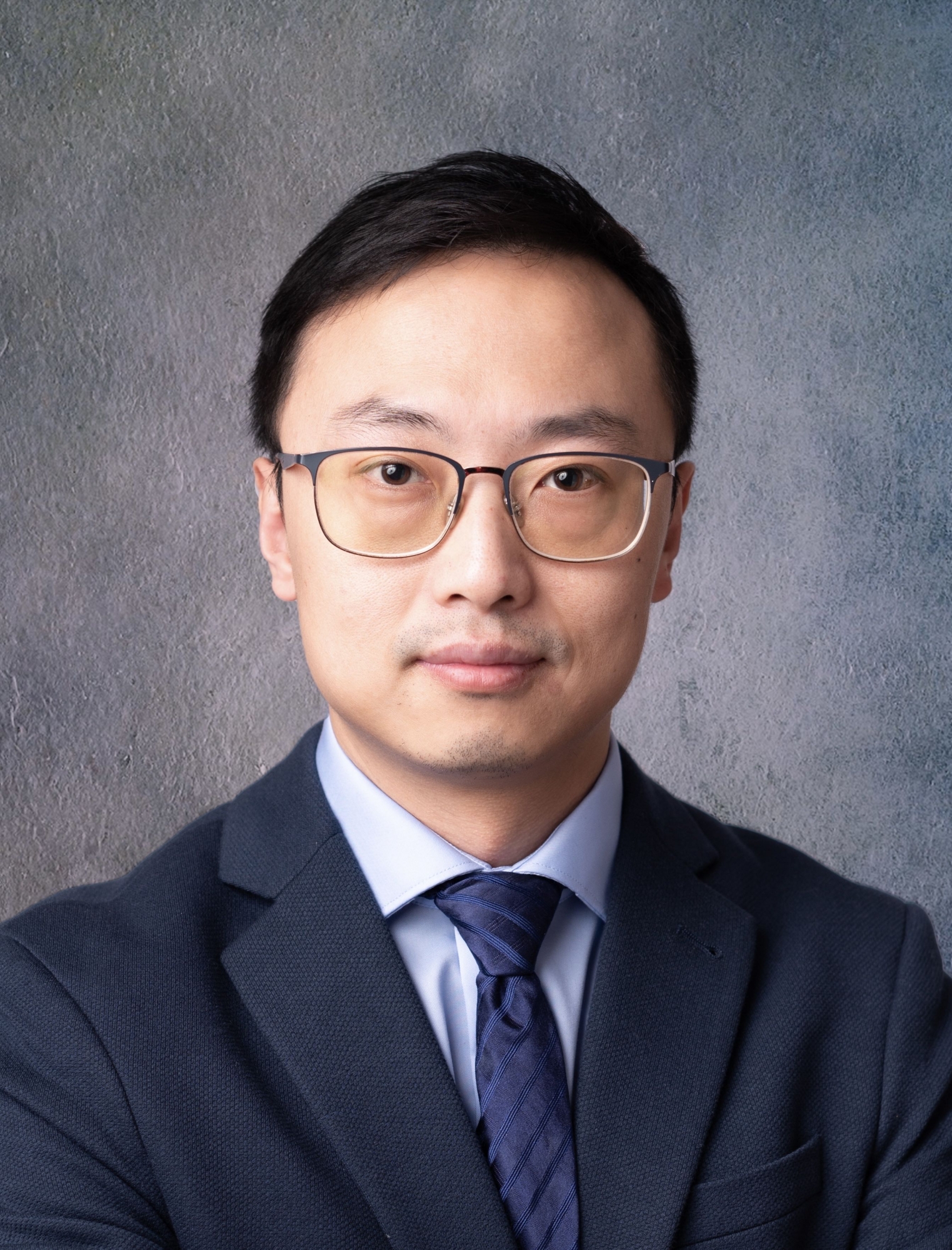}}]{Ning Zhang} (Senior Member, IEEE) received the Ph.D. degree in Electrical and Computer Engineering from the University of Waterloo in 2015. He is currently an Associate Professor and the Canada Research Chair of the Department of Electrical and Computer Engineering, University of Windsor. His research
interests include connected vehicles, wireless networking, and security. He is also a Distinguished Lecturer of IEEE ComSoc, a Highly Cited Researcher (Web of Science), and the Vice Chair of IEEE Technical Committee on Cognitive Networks and IEEE Technical Committee on Big Data. He serves/served as an Associate Editor for {\scshape IEEE Transactions on Mobile Computing}, {\scshape IEEE Communications Surveys And Tutorials}, {\scshape IEEE Internet of Things Journal}, and {\scshape IEEE Transactions on Cognitive Communications And Networking}. He also serves/served as the TPC/General/Symposium Chair for numerous conferences and workshops, such as {\scshape IEEE ICC}, {\scshape GLOBECOM}, {\scshape VTC}, {\scshape INFOCOM Workshop}, and {\scshape MOBICOM Workshops}.
\end{IEEEbiography}

\begin{IEEEbiography}[{\includegraphics[width=1in,height=1.25in,clip,keepaspectratio]{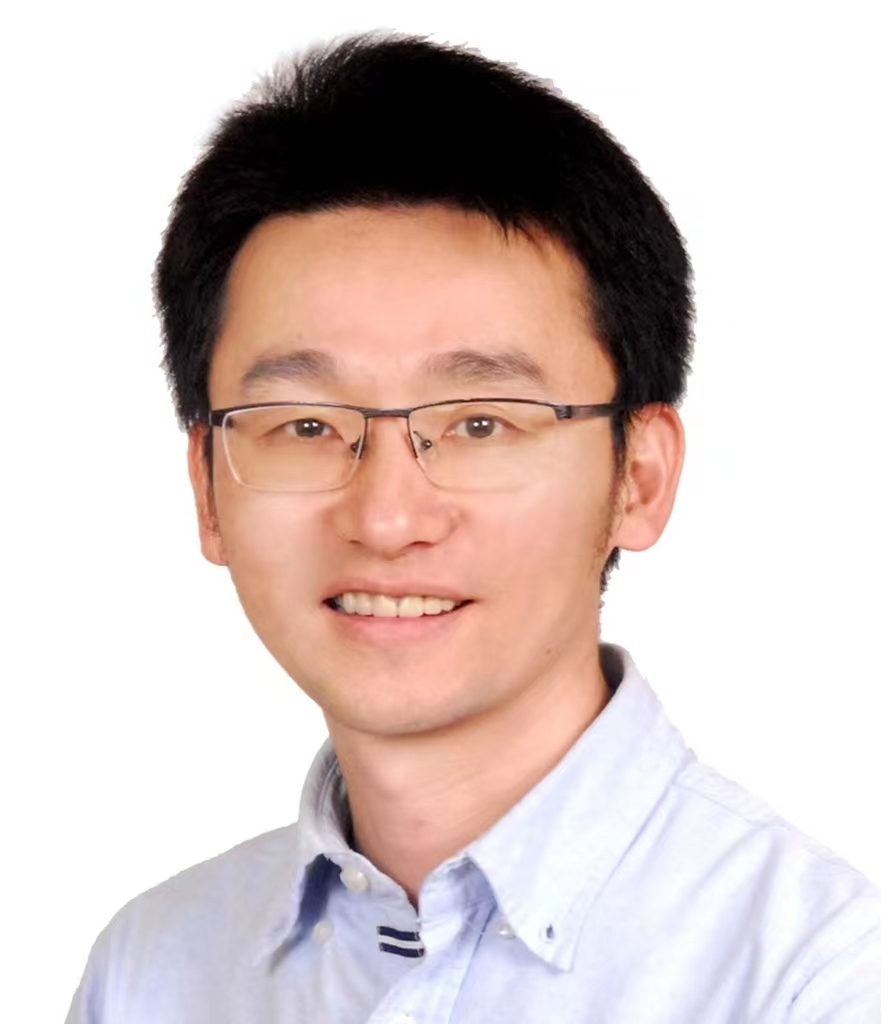}}]{Zhou Su} (Senior Member, IEEE)
has published technical papers, including top journals and top conferences, such as {\scshape IEEE Journal on Selected Areas in Communications}, {\scshape IEEE Transactions on Information Forensics and Security}, {\scshape IEEE Transactions on Dependable and Secure Computing}, {\scshape IEEE Transactions on Mobile Computing}, {\scshape IEEE/ACM Transactions on Networking}, and {\scshape INFOCOM}. Dr. Su received the Best Paper Award of the International Conference IEEE ICC2020, IEEE BigdataSE2019, and IEEE CyberSciTech2017. He is an Associate Editor of {\scshape IEEE Internet of Things Journal}, {\scshape IEEE Open Journal of the Computer Society}, and {\scshape IET Communications}. \end{IEEEbiography}

\begin{IEEEbiography}[{\includegraphics[width=1in,height=1.25in,clip,keepaspectratio]{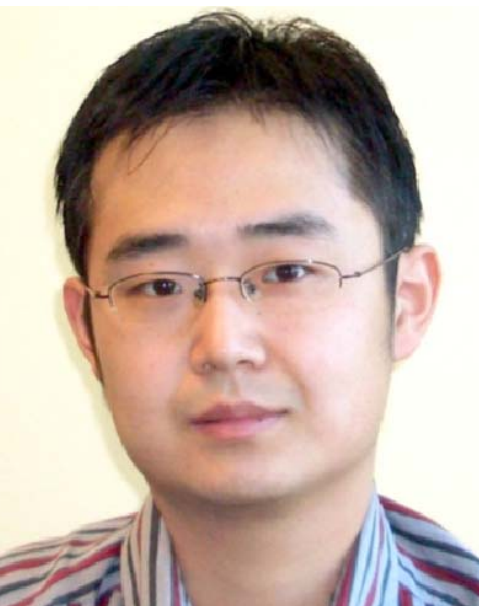}}]{Tom H. Luan} (Senior Member, IEEE) received the B.E. degree from the Xi'an Jiaotong University, China, the Master degree from the Hong Kong University of Science and Technology, Hong Kong, and the Ph.D. degree from the University of Waterloo, Canada, all in Electrical and Computer Engineering. During 2013 to 2017, Dr. Luan was a Lecturer in Mobile and Apps at the Deakin University, Australia. During 2017 to 2022, he was a Professor with the School of Cyber Engineering at Xidian University. Since 2022, he is with the School of Cyber Science and Engineering at Xi'an Jiaotong University, China, as a professor. Dr. Luan’s research mainly focuses on the LLM, AI-agent, Metaverse, and protocol and security design and performance evaluation of edge computing and digital twin network. Dr. Luan has published over 210 peer reviewed papers in journal and conferences, including {\scshape IEEE TON}, {\scshape TMC}, {\scshape TMM}, {\scshape TVT} and {\scshape INFOCOM}. He won the 2017 IEEE VTS Best Land Transportation Best Paper award and IEEE VTC 2023 best paper award.
\end{IEEEbiography}%\vspace{-5mm}

\begin{IEEEbiography}[{\includegraphics[width=1in,height=1.25in,clip,keepaspectratio]{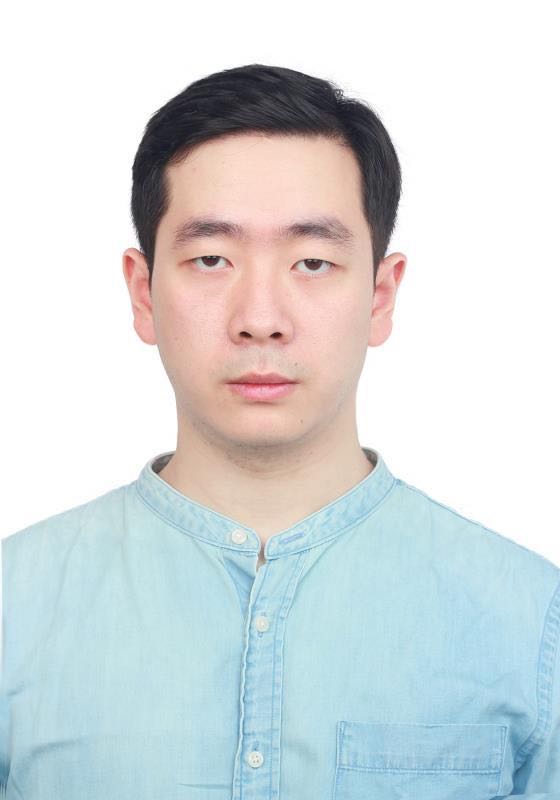}}]{Zhiyi Tian} (Member, IEEE) received the B.S. degree in information security and the M.S. degree in computer technology from Sichuan University, China, in 2017 and 2020, respectively. He received the Ph.D. degree in 2024 from University of Technology Sydney, Australia. His research interests include security and privacy issues in deep learning, semantic communications. He has been actively involved in the research community by serving as a reviewer for prestige journals, such as {\scshape ACM Computing Surveys}, {\scshape IEEE Communications Surveys and Tutorials}, {\scshape TIFS}, {\scshape TKDD}, and international conferences, such as {\scshape IEEE ICC} and {\scshape IEEE GLOBECOM}.
\end{IEEEbiography}%\vspace{-5mm}

\begin{IEEEbiography}[{\includegraphics[width=1in,height=1.25in,clip,keepaspectratio]{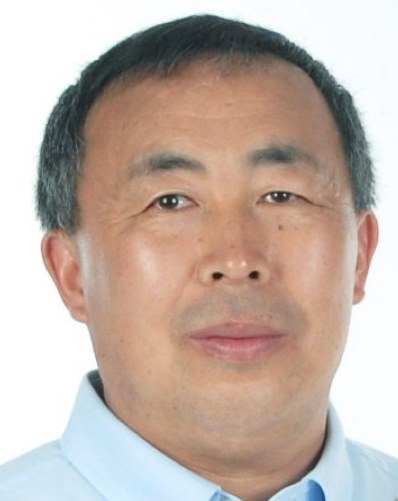}}]{Xuemin (Sherman) Shen}
(Fellow, IEEE) received the Ph.D. degree in electrical engineering from Rutgers University, New Brunswick, NJ, USA, in 1990. He is currently a University Professor with the Department of Electrical and Computer Engineering, University of Waterloo, Canada. His research interests include network resource management, wireless network security, the Internet of Things, AI for networks, and vehicular networks. He is a Registered Professional Engineer of Ontario, Canada, an Engineering Institute of Canada Fellow, a Canadian Academy of Engineering Fellow, a Royal Society of Canada Fellow, a Chinese Academy of Engineering Foreign Member, an International Fellow of the Engineering Academy of Japan, and a Distinguished Lecturer of the IEEE Vehicular Technology Society and Communications Society. He received the “West Lake Friendship Award” from Zhejiang Province in 2023, the President’s Excellence in Research from the University of Waterloo in 2022, Canadian Award for Telecommunications Research from Canadian Society of Information Theory (CSIT) in 2021, the R. A. Fessenden Award from IEEE, Canada in 2019, the Award of Merit from the Federation of Chinese Canadian Professionals (Ontario) in 2019, the James Evans Avant Garde Award from the IEEE Vehicular Technology Society in 2018, the Joseph LoCicero Award in 2015 and Education Award in 2017 from the IEEE Communications Society (ComSoc), and the Technical Recognition Award from the Wireless Communications Technical Committee in 2019 and the AHSNTechnical Committee in 2013. He also received the Excellent Graduate Supervision Award from the University of Waterloo in 2006 and the Premier’s Research Excellence Award (PREA) from the Province of Ontario, Canada, in 2003. He is the Past President of the IEEE Communications Society. He was the Vice President for Technical and Educational Activities, the Vice President for Publications, the Member-at-Large on the Board of Governors, the Chair of the Distinguished Lecturer Selection Committee, and a Member of the IEEE Fellow Selection Committee of the ComSoc. He served as the Editor-in-chief of {\scshape IEEE Internet of Things Journal}, {\scshape IEEE Network}, and {\scshape IET Communications}.

\end{IEEEbiography}

\end{document}